\documentclass[notitlepage,twocolumn,pre,tightenlines,superscriptaddress,showpacs,floatfix]{revtex4-1} 

\usepackage{amsmath,amssymb,amsfonts,latexsym}
\usepackage{ulem}
\usepackage{bbold}
\usepackage{graphicx}
\usepackage{epsfig}
\usepackage{color}
\usepackage[activate=normal]{pdfcprot}  
\usepackage{bm} 
\usepackage{psfrag}
\usepackage[caption = false]{subfig}

\newcommand{\be}{\begin{equation}}
\newcommand{\ee}{\end{equation}}
\newcommand{\ben}{\begin{equation*}}
\newcommand{\een}{\end{equation*}}
\newcommand{\ba}{\begin{eqnarray}}
\newcommand{\ea}{\end{eqnarray}}

\newcommand{\technionphy}{Department of Physics, Technion-IIT, 32000 Haifa, Israel}
\newcommand{\technionmech}{Department of Mechanical Engineering, Technion-IIT, 32000 Haifa, Israel}

\begin{document}

\title{Self-propulsion and self-navigation: Activity is a precursor to jamming}
\author{Mathias Casiulis}
\affiliation{\technionphy}
\email{casiulis@campus.technion.ac.il \\levine@technion.ac.il}
\author{Daniel Hexner}
\affiliation{\technionmech}
\email{danielhe@technion.ac.il}
\author{Dov Levine}
\affiliation{\technionphy}
\email{levine@technion.ac.il}

\date{\today}

\begin{abstract}
Traffic jams are an everyday hindrance to transport, and typically arise when many vehicles have the same or a similar destination.
We show, however, that even when uniformly distributed in space and uncorrelated, targets have a crucial effect on transport. 
At modest densities an instability arises leading to jams with emergent correlations between the targets.
By considering limiting cases of the dynamics which map onto active Brownian particles, we argue that motility induced phase separation is the precursor to jams.  That is, jams are MIPS seeds that undergo an extra instability due to target accumulation. This provides a quantitative prediction of the onset density for jamming, and suggests how jamming might be delayed or prevented.
We study the transition between jammed and flowing phase, and find that transport is most efficient on the cusp of jamming.
\end{abstract}

\maketitle

\section{Introduction}
The dynamics of autonomous entities navigating to specific positions in space is an important problem for understanding automobile traffic~\cite{KarlinPeres,Helbing2001,Orosz2010,Nakayama2016,Aoyama2020}, as well as collective navigation of animals~\cite{Benhamou2004,Benhamou2014}, pedestrians~\cite{Moussaid2009,Moussaid2012,Yajima2020}, or robots such as delivery drones~\cite{Nash2010,Standley2011,Morris2016,Ma2017,Surynek2020,Mai2020,Talamali2021}. 
Essentially, this problem is one of a large number of agents traveling, each to its own target while avoiding collisions, and trying to do this as efficiently as possible. 
As the density of the agents increases, direct motion and collision avoidance come into increasing conflict, which can lead to collective effects such as jamming.

Traffic jams can arise from different factors, both extrinsic and intrinsic. 
Extrinsic influences include poor road network geometry and topology or badly synchronized traffic regulation, which can be somewhat mitigated by careful planning (see for instance~\cite{Biham1992,KarlinPeres,Helbing2001}).
However, other factors may be intrinsic to the nature of dynamics, and it behooves us to examine how their effects can be reduced. 
The importance of intrinsic influences such as self-propulsion and seemingly benign interactions, is well-known in the field of active matter, where they may lead to phenomena like motility-induced phase separation~\cite{Fily2012} (MIPS).
It is tempting to ask whether these can also affect the nature of traffic flow, and perhaps even lead to phenomena such as so-called ``ghost jams"~\cite{Lighthill1955,Richards1956}.

In this paper we focus on phenomena of collections of self-propelled agents, each trying to get to its own target.
This setting contrasts with usual works on target searching, that consider particles whose dynamics are independent of the location of an absorbing boundary~\footnote{These dynamics are typically Brownian~\cite{Redner2001,Benichou2010,Grebenkov2015,Agranov2017,Agranov2019}, more recently advected~\cite{MejiaMonasterio2020} or self-propelled~\cite{Basu2018}.}, and with prey-predator models~\cite{Chakraborty2020}, in which all preys are equivalent to a predator.
It can rather be seen as an extension of crowd escape problems~\cite{Helbing2000}, but instead of navigating to a single location, agents all have distinct destinations.

Using numerical simulations, we obtain three main results.
\begin{itemize}
    \item First, we show that, even when targets are uniformly distributed and uncorrelated, they have a striking effect on transport, and may lead to jamming even at modest density.
    \item Second, we show that this jamming is an instability building up on top of Motility-Induced Phase Separation (MIPS), the phase separation observed in systems of self-propelled particles, even in the absence of explicit attraction~\cite{Fily2012,Cates2014}.
As small MIPS clusters form, they overlap with targets belonging to other particles.
This leads to a collective feedback whereby a growing cluster attracts more and more particles due to the growing number of targets it contains.
\item Finally, we show that the optimal strategy to reach targets in the shortest time possible changes discontinuously at the jamming density.
Below it, it is best to go straight to one's target, being deviated only by occasional collisions, while above it, one should instead deviate from the direct path, but just enough not to jam.
We report this feature using three different parameters of two different models, and suggest that being on the edge of jamming is generically the best strategy at high traffic densities.
\end{itemize}

In Sec.~\ref{sec:models}, we begin by defining two models of self-propelled and self-navigating particles in two dimensional continuous space.
We then characterise their phases in the noiseless case in Sec.~\ref{sec:noiseless}, which enables us to link the jamming transition to MIPS.
Introducing noise terms, in Sec.~\ref{sec:noisy}, we show that a finite amplitude of noise destroys jams, and that the value of the noise amplitude that makes target-reaching fastest changes from zero noise at low density, to the lowest value of the noise that unjams the system beyond it.
We then tune the degree of self-navigation of our model, and present their full phase diagram, that explicitly maps out MIPS and jamming.
We show that tuning the degree of self-navigation, just like varying the amplitude of noise terms, lead to optimal travel times on the cusp of the jamming transition at high densities, which suggests that it is a generic feature of self-navigating systems.
Finally, in Sec.~\ref{sec:Conclusions}, we present our conclusions, and  perspectives for self-navigating systems.

\section{Models\label{sec:models}}

We consider $N$ agents, placed in a 2D square box with side-length $L$ with periodic boundary conditions, with initial positions $\{\bm{r}_i(0)\}_{1\leq i\leq N}$ drawn uniformly in space.
To each agent we associate a ``target'' at a fixed position in space,  $\bm{r}_{T,i}$, also drawn uniformly.
When unhindered by other agents, each agent navigates at a speed $v_{0}$ towards its target.
As a result, agents are not only self-propelled, but also self-navigating, as their self-propulsion continuously re-orients towards a well-defined location.
Whenever an agent reaches its target, both are removed from the system, much like a car would exit the road network when parking.
At each such absorption event, we introduce a new agent-target pair with random positions drawn uniformly in space.

To avoid overlaps between them, we represent agents by particles that interact with a short range repulsive force $\bm {f}_{ij}\left(r\right)=k(a-r_{ji})\hat{\bm{r}}_{ji}$ when $(r\leq a)$; where $\hat{\bm{r}}_{ji}$ is the unit vector between the centers of the particles, $a$ is their repulsive diameters, and $k$ is the stiffness of repulsion.
For simplicity, we assume that targets are also disks of diameter $a$, that do not move once they are drawn.
Contact between a particle and its target then occurs when $\left\|\bm{r}_i(t) - \bm{r}_{T,i} \right\| = a$.

Finally, we allow the dynamics of particles to be noisy.
The noise term can either be random kicks in space that could be due to navigation on rough terrain for instance, or random kicks on the orientation of self-navigation, that could be due to imperfect geolocation. 
In order to study both noise terms separately, and to test the generality of our results, we introduce two separate models that both include self-propulsion at a constant velocity, a repulsive force and a noise term.
\\ \ \\
I. \textit{Homing Brownian Particles (HBPs): }
\begin{align}
    \dot{\bm{r}}_i &= v_{0}\hat{\bm{e}}(\theta_{i,T}) + \sum\limits_{j\neq i} \bm{f}_{ji} +\sqrt{2D_0}\bm{\xi}_{i} \label{eq:MinimalEOM}
\end{align}

\noindent II. \textit{Homing Active Brownian Particles (HABPs): }
\begin{align}
\dot{\bm{r}}_{i} & =v_{0}\hat{\bm{e}}(\theta_i)+\sum\limits _{j\neq i}\bm{f}_{ji}\nonumber \\
\dot{\theta_{i}} & =\omega_{r}\left(\theta_{i,T}-\theta_{i}\right)+\sqrt{2D_{r}}\xi_{i},\label{eq:HABPDynamics}
\end{align}
Here $\hat{\bm{e}}(\theta)$ is the unit vector pointing at an angle $\theta$, and ${\theta}_{i,T}$ is the angle from the agent $i$ to its target. 
In the first model the self-propulsion always points toward the target while in the second the direction is dynamic, relaxing towards the target at a rate $\omega_{r}$.
We note that the relaxation term in Eq.~(\ref{eq:HABPDynamics}) should be understood to be modulus $2\pi$, such that $\theta_{i,T}-\theta_{i}\in\left[-\pi,\pi\right[$.
This choice of relaxation term ensures that only $\theta_i = \theta_{i,T}$ is an equilibrium position for the angle, so that free particles cannot easily travel away from their destination.
In the two models the noise also enters differently, although both are taken to be of unit variance and zero mean.
$D_0$ and $D_r$ are translational and rotational diffusion constants that can be tuned to adjust the noise levels.

After rescaling space and time units according to $r \to r/a$ and $t \to t v_0 /a$, the dimensionless parameters that can be tuned are the number of particles $N$, the area fraction $\phi  \equiv N \pi a^2 / 4 L^2$ (identical for particles and targets), and the dimensionless force amplitude $\tilde{F} \equiv k a / v_0$.
We set $\tilde{F} = 1000 \gg 1$ so that particles essentially never overlap.
These models are reminiscent of minimal models of pedestrians trying to reach the exit of a room~\cite{Helbing2000}, only with distinct individual goals.

\section{Noiseless Case\label{sec:noiseless}}

\subsection{Homing Brownian Particles: A Jamming Transition\label{sec:noiselessHBPs}}

\begin{figure}[b]
    \centering
    \includegraphics[height = 0.46\columnwidth]{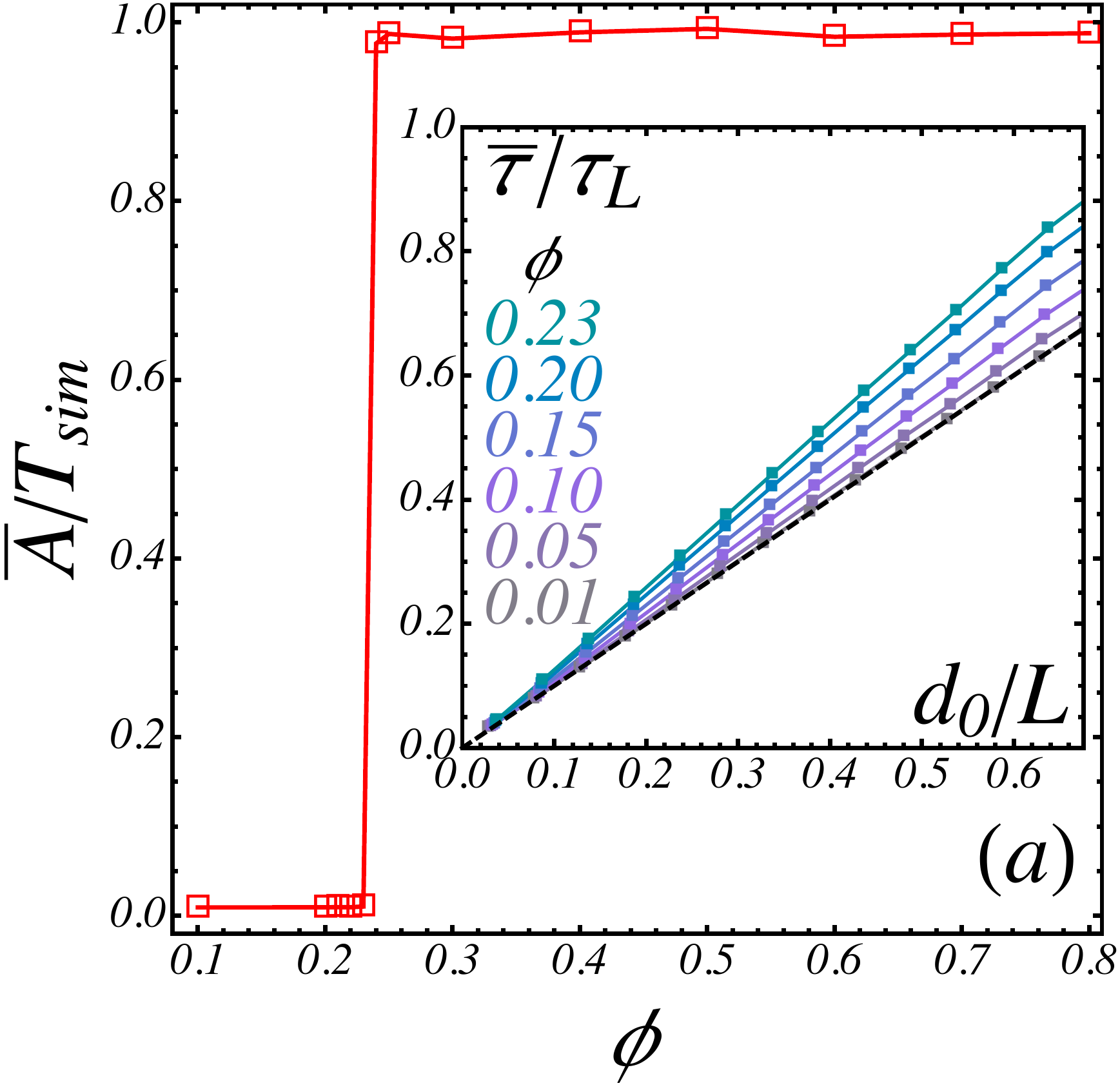}
    \includegraphics[height = 0.46\columnwidth]{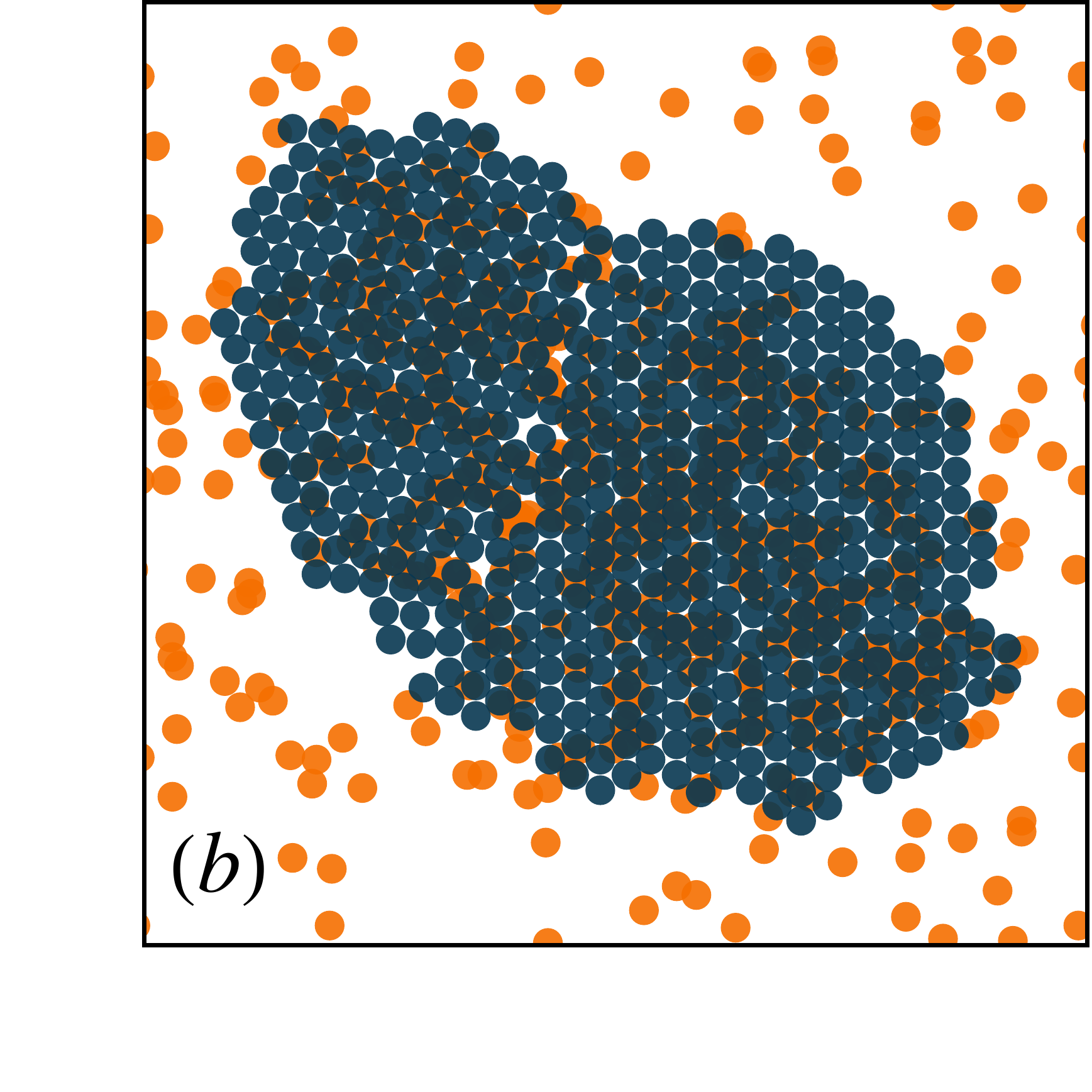}
    \caption{\textbf{Jamming transition.}
    $(a)$ Steady-state average age over time $\overline{A}/T_{sim}$ versus $\phi$ for $N = 2048$ particles in noiseless HBPs.
    Inset: Average arrival  time $\overline{\tau}$, divided by the ballistic arrival time $\tau_L = L / v_0$, against $d_0/L$, for a few densities. 
    $(b)$ Snapshot of a jam ($N = 512$, $\phi = 0.4$), with particles (blue) and targets (orange).
    }
    \label{fig:Fig1}
\end{figure}
To get a feel of the behavior, we first consider Model I in the absence of noise, by setting $D_0 = 0$.
In this model, particles always deterministically travel towards their targets, and are only deviated by collisions with other particles.
Simulating the system at different $\phi$, we observe two regimes shown in Fig.~\ref{fig:Fig1}$(a)$, where we plot the average age of particles $\overline{A}$ (\textit{i.e.} the time elapsed since their birth, either at $t=0$ or after an absorption/creation event) divided by the total simulation time $T_{sim}$ in steady state ($T_{sim} \sim 10^4$ in rescaled units) versus $\phi$.
This quantity tends to $0$ at long times if particles reach their targets in a finite time, and $1$ if they get stuck.
At low $\phi$, the system is in a fluid state where particles reach their targets in a finite time.
Above a critical density $\phi_J \approx 0.23$ the system ends up in traffic jams characterised by frozen dynamics and a dense crystalline structure, with no fluid phase coexisting with the jam.
An example snapshot of a jam is shown in Fig.~\ref{fig:Fig1}$(b)$, and videos of jamming events for larger systems ($N = 2048$ and $N = 32768$) are shown in the SI (see captions in App.~\ref{app:VideoCaptions}).

Inside traffic jams, targets accumulate, as they are shielded by the surrounding jammed particles. 
We note that $\phi_J$ is similar to the packing fraction of the lower branch MIPS in ABPs at zero noise~\cite{Patch2017,Bruss2018,VanDamme2019,Nie2020} and that, within numerical errors, we observed the jamming transition at $\phi_J \approx 0.23$ for system sizes up to $N=32768$ (see App.~\ref{app:FiniteSizeEffectsJamming}).

However, the MIPS observed in infinitely persistent ABPs is markedly different from the jams we describe here.
Indeed, our jams are dynamically frozen and crystalline at all densities larger than $\phi_J$, and surrounded solely by vacuum, while MIPS in zero-noise ABPs are slow but not dynamically frozen, and surrounded by a finite-density gas\footnote{ABPs also have a strictly frozen phase, but at packing fractions considerably higher than what we observe~\cite{Henkes2011,Digregorio2018,VanDamme2019,Mandal2020}. }.
This difference in phenomenology is tied to the fact that the present model is \textit{not} a usual model of infinitely persistent ABPs: self-propulsion here constantly changes its orientation towards the target.
We elucidate the precise link between MIPS and these jams in Sec.~\ref{sec:noiselessHABPs}.

In the inset of Fig.~\ref{fig:Fig1}$(a)$, we show the (normalized) average time to arrive at a target as a function of the initial separation of the particle from its destination.
In the fluid phase, the particles reach their targets ballistically, $\overline{\tau} \propto d_0$.
As $\phi$ increases, the average speed decreases, but remains finite up to the jamming transition.
The jamming transition is discontinuous, showing no structural or dynamical signature at densities $\phi < \phi_J$.

\subsection{Homing Active Brownian Particles: Jams Build up on MIPS\label{sec:noiselessHABPs}}

We now turn our attention to or Model II, Homing Active Brownian Particles, in the absence of noise.
The two dimensionless parameters of the problem are then the packing fraction $\phi$, and the dimensionless relaxation rate $\Omega_r \equiv \omega_r a / v_0$, which may be thought of as an angular ``stiffness".
In the limit of $\Omega_r \to 0$, the model reduces to a model of ABPs.
In ABPs, previous works have shown that the self-propulsion leads to an effective attraction~\cite{Farage2015,Cates2014,Solon2016,Turci2021} that causes MIPS beyond a critical value of the dimensionless activity parameter, usually called a (rotational) Péclet number, $Pe_{r}\equiv v_0 / (a D_r)$.
To understand the role of $\Omega_r$, we first establish the phase diagram of ABPs for our choice of repulsive interaction, which is shown in App.~\ref{app:ABPMIPS}.
A critical point is observed at $(Pe_{r,C},\phi_C) \approx (4.5, 0.55)$, and a phase separation domain grows from it at larger $Pe_r$, in agreement with previous results on ABPs with harmonic repulsion~\cite{Patch2017,Nie2020}.

\begin{figure}
    \centering
    \includegraphics[width =0.48\columnwidth]{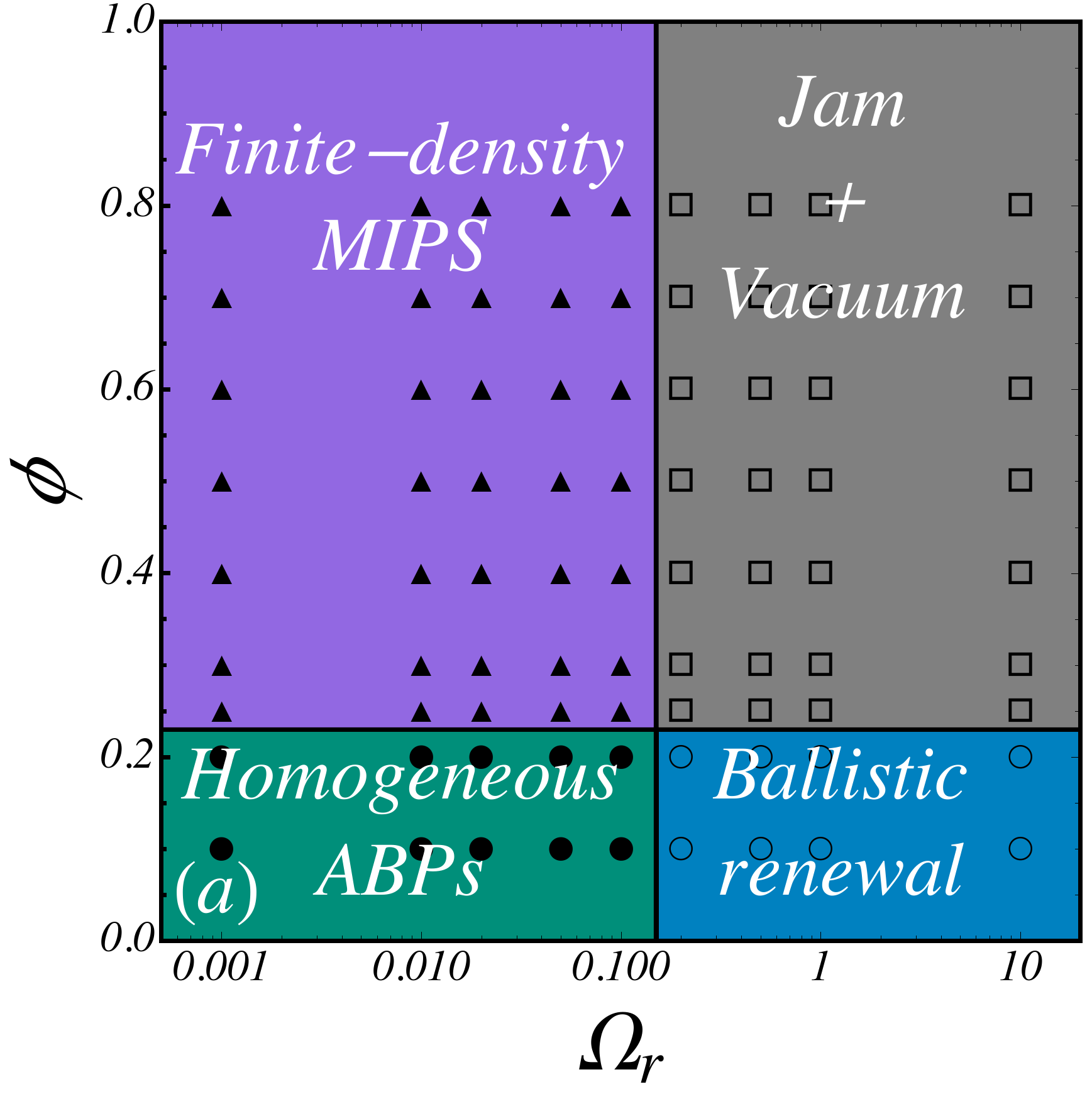}
    \includegraphics[width =0.48\columnwidth]{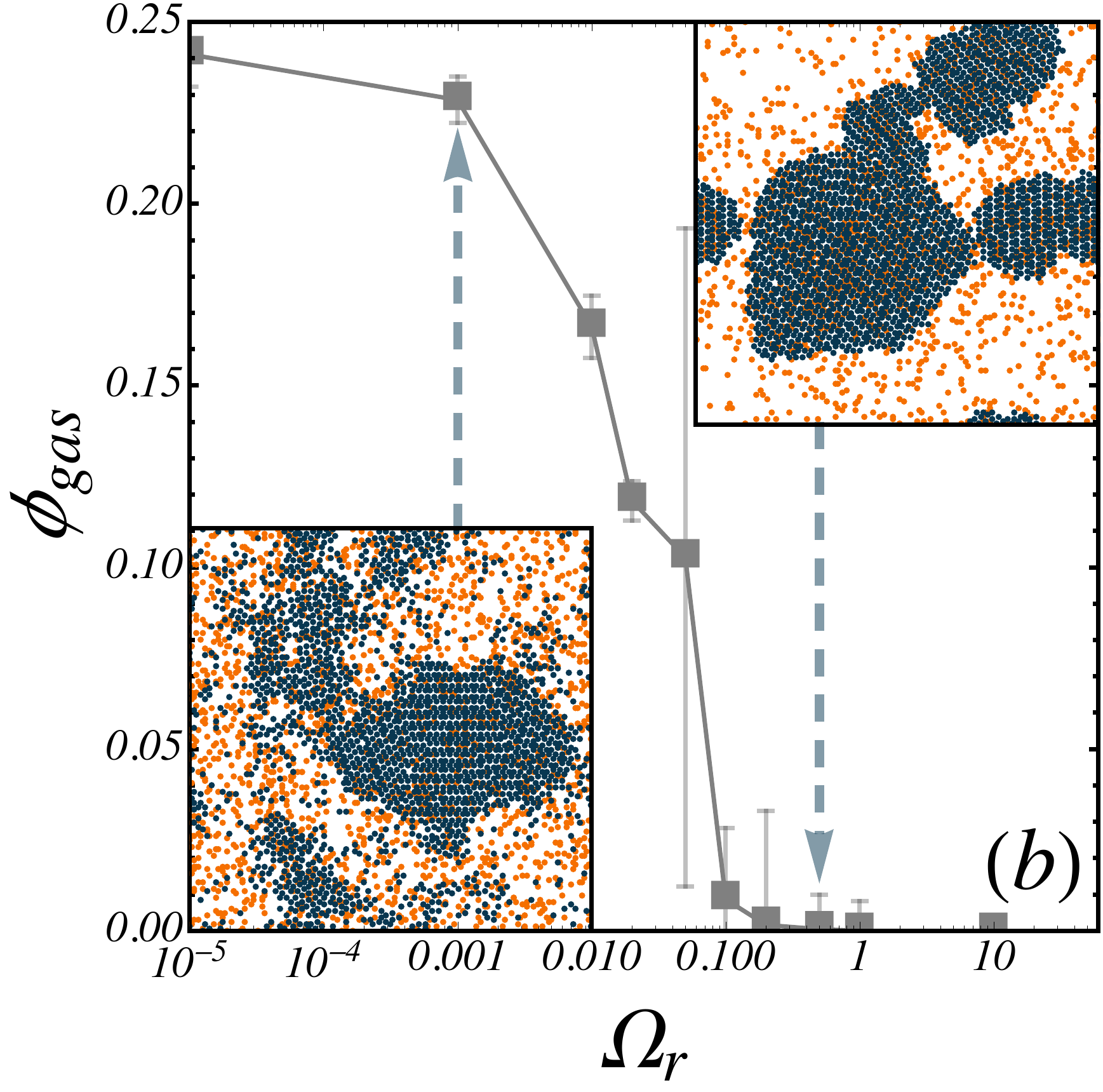}
    \caption{\textbf{From MIPS to jams.}
    $(a)$ Observed phases, in the $(\Omega_r,\phi)$ plane of noiseless HABPs.
    $(b)$ Packing fraction of the gas, $\phi_{gas}$ against the relaxation rate $\Omega_r$ at $\phi = 0.4$ (error bars: standard deviations across $10$ sets of $10^2$ snapshots).
    }
    \label{fig:Fig4}
\end{figure}

The lower branch of the phase separation domain of MIPS is known to saturate at a non-zero packing fraction $\phi_{gas}$ as $Pe_r \to \infty$~\cite{Reichhardt2014,Solon2015d,Bruss2018,Nie2020}.
This high-Péclet limit density seems to coincide with the jamming density of Fig.~\ref{fig:Fig1}$(a)$.
To probe the link between the two, we vary the relaxation rate $\Omega_r$ continuously.
In Fig.~\ref{fig:Fig4}$(a)$, we show the  phase diagram in the $(\Omega_r,\phi)$ plane.
For small $\Omega_r$, the system is equivalent to one of ABPs, and features a phase separation at high enough densities.
At densities high enough that MIPS is observed at slow relaxation rates, increasing $\Omega_r$ progressively empties the gas phase until it becomes vacuum, and leads to freezing of the dense phase.
This is illustrated in Fig.~\ref{fig:Fig4}$(b)$, where we show that the density of the gas, $\phi_{gas}$ goes to zero as $\Omega_r$ increases (illustrative snapshots are shown in the insets).
\begin{figure*}
    \centering
    \includegraphics[width=0.32\textwidth]{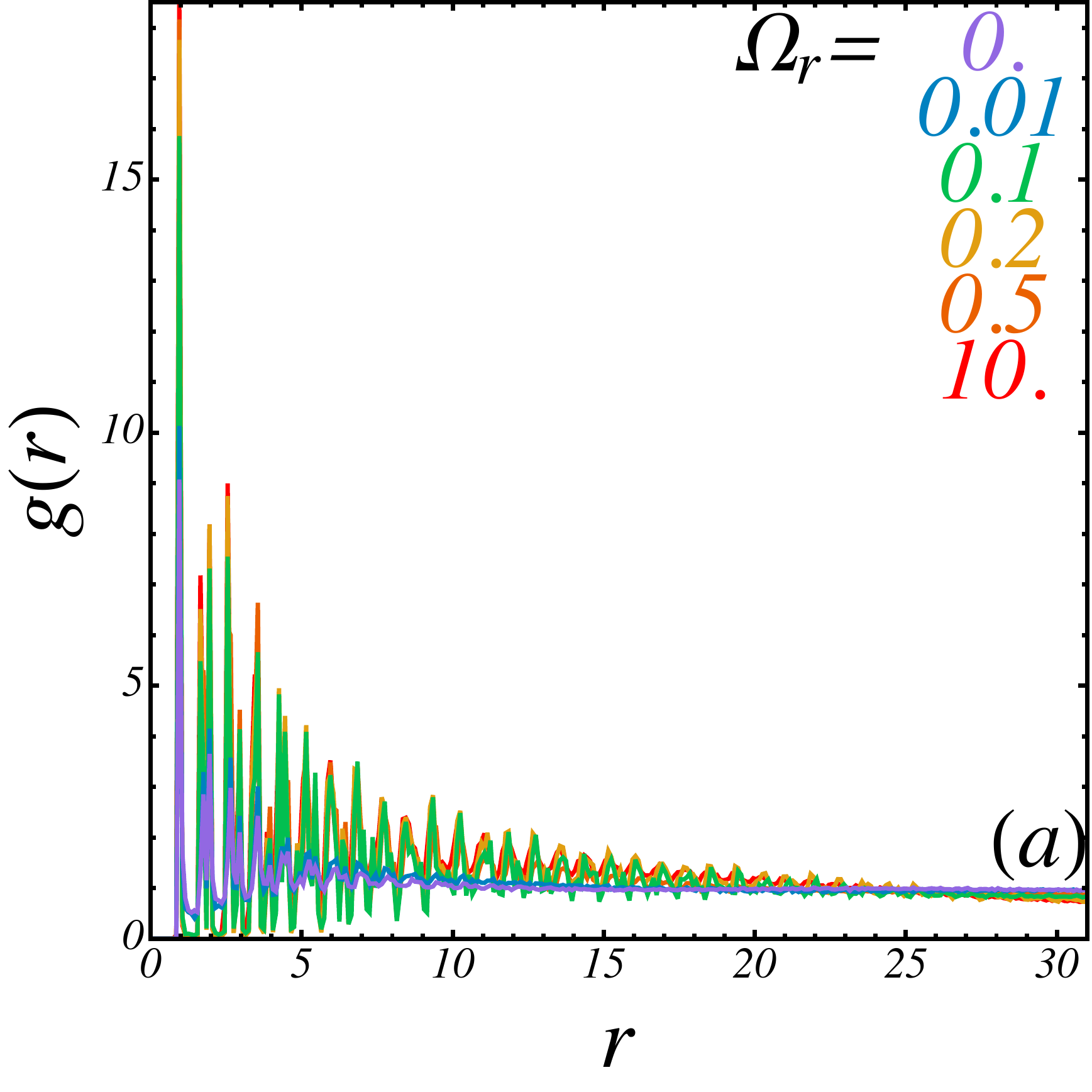}
    \includegraphics[width=0.32\textwidth]{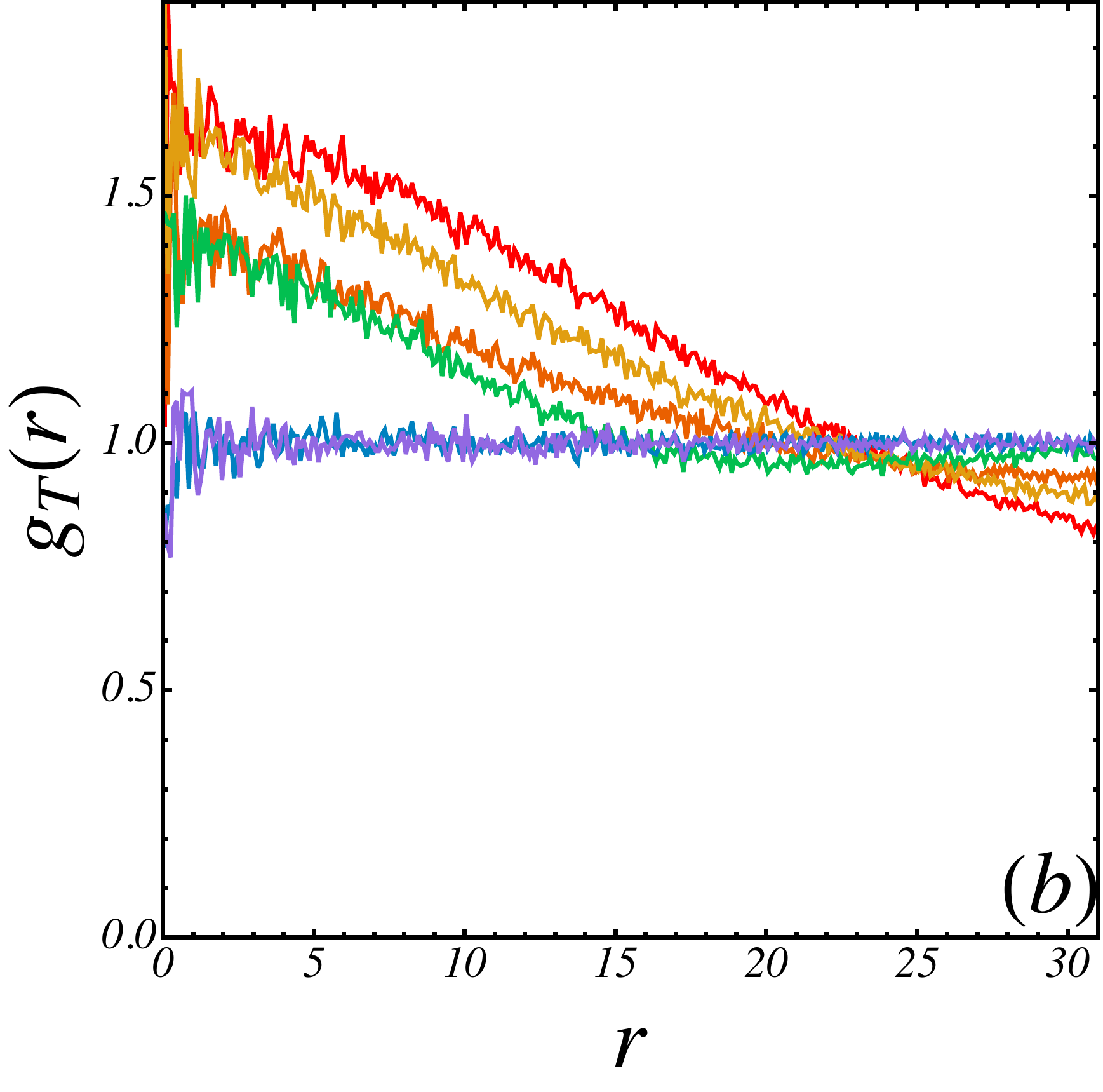}
    \includegraphics[width=0.32\textwidth]{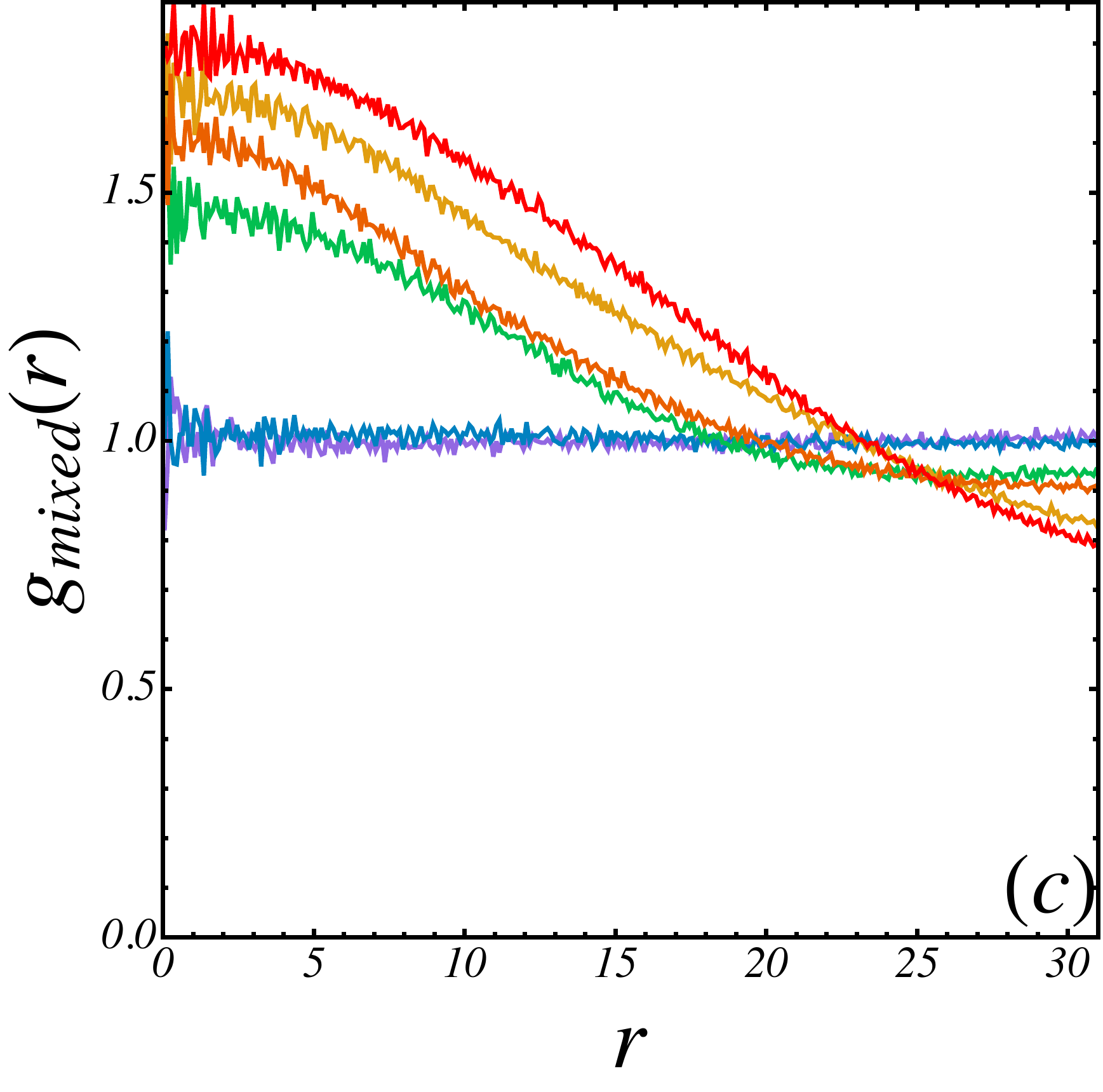}
    \caption{\textbf{Density-density correlations.}
    $(a)$ Particle-particle, $(b)$ target-target, and $(c)$ particle-target pair correlation functions in steady state, and for a few relaxation rates coded by the colour of each curve (see panel $(a)$).
    All curves were obtained for $\phi = 0.4$, $N = 2048$, and were averaged over 10 independent realisations of the steady state.
    Note the factor of $10$ in the vertical scale between panel $(a)$ and the other two panels.
    }
    \label{fig:crossgs}
\end{figure*}
This sheds light on the connection between the lower-branch of MIPS and the jams of Fig.~\ref{fig:Fig1}$(a)$: at high enough densities, particles start forming clusters, following the MIPS scenario.
As time goes by, the faster arrivals outside of clusters is a selective bias.
As accessible targets in the gas phase are reached faster, newly created targets become more likely to be drawn behind, or inside of clusters.
This attracts more particles to clusters, which therefore grow, thus becoming even more likely to hide other particles' targets.

One can check that targets do accumulate inside of clusters while the system jams, as shown in Fig.~\ref{fig:crossgs}.
In panel $(a)$, we plot the usual pair correlation function $g(r)$, in steady state averaged over 10 independent runs, for systems of $N = 2048$ particles at a packing fraction $\phi = 0.4$, but for various relaxation rates that cross the jamming value $\Omega_r \approx 0.1$.
This panel shows that jams are characterised by a very sharp ordering transition, with peaks that are both higher and longer-ranged than in the persistent ABP case $\Omega_r = 0$.
In panel $(b)$, we show the corresponding pair correlation functions $g_T(r)$ between targets.
We show that, while in the MIPS regime targets remain completely uncorrelated in steady state, in jams they become correlated at short range and anticorrelated beyond a range that roughly corresponds to the size of jams.
In other words, targets accumulate together, at a range similar to the extent of jams, which can be read from panel $(a)$.
Notice, however, that the amplitude of these correlations is much smaller than that of the correlations between particles: the accumulation of targets is far from being as sharp as that of particles.
Finally, in panel $(c)$, we plot a mixed particle-target pair correlation function.
This function, called $g_{mixed}(r)$, is computed following
\begin{align}
    g_{mixed} &\equiv \frac{1}{c_0}\sum\limits_{i,j = 1}^{N} \hat{\delta}\left(\left\|\bm{r}_i - \bm{r}_{T,j} \right\| - r\right),
\end{align}
where $\hat{\delta}$ is a binning function, and $c_0$ is a normalisation that ensures that $g_{mixed}(r) = 1$ for independent, uniform densities.
This function roughly has the same shape as the correlation between targets: it is completely flat in the MIPS regime, but develops positive correlations at short range and negative correlations at long range, with a typical range that is roughly the size of jams.
This evolution indicates that, not only are target accumulating together, they also typically accumulate at the same place as particles.
Like the correlation between targets, the amplitude of these correlations is much smaller than that of the correlations between particles.

Altogether, these density-density correlations prove that jam formation is associated to some accumulation of targets together inside of jams.
While the sharpest correlation is that between particles, the values and ranges of the target-target and particle-target correlations support the scenario that we propose according to which jams are MIPS seeds that undergo an extra instability due to target accumulation.
This scenario is further confirmed by comparing the formation of MIPS clusters to that of a jam starting from a uniform density (see App.~\ref{app:JammingvsMIPS}, and the video captioned in App.~\ref{app:VideoCaptions}).
The dynamics of ABPs and HABPs are indistinguishable at short times, until targets and particles start co-accumulating on MIPS-like clusters which then play the role of seeds to form jams.
To strengthen this hypothesis, we have checked that the MIPS and jamming densities are also similar in $3d$ (see App.~\ref{app:3dext}).
Note that the role of MIPS seeds in jamming is reminiscent of the fact that usual MIPS is also replaced by a more violent crystallization in the presence of randomly-placed hard obstacles, provided that the density of said obstacles is high enough~\cite{Reichhardt2014}: one might argue that, due to the presence of targets, MIPS seeds play the role of emergent still obstacles.

\begin{figure}
    \centering
    \includegraphics[height =0.64\columnwidth]{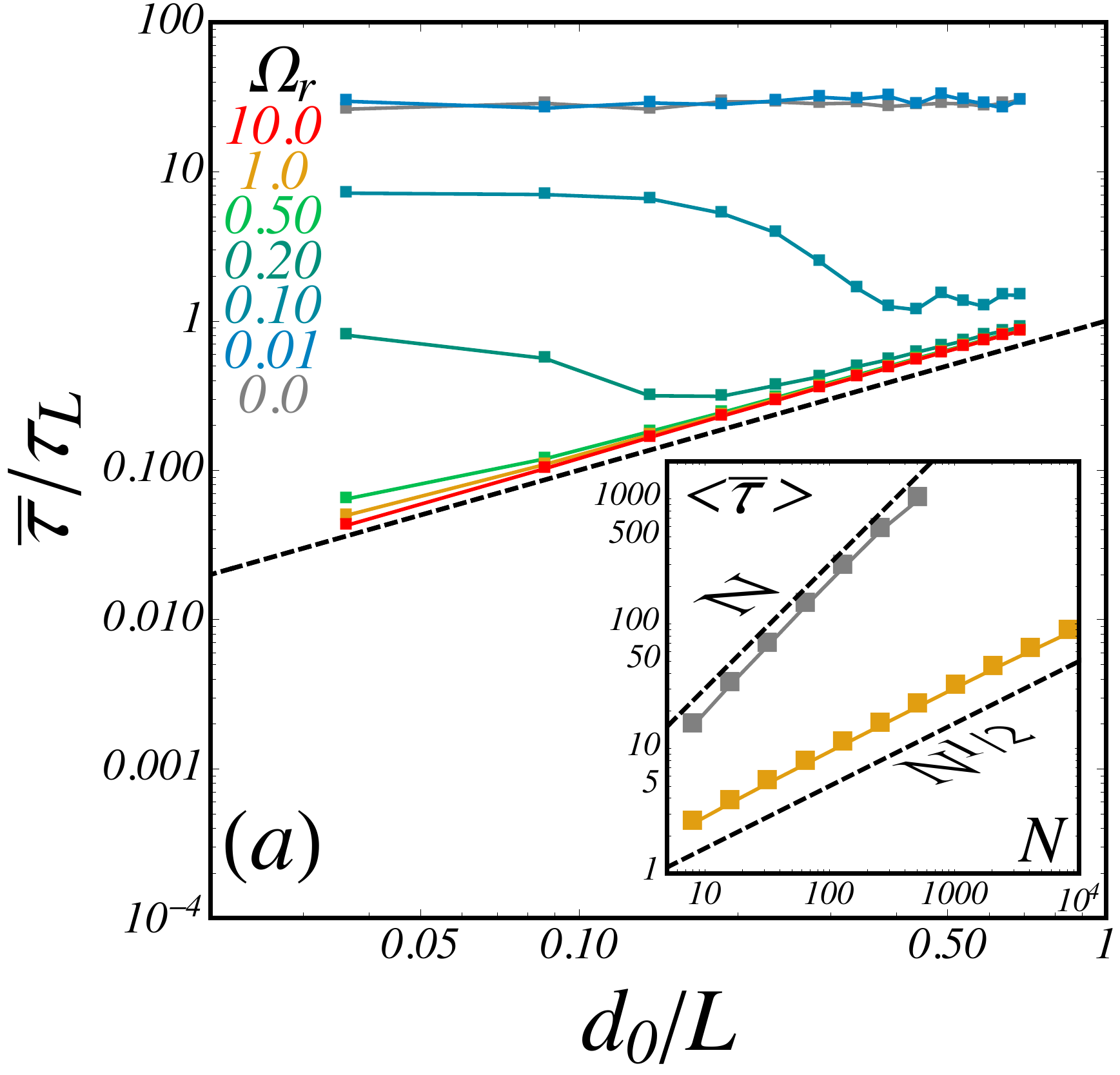}
    \includegraphics[height =0.64\columnwidth]{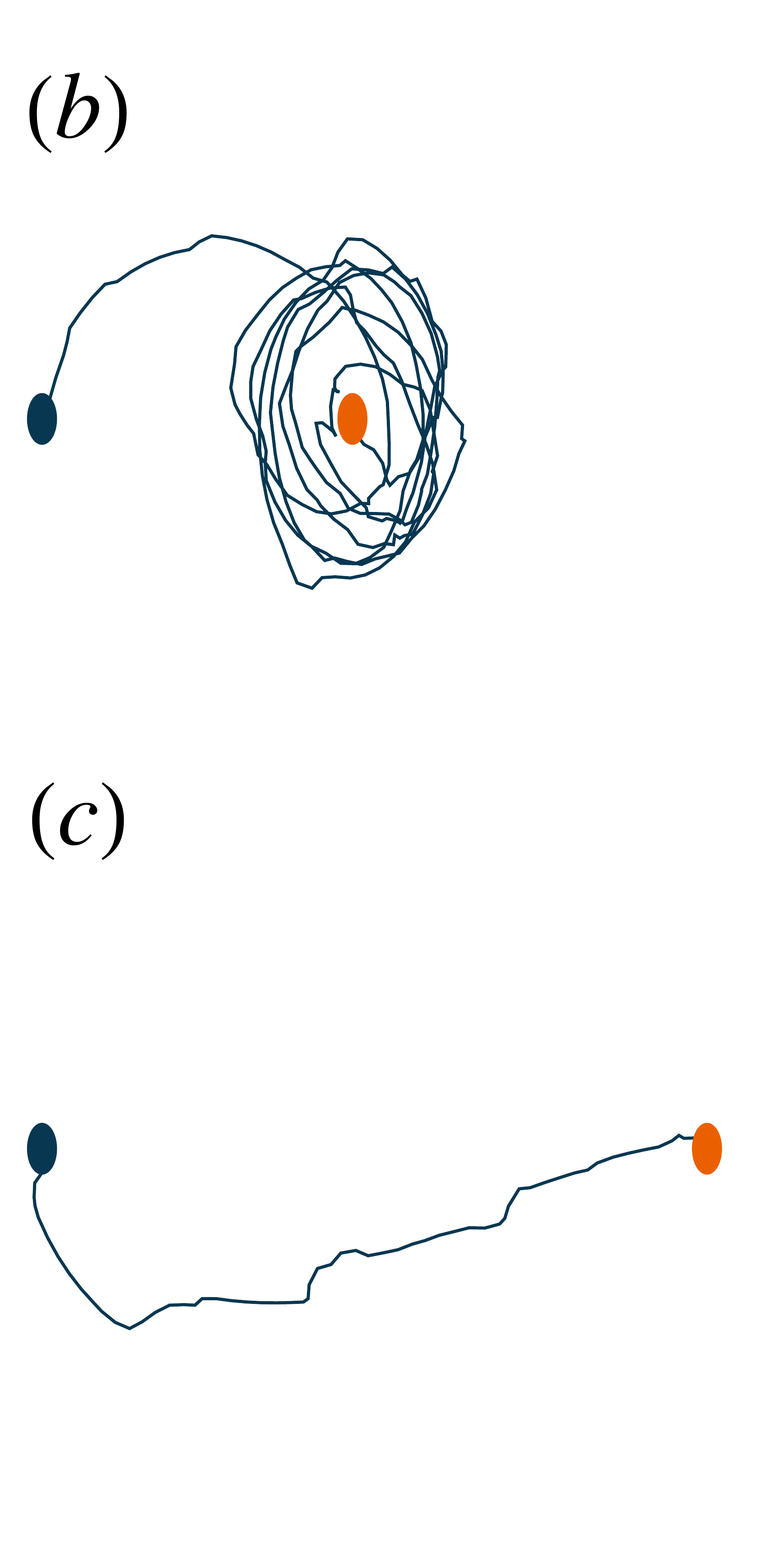}
    \caption{\textbf{Relaxation rate and target reaching.}
    $(a)$ $\overline{\tau}$, divided by $\tau_L = L/v_0$, against $d_0$ for $Pe_r = 1, \phi = 0.2$, $N=2048$ and a few $\Omega_r$, at zero noise amplitudes.
    Inset: Mean escape time $\langle\overline{\tau}\rangle$ (averaged over distances) against the number of particles $N$, in log-log scale, at $\phi = 0.2$ and for $\Omega_r = 0.0$ (gray) and $1.0$ (yellow). Dashed lines indicate the ballistic ($\overline{\tau} \propto L \propto N^{1/2}$) and diffusive ($\overline{\tau} \propto L^2 \propto N$) scalings.
    $(b)-(c)$ Sample trajectories from the ensemble used in $(a)$, obtained for $\Omega_r = 0.1$ and for $d_0 \approx 0.2 L$ $(b)$ and $d_0 \approx 0.5 L$ $(c)$. 
    The initial position of the particle is represented by a blue disk, its trajectory by a blue line, and its target by an orange disk.
    }
    \label{fig:Zeronoiserelaxtimes}
\end{figure}
How does the relaxation rate $\Omega_r$ impact on the ability to reach targets?  
In panel $(a)$ of Fig.~\ref{fig:Zeronoiserelaxtimes}, we show the mean arrival times as a function of the corresponding initial distances in the case of noiseless HABPs.
Fast relaxation yields ballistic renewal at all distances.
As $\Omega_r$ decreases, the arrival times start increasing from the short-distance side: this is a sign that particles need a finite time to re-orient towards their target.

As a result, if a particle ends up close to its target but with its self-propulsion vector pointing off-target, for instance due to a collision, it is likely to overshoot, end up behind the target, and thus to end up in a spiraling trajectory converging to the target.
This is illustrated by a sample trajectory in Fig.~\ref{fig:Zeronoiserelaxtimes}$(b)$.
If a particle starts far away from its target, however, it has ample time to bring its self-propulsion vector back on track and to follow an almost straight line to its target, as shown in Fig.~\ref{fig:Zeronoiserelaxtimes}$(c)$.
At very low values of $\Omega_r$ and in finite size, particles collide much faster than they can re-orient, and essentially never manage to point towards their target.
Arrival times are then much longer, and almost independent of the initial distance.
This suggests that particles sweep the whole box before reaching their targets, which takes a typical time $\tau_{box} \sim L^2/v_0 a$.
This is confirmed in the inset of Fig.~\ref{fig:Zeronoiserelaxtimes}$(a)$ the mean time $\langle\overline{\tau}\rangle$, where $\langle \cdot\rangle$ stands for averaging over initial distances to targets, against the system size $N$. 
We indeed find that $\langle\overline{\tau}\rangle\sim N \sim L^2$ in the ABP case while $\langle\overline{\tau}\rangle\sim N^{1/2} \sim L$ for fast relaxation rates.

\begin{figure}[b]
    \centering
    \includegraphics[width=0.48\textwidth]{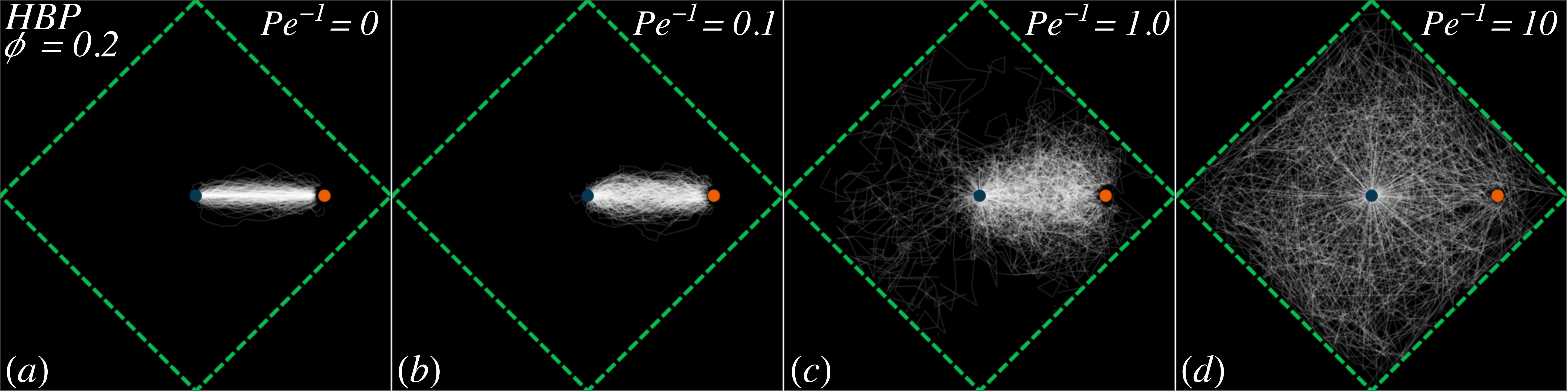}\\
    \includegraphics[width=0.48\textwidth]{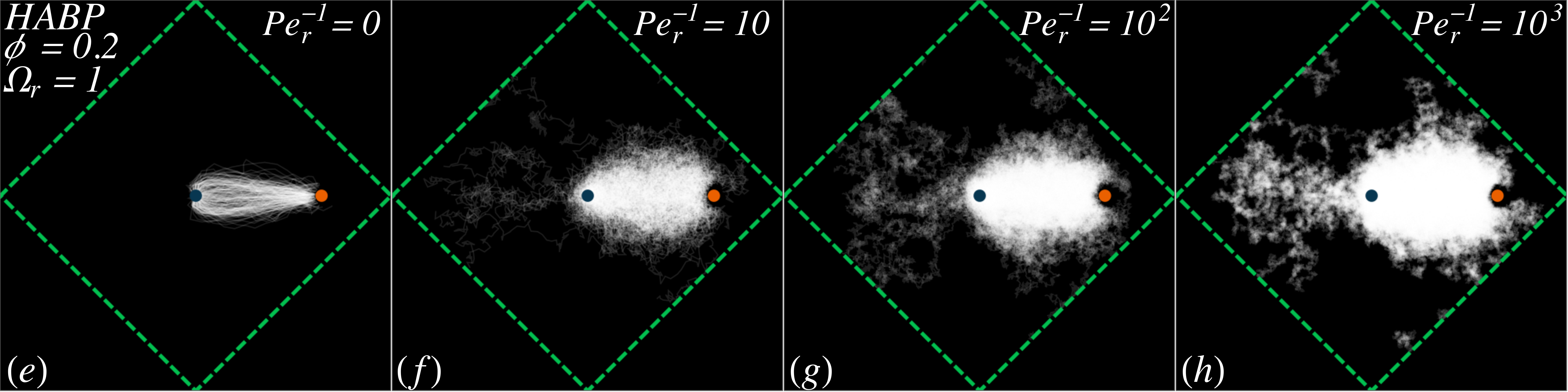}\\
    \includegraphics[width=0.48\textwidth]{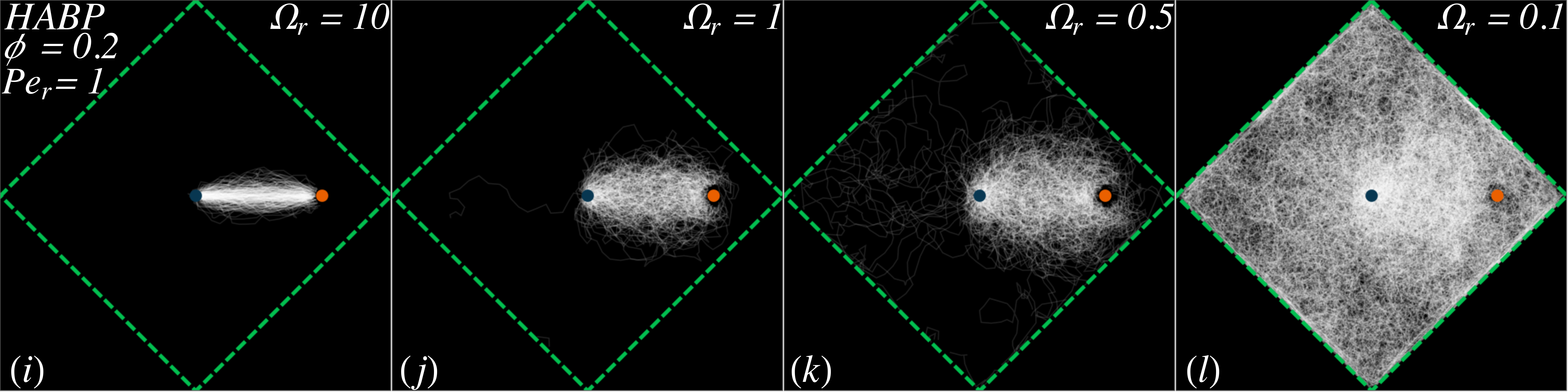}
    \caption{\textbf{Sample trajectories from initial position to target.}
    In this figure, $N = 128$ particles and $\phi = 0.2$.
    In each panel, we show $256$ trajectories, selected so that the target is initially at $d_0 = L/2 - a \pm a/2$.
    Each trajectory (white) is re-centered on its place of birth (blue disk), and rotated so that the target (orange disk) lies on the $x>0$ semi-axis.
    A green dashed line shows the bounds of the periodic simulation box.
    Trajectories are plotted with a finite transparency, so that brighter regions correspond to longer residency times.
    $(a)-(d)$ HBP trajectories for a few values of $Pe$.
    $(e)-(h)$ HABP trajectories at a fixed $\Omega_r = 1$ for a few values of $Pe_r$.
    $(i)-(l)$ HABP trajectories at a fixed $Pe_r = 1$ for a few values of $\Omega_r$.}
    \label{fig:Fig2}
\end{figure}
\section{Effects of Noise\label{sec:noisy}}

We now explore the influence of noise and relaxation on the jamming transition, and on the typical times to reach targets. 
Using the same rescalings as before, there are three independent dimensionless parameters: the Péclet number $Pe = v_0 a/D_0$, its rotational equivalent $Pe_r = v_0/(a D_r)$, and the dimensionless relaxation rate $\Omega_r=\omega_r a / v_0$.
The effect of changing parameters on individual trajectories is shown in Fig. \ref{fig:Fig2}, where in all cases $\phi = 0.2 < \phi_J$. For weak noise or large $\Omega_r$, the agents proceed to their targets in a rather direct fashion, with their trajectories broadening slightly as noise increases or $\Omega_r$ decreases.
Ultimately, when the noise is large or $\Omega_r$ is small, the trajectories become diffusive in nature (see App.~\ref{app:IndividualTrajectories} for details on the construction of the figure).

\subsection{Effects of noise terms}

\begin{figure}
    \centering
    \includegraphics[width =0.46\columnwidth]{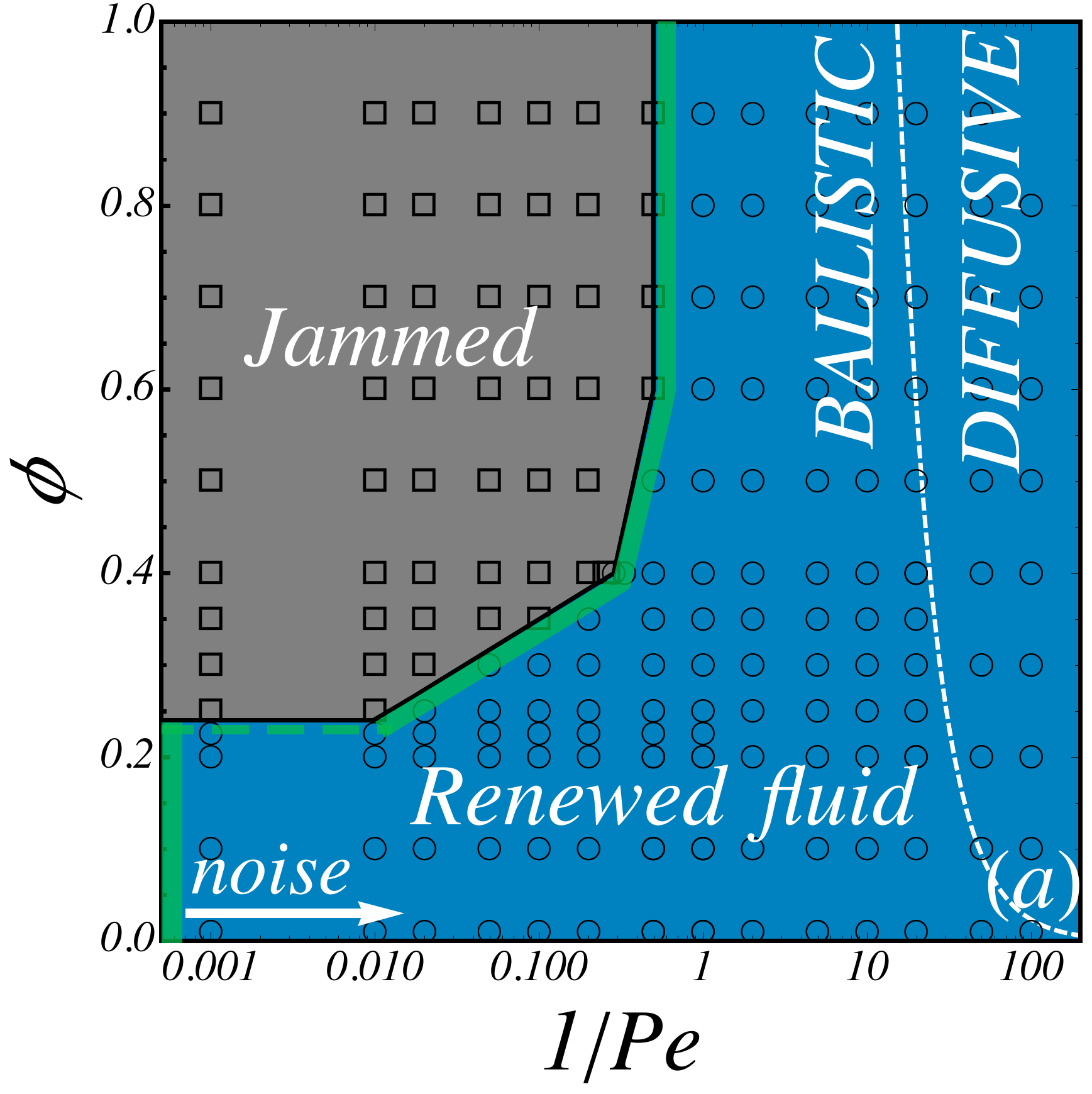}
    \includegraphics[width =0.46\columnwidth]{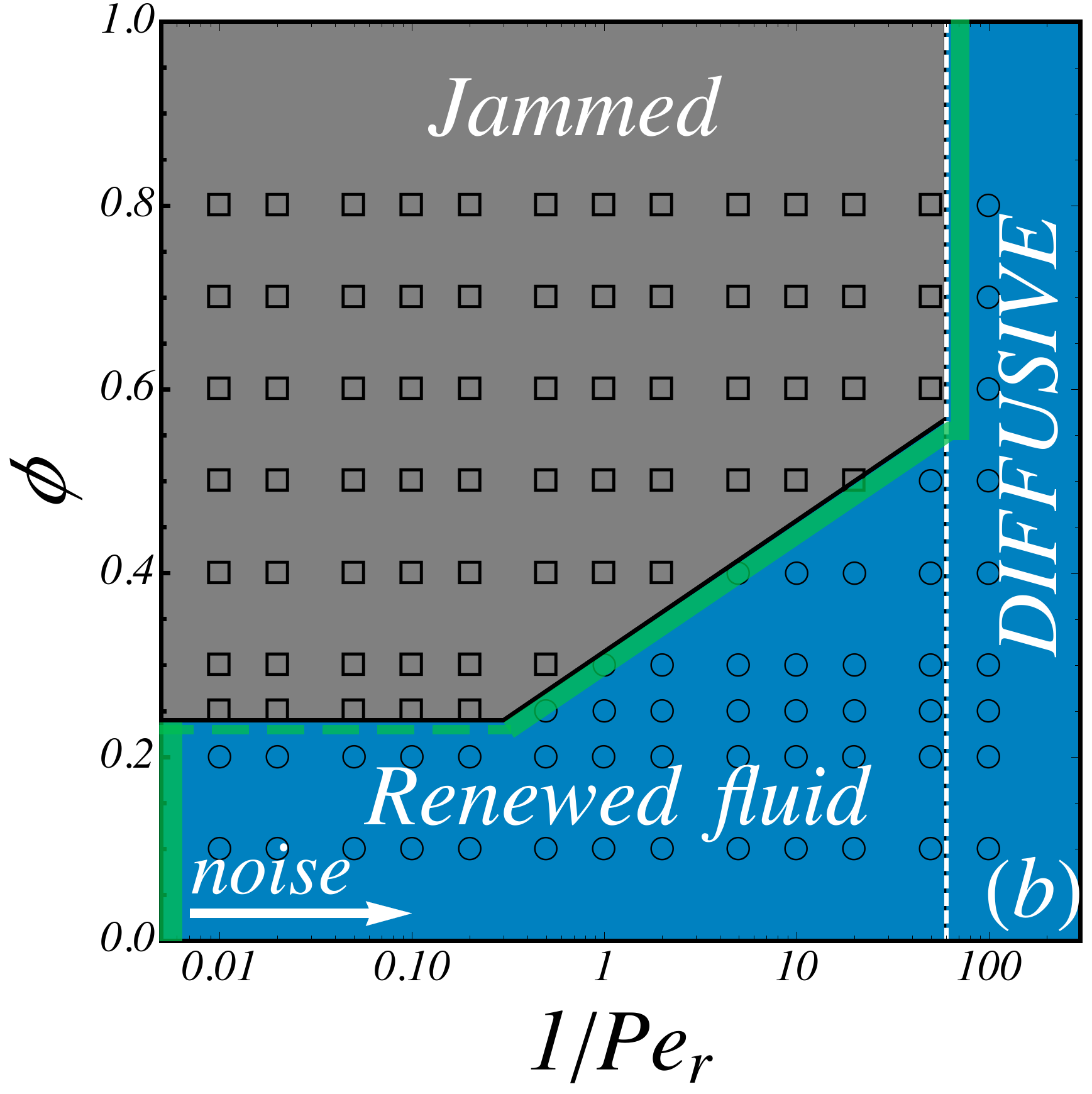}
    \vspace{-3mm}
    \caption{\textbf{Melting jams with noise.}
    $(a)$ Observed states in the $(1/Pe, \phi)$ plane for HBPs.
    Empty squares: Jammed states. Empty circles: Renewed fluid states.
    The white line indicates $\overline{d_0^2}/D_0 = \overline{d_0}/v_0$.
    Green lines highlight the fastest ballistic arrivals at each density (see Fig.~\ref{fig:FigOpti}).
    $(b)$ Same as $(a)$ but for HABPs at $\Omega_r = 1$, and tuning the rotational noise.
    The dashed white line indicates empirical observation of diffusive renewal.
    }
    \label{fig:Fig3}
\end{figure}
We first focus on noise terms.
In Fig.~\ref{fig:Fig3}$(a)$, we show the effect of finite noise on the phases of HBPs (Model I), by mapping out the different regimes seen in the $(Pe^{-1},\phi)$ plane.
We see that at high enough values of the noise, $Pe^{-1} \geq Pe_c^{-1} \sim 0.3$, the jam-vacuum states are destroyed and replaced by a homogeneous fluid of renewed particles, which is reminiscent of the melting of granular jams~\cite{Liu1998}.
In Fig.~\ref{fig:Fig3}$(b)$, we show the equivalent diagram obtained for HABPs (Model II), when setting the relaxation rate to $\Omega_r = 1.0$ and tune $1/Pe_r$.
This diagram is very similar to Fig.~\ref{fig:Fig3}$(a)$: beyond a finite noise amplitude $1/Pe_r \approx 50$, the steady state is not a jam but a homogeneous fluid, in which particles reach their targets in finite times.
We check in App.~\ref{app:FiniteSizeEffectsJamming} that this unjamming transition with noise is preserved in larger systems.

\begin{figure}
    \centering
    \includegraphics[width =0.46\columnwidth]{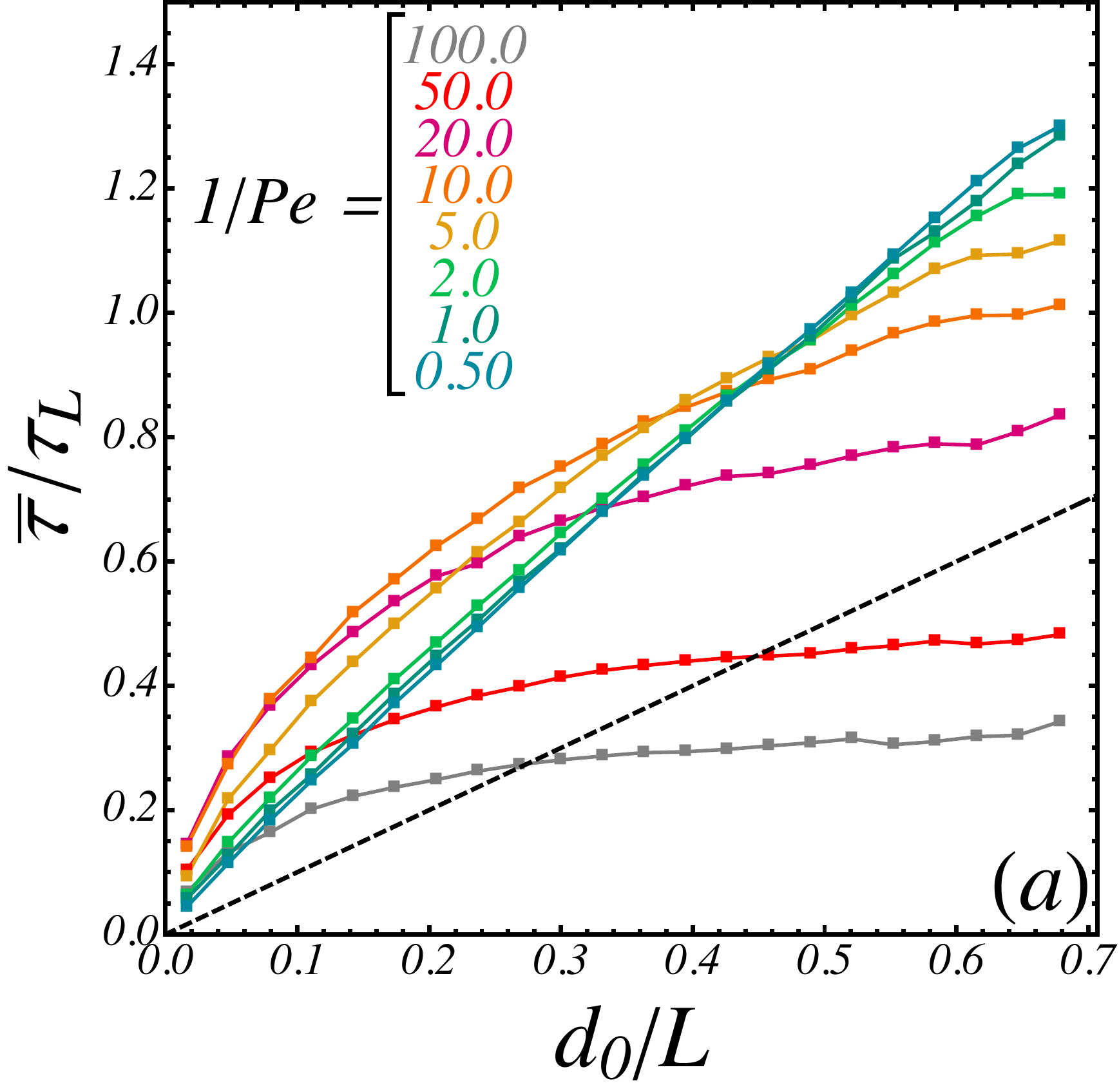}
    \includegraphics[width =0.46\columnwidth]{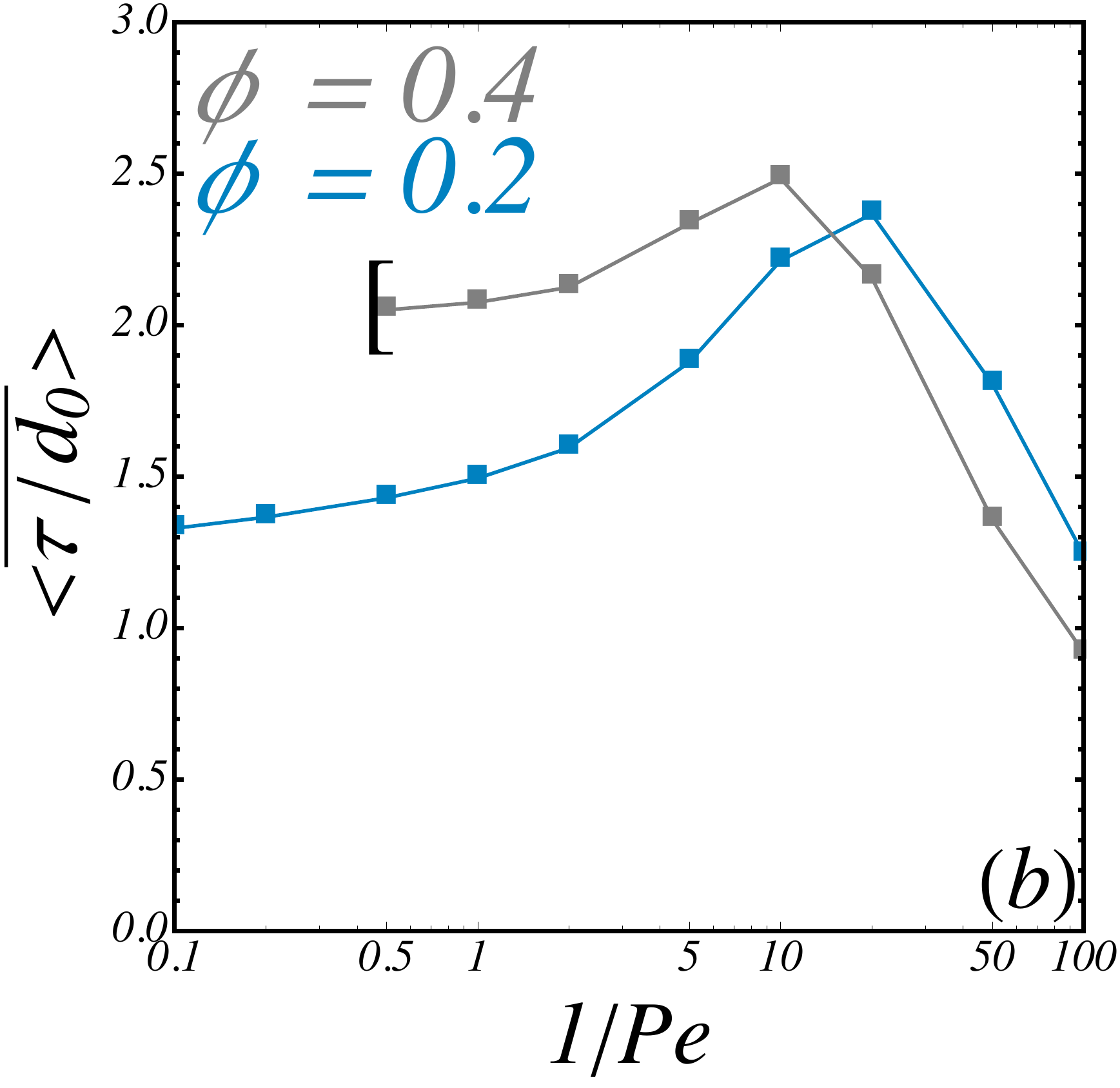} \\
    \includegraphics[width =0.46\columnwidth]{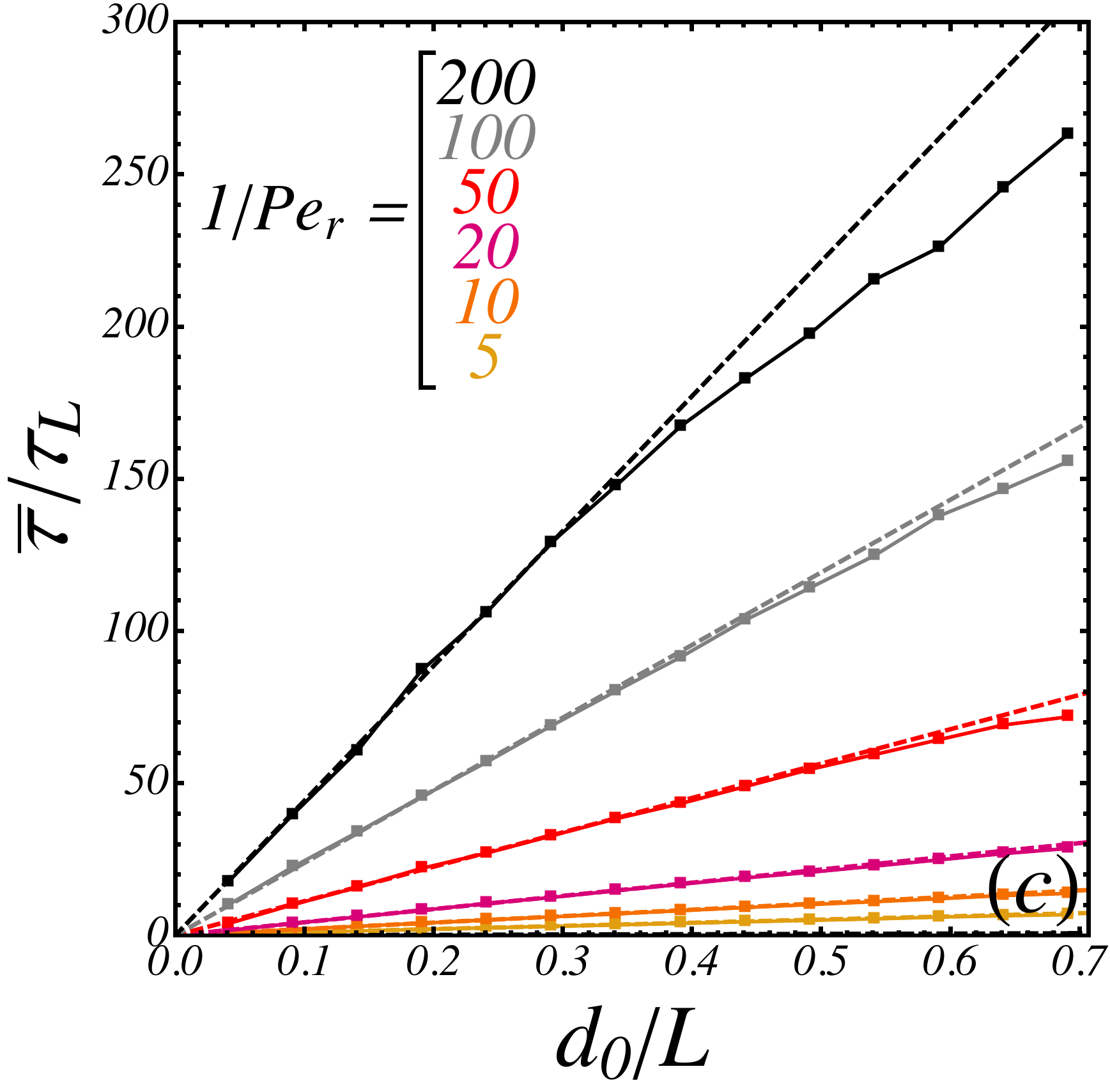}
    \includegraphics[width =0.46\columnwidth]{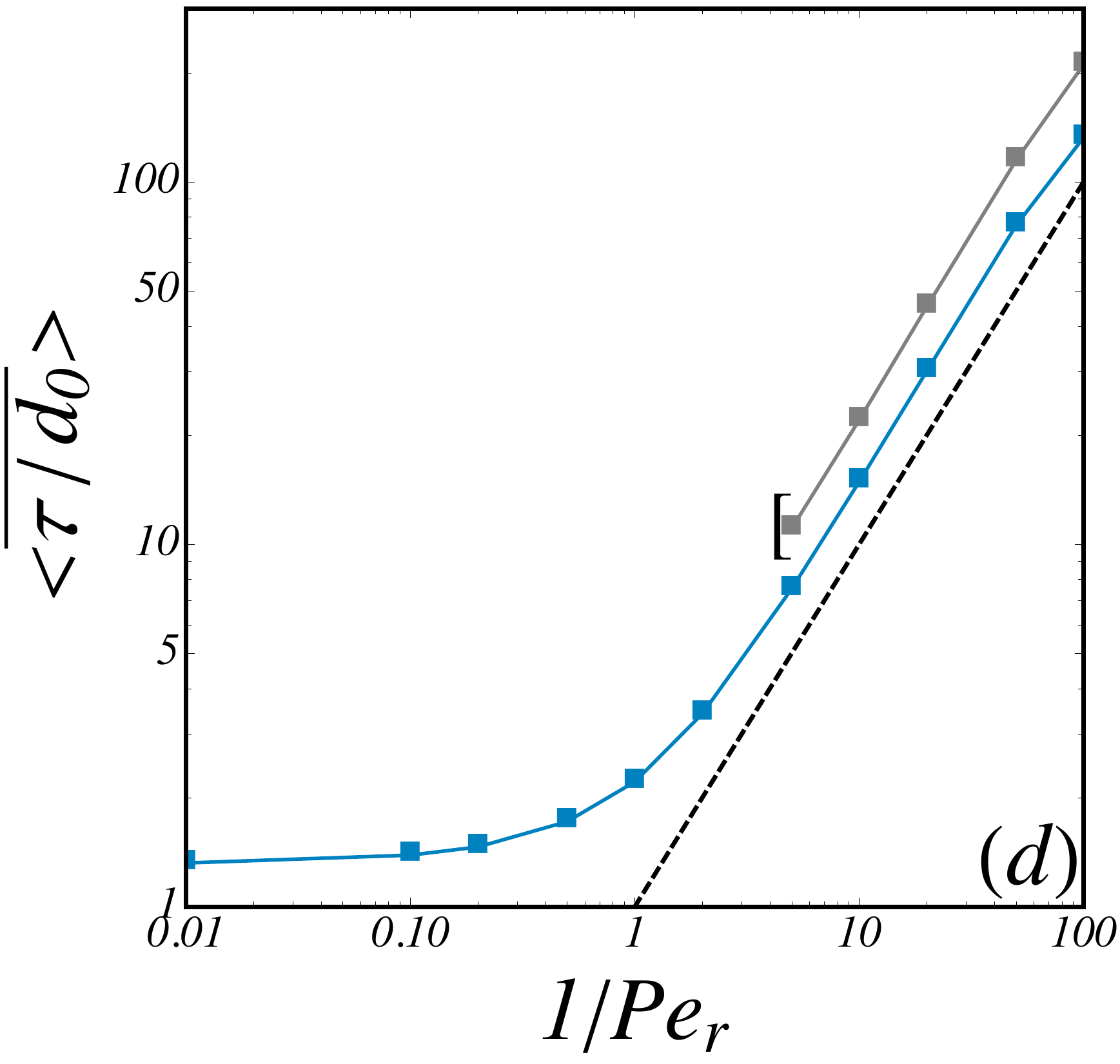}
    \vspace{-3mm}
    \caption{\textbf{Arrivals are fastest on the cusp of jamming.}
    $(a)$ Average arrival time $\overline{\tau}$, divided by $\tau_L = L / v_0$, against $d_0/L$, for $\phi = 0.4$ and a few $Pe$. The dashed line shows $\tau = d_0 / v_0$. 
    $(b)$ Average inverse travel speed $\langle\overline{\tau / d_0}\rangle$ against dimensionless noise amplitude, in log-lin scale, at $\phi = 0.2$ (blue) and $\phi = 0.4$ (gray). The bracket indicates jamming.
    $(c-d)$ Same as $(a-b)$ but for HABPs at $\Omega_r = 1$, and tuning the rotational noise.
    In $(c)$, the dashed white line indicates empirical observation of diffusive renewal.
    In $(c)$, the coloured dashed lines show the best short-distance ballistic fits, and in $(d)$, which is in log-log, the dashed line indicates $\langle\overline{\tau/d_0}\rangle \propto 1/Pe_r$.
    In the whole figure, $N = 2048$.
    }
    \label{fig:FigOpti}
\end{figure}
Having mapped out jammed states, in Fig.~\ref{fig:FigOpti}, we focus on the effect of noise terms on arrival times in the unjammed phase.
First, in Fig.~\ref{fig:FigOpti}$(a)$ we show the mean arrival time, conditional on initial distance to the target, against that initial separation at $\phi = 0.4$ and for noise intensities that melt the system in HBPs.
At low enough noise values, the motion is ballistic at some effective speed for all distances, even though the system is dense.
When the noise is strong, the motion becomes diffusive, resulting in a concave curve.
We checked (see App.~\ref{app:sizescalings}) that the linear regime corresponds to ballistic target reaching as the system size is tuned, $\overline{\tau} \propto L$, whereas the sub-linear regime displays a diffusive regime $\overline{\tau} \propto L^2$.
In Fig.~\ref{fig:FigOpti}$(b)$, we plot the mean of the effective inverse speed $\langle\overline{\tau/d_0}\rangle$, averaged over all initial distances, against the noise amplitude, and at two different densities: one below jamming, $\phi = 0.2$, and one above it, $\phi = 0.4$.
Both below and above the jamming density, increasing the noise intensity leads first to an increase of the effective inverse speed (slower travels), then to a decrease at large values (faster travels).
That decrease is observed when the typical diffusive time to reach the mean distance to a target, $\tau_{diff} = \langle d_0 \rangle^2/D_0$ becomes shorter than the ballistic time $\tau_0 = \langle d_0 \rangle/v_0$.
This line, shown in Fig.~\ref{fig:Fig3}$(a)$, separates the ballistic and the diffusive regimes.
While increasing the translational noise indefinitely does make travel times shorter, note that since $\langle d_0 \rangle \propto L$ (see App.~\ref{app:NumericalMethods}) the crossover between ballistic and diffusive travel is sent to larger and larger noise amplitudes as the system size increases.
As a result, the infinite-density minimum of travel times becomes unreachable in the $N\to\infty$ limit.
Therefore, in the limit of large systems, the optimal travel strategy at any given density is the local minimum found at small noises.
This optimal value is zero noise at small densities, and jumps discontinuously to the smallest value that unjams the system at the jamming density: it is highlighted in green in Fig.~\ref{fig:Fig3}$(a)$.
We checked that the optimality near jamming is observed regardless of the value of the density (see App.~\ref{app:MoreDensityLines}).

In Fig.~\ref{fig:FigOpti}$(c)-(d)$, we show the curves equivalent to those of Fig.~\ref{fig:FigOpti}$(a)-(b)$, but this time in HABPs at $\Omega_r = 1$, and tuning the amplitude of the rotational noise.
In Fig.~\ref{fig:FigOpti}$(c)$, we show mean arrival times against the initial distance to the target at density $\phi = 0.4$ for a few values of $Pe_r$.
As in the case of translational noise, motion is ballistic even at high densities, provided that the noise amplitude is small enough.
At larger noise amplitudes, however, target-reaching becomes diffusive, which is here encoded by a sub-linear trend, here emphasized by plotting the initial tangent to each curve (see App.~\ref{app:sizescalings} for size scalings of travel times).
The locations of the ballistic and diffusive regimes, which are this time deduced by inspection of the time against distance, are reported in the diagram of Fig.~\ref{fig:Fig3}$(b)$.
Finally, in Fig.~\ref{fig:FigOpti}$(d)$, we plot the mean inverse travel speeds $\langle\overline{\tau/d_0}\rangle$ against the rotational noise amplitude at the same two densities as in Fig.~\ref{fig:FigOpti}$(b)$ (this time in log-log scale).
In contrast with the case of translational noise, the inverse arrival speed (and, therefore, the average arrival time) grows monotonically with noise, with $\overline{\tau} \sim 1/Pe_r$, as indicated by a dashed black line.
The reason is that in the high-noise regime ($1/Pe_r > 100$), motion becomes slower as noise washes out the memory of the target location, so that particles are essentially ABPs, diffusing in space with an effective coefficient~\cite{Fily2012} $D_{eff} = v_0^2 / 2 D_r$, that decreases as rotational noise increases.
Once again, we observe that the optimal strategy to shorten travels is to have zero noise below jamming, but the smallest amount possible that ensures there is no jam above the jamming density (see green line in Fig.~\ref{fig:Fig3}$(b)$).

In short, in both models,  for large systems, {\it optimal travel times averaged over initial distances are achieved at the lowest noise amplitude for which the system is not jammed}.
In smaller systems, the diffusive regime of HBPs obtained at very large noise amplitudes may be more efficient, but this regime is sent to infinite noise in the large system limit. 
Seeing the noise amplitude as a travel strategy, it means that the optimal travel strategy changes discontinuously at the jamming transition, from zero noise at $\phi < \phi_J$ to a finite value of the noise beyond $\phi_J$.

\subsection{Full Phase Diagram of HABPs\label{sec:FullHABP}}

\begin{figure}
    \centering
    \includegraphics[width=0.48\columnwidth]{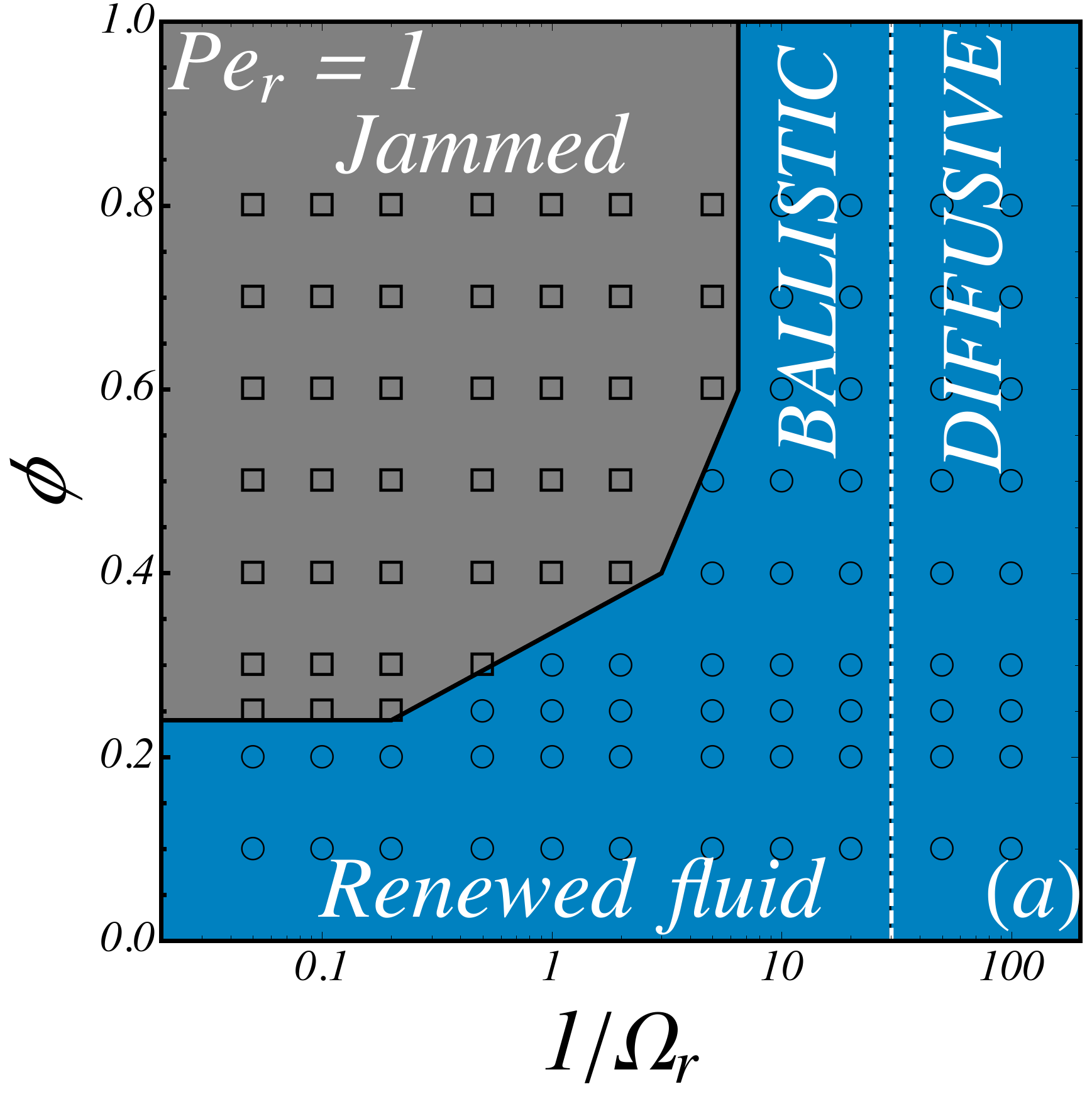}
    \includegraphics[width=0.48\columnwidth]{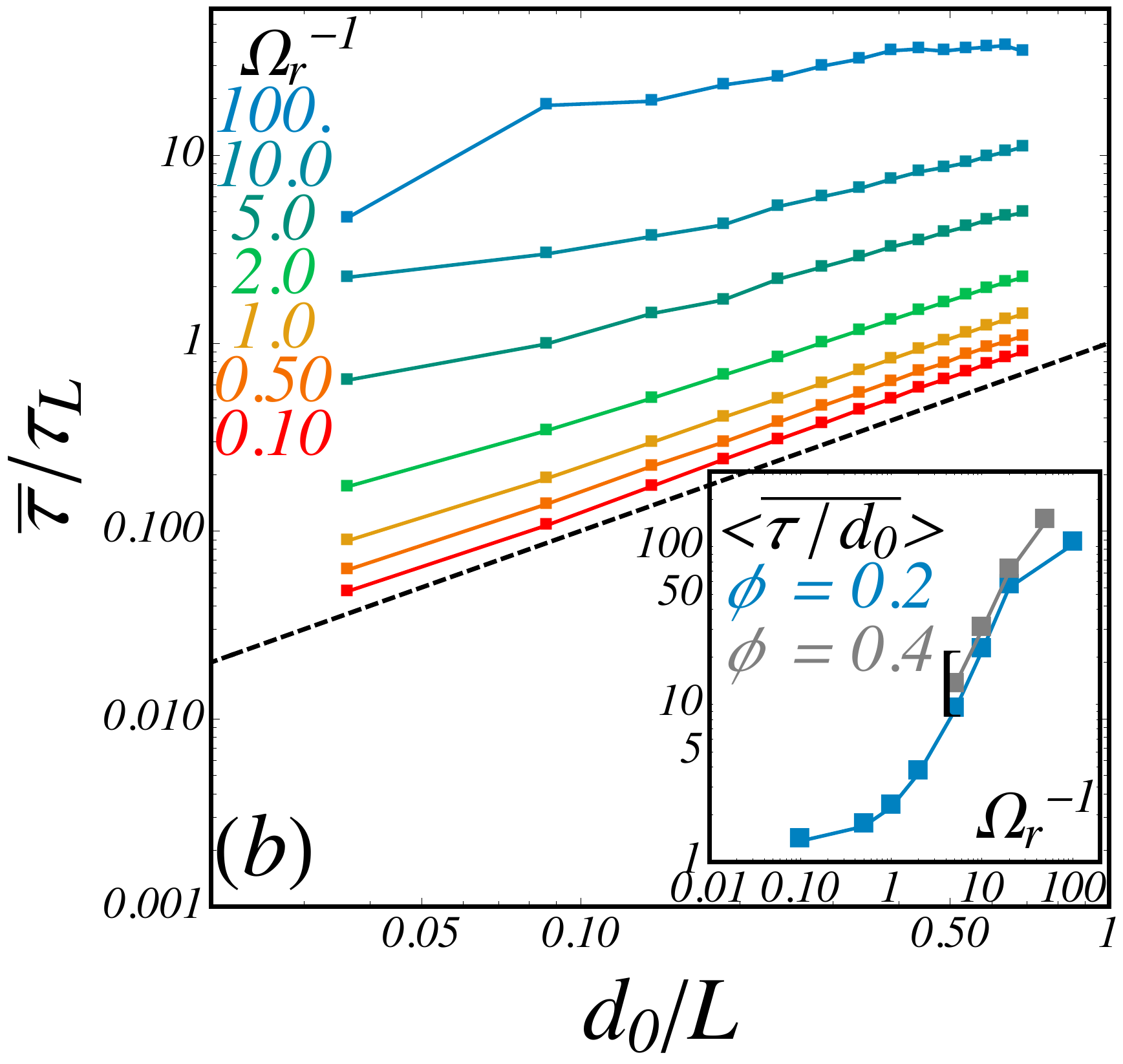}
    \caption{\textbf{Varying the relaxation rate at finite noise.}
    $(a)$ Phase diagram in the $(1/\Omega_r,\phi)$ plane at $Pe_r = 1$ for $N=2048$.
    $(b)$ Corresponding $\overline{\tau}$ against $d_0$ for $Pe_r = 1, \phi = 0.2$ and a few $\Omega_r$.
    Inset: $\langle\overline{\tau/d_0}\rangle$ against $\Omega_r^{-1}$, in log-log scale, at densities $\phi = 0.2$ and $0.4$. 
    The bracket indicates jamming.
    }
    \label{fig:HABPRelax}
\end{figure}
We now extend our description of jamming and target attainment to the full diagram of HABPs at finite noise and relaxation rates.
In Fig.~\ref{fig:HABPRelax}$(a)$, we show the observed phases in a system of HABPs at a finite amplitude of the noise, $Pe_r = 1$, as a function of the relaxation rate towards targets.
The diagram is very similar to the phase diagrams of Fig.~\ref{fig:Fig3}, observed when tuning the noise amplitudes: the system is jammed when the relaxation rate is large, and is unjammed by a finite relaxation rate.
Arrival times in the unjammed phase are ballistic provided that the relaxation rate is fast enough, and become diffusive if it is too slow (here, the diffusive domain is estimated empirically from the aspect of the time versus distance curves).

In Fig.~\ref{fig:HABPRelax}$(b)$, we plot the corresponding mean escape time against the initial distance at $\phi = 0.2$ and for a few values of $\Omega_r$.
As in the noiseless case, one finds that faster relaxation leads to a decrease of arrival times, and to ballistic travel, while slower relaxation leads to diffusive behaviour characterised by non-linear time against distance curves.
Note that finite noise washes out the effect of larger arrival times starting from shorter distances shown in Fig.~\ref{fig:Zeronoiserelaxtimes}, indicating that spiraling trajectories are likely to be an artifact of deterministic dynamics.
In the inset, we plot the effective inverse speed $\langle\overline{\tau/d_0}\rangle$ against the inverse relaxation rate at two densities, one below jamming $\phi = 0.2$ and one above $\phi = 0.4$.
We show that arrivals become slower as the relaxation rate becomes slower.
In particular, it means that the fastest arrivals at a fixed density are once again located right at the edge of the jammed phase for $\phi>\phi_J$, and at infinitely fast relaxation for $\phi<\phi_J$.

\begin{figure}
    \centering
    \includegraphics[width =0.60\columnwidth]{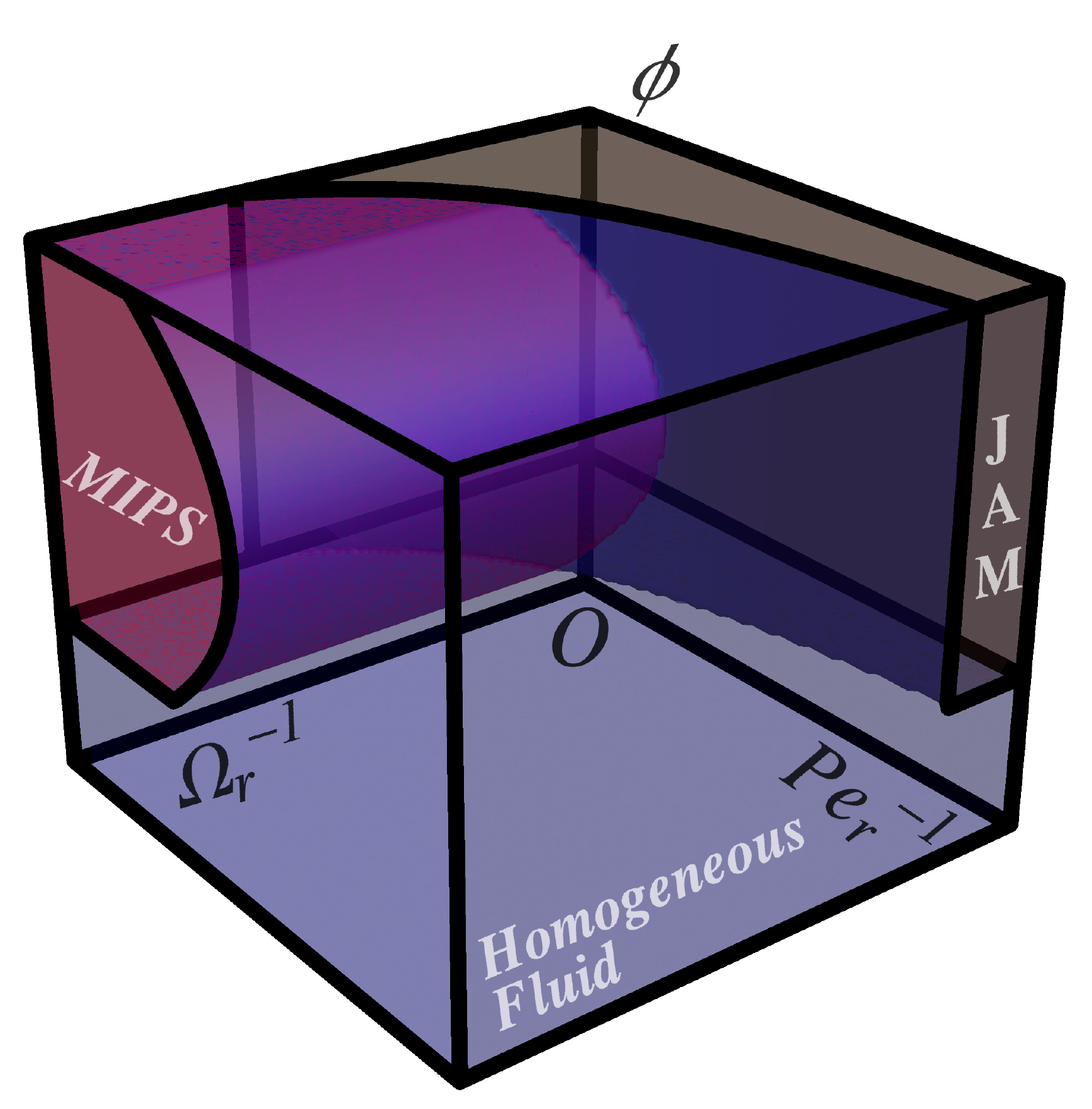}
    \caption{\textbf{Full HABP phase diagram.}
    Sketch of the $3d$ phase diagram of HABPs in the $(1/\Omega_r, 1/Pe_r, \phi)$ basis.
    }
    \label{fig:FullHABP}
\end{figure}

Finally, putting together the phase diagrams of Figs.~\ref{fig:Fig4}, ~\ref{fig:Fig3}, and~\ref{fig:HABPRelax}, as well as a few extra slices at constant $Pe_r$ shown in App.~\ref{app:OmegarPhi}, we are able to infer the full phase diagram of HABPs.
The result is shown in Fig.~\ref{fig:FullHABP}, as a full $3d$ phase diagram in the $(1/Pe_r, 1/\Omega_r, \phi)$ space.
The MIPS phase, observed at low noise and slow relaxation, meets a jammed phase that starts at the same density.
This jammed phase reaches all the way to infinite noise provided that relaxation remains faster than rotational diffusion.
In other words, {\it at any large value of the noise, one can decrease the relaxation rate to unjam the system and recover a homogeneous fluid, and fastest arrivals are again achieved right next to jamming}.

\section{Conclusions\label{sec:Conclusions}}

We have introduced and studied a model of repulsive searchers that self-propel towards individual targets in continuous $2d$ space.
If particles go straight to their targets, this model reaches a frozen steady state at rather low densities compared to other systems of self-propelled particles~\cite{Henkes2011,Digregorio2018,VanDamme2019,Merrigan2019,Mandal2020}.
We show that these jams are an extra instability on top of MIPS such that the gas phase is emptied by target-reaching.
This suggests that precise destinations in space generically facilitate traffic jams, which is reminiscent of recent reports of phase separation in self-propelled particles that turn towards denser regions~\cite{Zhang2021}.

By tuning two different kinds of noise terms, and a relaxation rate of the self-propulsion orientation, we show that taking less direct routes to the target can unjam the system.
In all three scenarios, the shortest target-reaching times at high densities lie right outside of the jammed phase, where particles manage to reach their targets ballistically, even up to rather high densities.
This result implies that, at $\phi_J$, there is a discontinuous change in the optimal strategy to reach targets.
At low densities it is best to go straight to targets, as interactions between particles are uncommon.
However, at high densities, this becomes the worst strategy, as it leads to jamming.
Instead, one should fine-tune the velocity-to-noise-ratio to be close to the melting value.
This is reminiscent of slower-is-faster effects \cite{Helbing2000,Gershenson2011}: beyond a threshold density, it is more efficient for the collective to have each agent proceed to its objective at slower individual speeds.

Similar observations of an optimal noise amplitude or shape achieving optimal transport efficiency have also been described in models of random walkers, chemotaxis and flocking in complex environments~\cite{Chepizhko2013,Azimzade2017,Volpe2017,Bertrand2018}.
However, these optimal parameters typically evolve continuously as parameters of the problem are tuned: the discontinuity that we report is seemingly a non-trivial result of the presence of individual targets.
This particularity might be a critical ingredient to understand in order to optimise traffic problems in which agents have distinct goals in space.

\begin{acknowledgements}
We thank Yariv Kafri and Olivier Dauchot for useful and insightful discussions.  D.~L. and M.~C. were supported by the Israel Science Foundation under grant No. 1866/16. D.~H.  wishes to thank the Israel Science Foundation (grant No. 2385/20) and the Alon fellowship.
\end{acknowledgements}

\appendix
\section{Captions of the Videos\label{app:VideoCaptions}}

We here detail the content of the various ancillary video files.

\textit{Comparison.mp4 --} Short-time dynamics of a system of noiseless active Brownian particles, or ABPs, (left), and of noiseless homing active Brownian particles, or HABPs, with $\Omega_r = 10$ (right), starting from a uniform local density of particles in space, and a uniform distribution of self-propulsion orientations.
In both panels, the number of particles is $N = 2048$, and the overall packing fraction $\phi = 0.4$.
Each video is animated at 30 frames per second, and the time between two frames is $\delta t = 0.5 a / v_0$, with $a$ the repulsive diameter of particles and $v_0$ the self-propulsion speed.
The colour of each particle codes for the phase of the local hexatic order parameter (mapped onto a colour wheel), and the arrow inside them shows the instantaneous orientation of their self-propulsion.

\textit{32k\_jamming.mp4 --} Example of a larger-scale realisation of jamming of HABPs than those presented in the main text. In this simulation, the number of particles is $N = 32768$, the density $\phi = 0.4$, the noise amplitudes are all set to zero, and the relaxation rate is $\Omega_r = 100$.
The video contains 30 frames per second, and the time between two frames is $\Delta t = 50 a / v_0$.
The colour code is the same as in the previous video.

\textit{2048\_jammingdynamics.gif --} Example of jamming dynamics in the 3d extension of the model of HABPs. In this simulation, $N = 2048$ particles, $\phi = 0.30$, the noise is  set to zero, and $\Omega_r = 100$. The time between two frames of the gif is $\Delta t = 50 a/v_0$. Each particle is represented as a tinted metallic sphere, rendered with a ray-tracing algorithm. Taking advantage of the periodic boundary conditions (omitted in this representation), the system is centered on its long-time centre of mass. The camera is looking at the origin from the point $(2L, 2L, 2L)$.

\textit{2048\_jamoverview.gif --} Overview of a 3d jam of HABPs. Here, $N = 2048$ and $\phi = 0.3$. Each particle is represented in the same way as in the previous video. We here rotate the camera around the vertical axis, starting from $(2L, 2L, 2L)$, to show a steady state from various angles.

\section{Numerical methods}

\subsection{Simulation design\label{app:NumericalMethods}}

All the results presented in the main text are obtained via molecular dynamics simulations with the simplest possible order-1 integrator.
Namely, we write the equation of motion of any Cartesian component of the position of a particle symbolically as
\begin{align}
    dx = x(t+dt) - x(t) = v_{det} dt + v_{stoch}dt^{1/2},
\end{align}
where $dt$ is a fixed time step, $v_{det}$ is the deterministic part of the velocity that comes from self-propulsion and interactions with other particles, and $v_{stoch}$ is the stochastic part of the velocity that appears when we introduce noise.
In the case with noise, the stochastic part of the velocity simply reads $v_{stoch} = \sqrt{2 D_T} \eta_x$, with $\eta_x$ drawn from a unit-variance centered normal distribution, and it is zero otherwise.
The computation of the interaction part of $v_{det}$ is accelerated by introducing a partition of space into square cells twice as wide as the longest-range interaction in the system, and labelling at all times each particle with its cell number.
In practice, we set the time step to, at most, $dt = 10^{-4}$, or to the largest power of ten that ensures that no update $dx$ can be larger than $0.01$ in simulation units.
This choice ensures that even high noise amplitudes cannot simply bypass repulsive interactions, e.g. jump to the other side of a neighbouring particle, due to the choice of discretisation of time.
For instance, if $\sqrt{2 D_T} = 100$, corresponding in the main text to an inverse Péclet number $50$, we set $dt = 10^{-6}$.

The initial positions of particles and targets are each drawn uniformly in a periodic square simulation box with linear size $L$, only rejecting pairs such that particles are absorbed at drawing time.
When the relaxation rate of the self-propulsion orientation towards the target is finite, we also draw the initial polarity of each particle uniformly on the circle.

\begin{figure}
    \centering
    \includegraphics[width=.48\columnwidth]{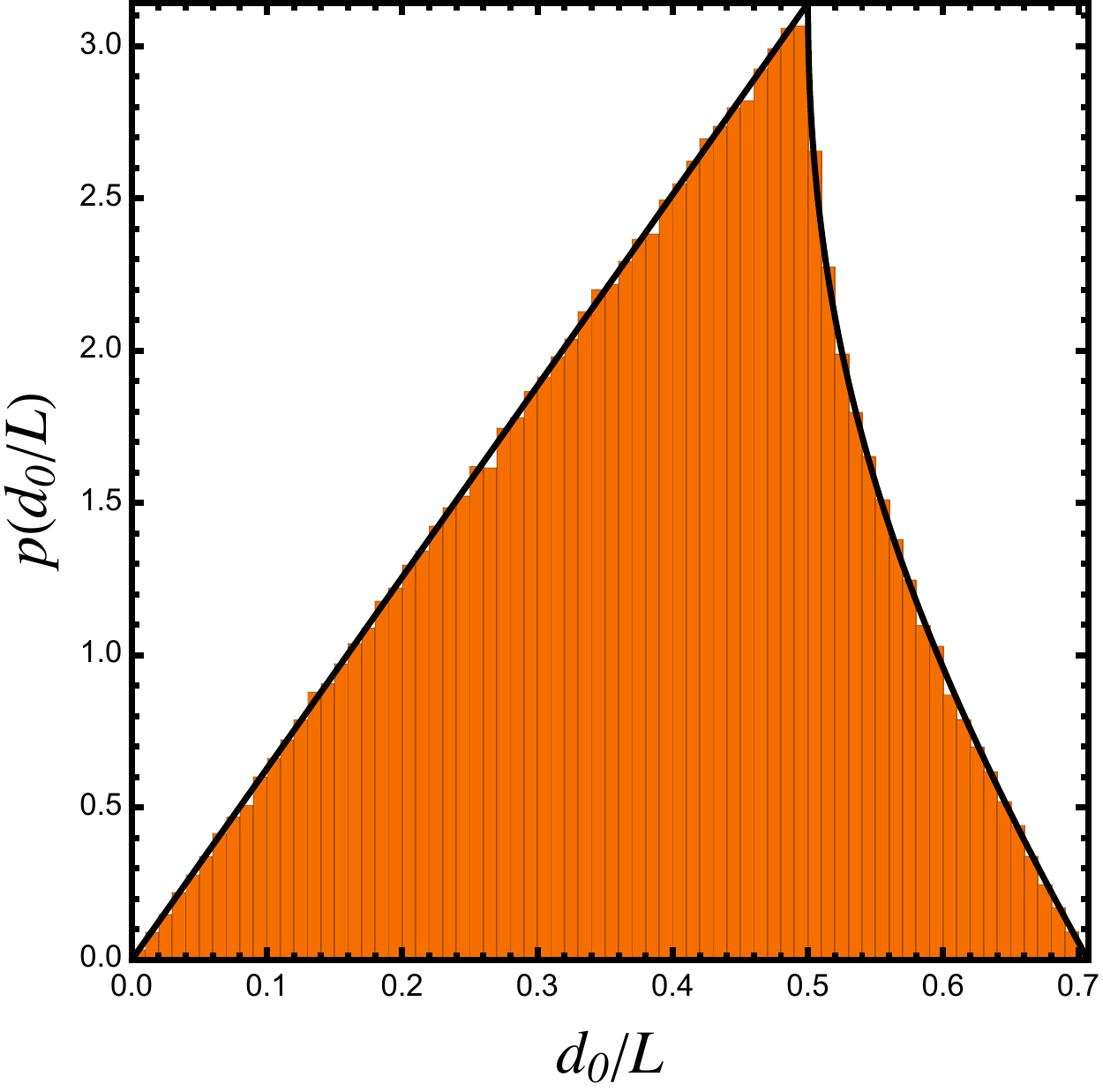}
    \includegraphics[width=.48\columnwidth]{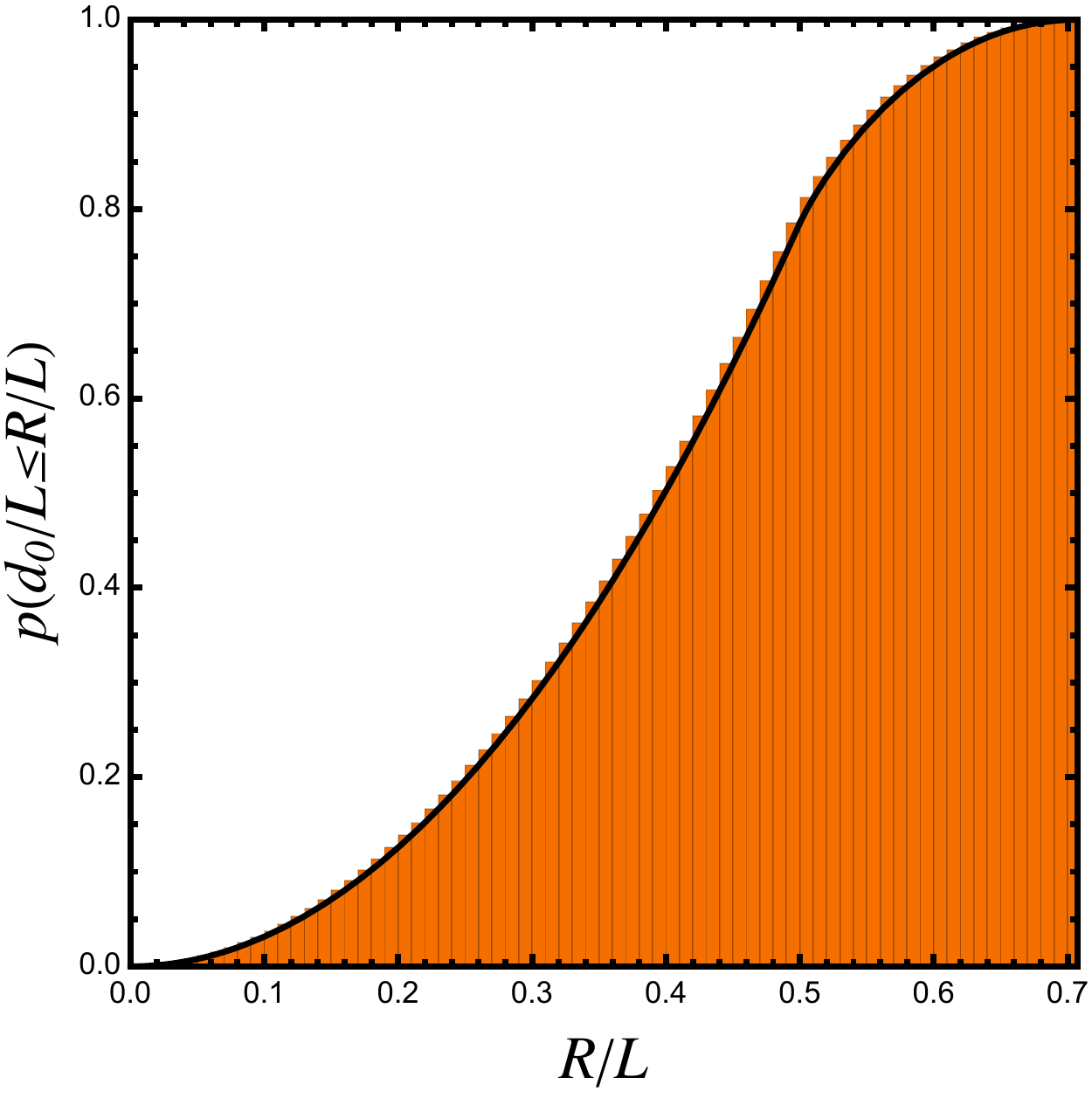}
    \caption{\textbf{Distribution of initial particle-target distances.} Left: Predicted PDF of initial particle-target distances when drawing both of them uniformly in a square periodic box (black line), and observed histogram of initial distances obtained 500000 values (orange).
    Right: same plot for the CDF instead of the PDF.}
    \label{fig:distancedistrib}
\end{figure}

As a result of this initialization procedure, the distribution of initial distances to targets is not flat.
Indeed, the targets and particles are both drawn uniformly in space, so that the distribution for the vector $\bm{d}_0$ linking a particle to its target at birth is
\begin{align}
    p(\bm{d}_0) = \frac{1}{L^2} \mathbb{1}\left( - \frac{L}{2} \leq d_{0,x}\leq \frac{L}{2}\right) \mathbb{1}\left( - \frac{L}{2} \leq d_{0,y} \leq \frac{L}{2}\right)
\end{align}
where we use the periodicity of the box to place the particle at the center of the box.
The probability distribution function of initial distances is then given by
\begin{align}
    p(d_0) = \int\limits_{box} d\theta d_0 p(\bm{d}_0).
\end{align}
Since the box is a periodic square, there are two integration domains: a first one corresponding to radii such that the circle with radius $d_0$ fits in the square box, and circles too large to fit.
In the latter case, the integral decomposes into 4 equal parts (one per corner of the square), so that the full integral reads:
\begin{align}
    p(d_0) = \frac{d_0}{L^2} &\left[ \vphantom{\int\limits_{\theta_{min}}^{\theta_{max}}} 2 \pi\, \mathbb{1}\left( d_0 \leq \frac{L}{2}\right) \right. \nonumber\\
    &\left.+ 4 \, \mathbb{1}\left( \frac{L}{2} \leq d_0 \leq \frac{\sqrt{2} L}{2}\right) \int\limits_{\theta_{min}}^{\theta_{max}} d\theta   \right],
\end{align}
where some simple geometry yields, in the upper right corner,
\begin{align}
    \theta_{min} &= \arccos\frac{L}{2 d_0}, \\
    \theta_{max} &= \frac{\pi}{2} - \theta_{min}.
\end{align}
All in all, one gets,
\begin{align}
    p(d_0) = &\frac{2 \pi d_0}{L^2} \mathbb{1}\left(d_0 \leq \frac{\sqrt{2} L}{2}\right) \nonumber \\
    &- \frac{8 d_0}{L^2} \arccos\frac{L}{2d_0} \mathbb{1}\left( \frac{L}{2} \leq d_0 \leq \frac{\sqrt{2} L}{2}\right).
\end{align}
It is also useful, for intuition, to know the CDF associated to this PDF, or namely the probability for the distance to the target to be smaller than some value $R$.
It is found through a straightforward integration:
\begin{align}
    &p(d_0 \leq R) = \frac{\pi R^2}{L^2} \mathbb{1}\left(R \leq \frac{\sqrt{2} L}{2}\right) \nonumber \\
    &+ \left( \frac{4R^2}{L^2}\text{arcsec}\frac{2R}{L} - \sqrt{\frac{4 R^2}{L^2} - 1} \right) \mathbb{1}\left( \frac{L}{2} \leq R \leq \frac{\sqrt{2} L}{2}\right).
\end{align}
The shape of both this PDF and CDF are shown in Fig.~\ref{fig:distancedistrib}, where they are superimposed to histograms obtained from simulations, which are a perfect match.

Using these results, one can measure the various statistical values of interest for this distance.
The mode of distances is $d_0^{max} = L/2$, and the cumulative, evaluated at the mode, is $p(d_0 \leq d_0^{max}) = \pi/4 \approx 0.79$.
The mean is given by
\begin{align}
    \langle d_0 \rangle = \frac{L}{12}\left(2\sqrt{2} + \log\left(3 + 2\sqrt{2}\right) \right) \approx 0.38 L,
\end{align}
and the standard deviation is
\begin{align}
    \sigma_{d_0} = \frac{L}{12}\sqrt{16 + \arccos^2 3 - 4 \sqrt{2} \text{arccosh} 3} \approx 0.14 L.
\end{align}

Finally, note that this distribution of distances is used in the main text to deduce the shape of escape times in the ballistic regime.
Indeed, as long as escapes are typically ballistic, the escape time $\tau_{d_0}$ from an initial distance $d_0$ is $\tau_{d_0} = d_0 / v^\star$ with some (measured) effective speed that \textit{a priori} depends on the noise amplitudes, relaxation rate, density and system size.
Assuming that $v^\star$ does not fluctuate too much between particles, one can assume that the conditional probabilities $\tilde{p}(\tau_{d_0})$ to observe an escape time $\tau_{d_0}$ for an initial distance $d_0$, and $\hat{p}({d_0}_\tau)$ that the initial distance was $d_0$ given that the escape time was $\tau_{d_0}$, are both Dirac deltas.
Consequently, the PDF $P(\tau)$ of arrival times integrated over initial distances is simply given by a dilatation of the distribution of initial distances, $P(\tau) = p(v^\star \tau)/v^\star$.

\subsection{3d extension\label{app:3dext}}

In the main text, we briefly discuss results obtained for the $3d$ extension of an HABP model.
While the update of positions presented above can easily be generalised to three-dimensional space by adding one axis and a third translational noise component, the angular dynamics are integrated slightly differently.
Indeed, updating $3d$ unit vectors via their representation as two angles on a sphere can easily lead to numerical instabilities, due to the presence of poles.
It is therefore convenient to represent the polarity of each particle as a full $3d$ vector, but to update it in such a way that its modulus remains (up to small numerical errors) unity.
A trick to do so is to rewrite the dynamics of the polarity of each particle as
\begin{align}
    d\hat{\bm{e}}_\theta &\equiv \hat{\bm{e}}_\theta(t+dt) - \hat{\bm{e}}_\theta(t)  \nonumber \\
    &= \bm{F}_{det} dt + \bm{\omega}_{stoch}\times \hat{\bm{e}}_\theta(t) dt^{1/2} - 2 D_r  \hat{\bm{e}}_\theta(t) dt,
\end{align}
where $\times$ represents a vector product, and $D_r$ is the rotational diffusion constant such that $\left\langle \omega_{stoch,i}(t) \omega_{stock,j}(t')\right\rangle = 2 D_r \delta_{ij}\delta(t - t')$.

Regarding the stochastic part, while a Wiener process on the sphere is naturally expressed as two independent white noise terms $\eta_\theta$ and $\eta_\phi$ acting on the spherical representation of the unit vector~\cite{Raible2004}, it can equivalently be represented as a Cartesian stochastic rotation vector $\bm{\omega}_{stoch}$, that reads~\cite{Winkler2015}
\begin{align}
    \bm{\omega}_{stoch} = \sqrt{2 D_r} \bm{\eta},
\end{align}
with $\bm{\eta}$ a 3-dimensional Wiener process with independent Cartesian components.

As for the deterministic part, by analogy to the $2d$ coupling to targets, we choose to enforce a harmonic coupling between the orientation of the particle-target vector and the internal polarity of particles on the sphere.
Namely, we subject each polarity to the potential
\begin{align}
    U(\theta,\phi) = \frac{\omega_r}{2}\left[\left(\theta - \theta_T\right)^2 + \left(\phi - \phi_T \right)^2 \right],
\end{align}
where $(\theta,\phi)$ are the spherical-coordinate representation of the polarity of the particle, and $(\theta_T,\phi_T)$ are the two angular components of the vector $\bm{r_T} - \bm{r}$ linking a particle to its target, in the spherical basis centered on the particle.
This potential results in a harmonic force on the sphere that can be expressed as
\begin{align}
    \bm{F} &= -\partial_\theta U \hat{\bm{u}}_\theta - \partial_\phi U \hat{\bm{u}}_\phi, \\
           &= \omega_r \left[ \left(\theta_T - \theta\right)\hat{\bm{u}}_\theta + \left(\phi_T - \phi \right)\hat{\bm{u}}_\phi \right] 
\end{align}
with $(\hat{\bm{u}}_\theta, \hat{\bm{u}}_\phi)$ the two orthoradial basis vectors associated to the spherical coordinates $\theta,\phi$.
Both of them can be expressed in Cartesian coordinates for the update,
\begin{align}
    \hat{\bm{u}}_\theta &= \cos\theta \cos \phi \, \hat{\bm{e}}_x + \cos\theta\sin\phi \, \hat{\bm{e}}_y - \sin\theta \, \hat{\bm{e}}_z,\\
    \hat{\bm{u}}_\phi &= - \sin\phi \, \hat{\bm{e}}_x + \cos\phi \, \hat{\bm{e}}_y.
\end{align}

As a failsafe, we check during updates that the modulus of the polarity vector does not deviate from unity by more than one percent.

\subsection{Construction of the phase diagrams \label{sec:PhDConstruction}}

\begin{figure}
    \centering
    \includegraphics[height = .35\columnwidth]{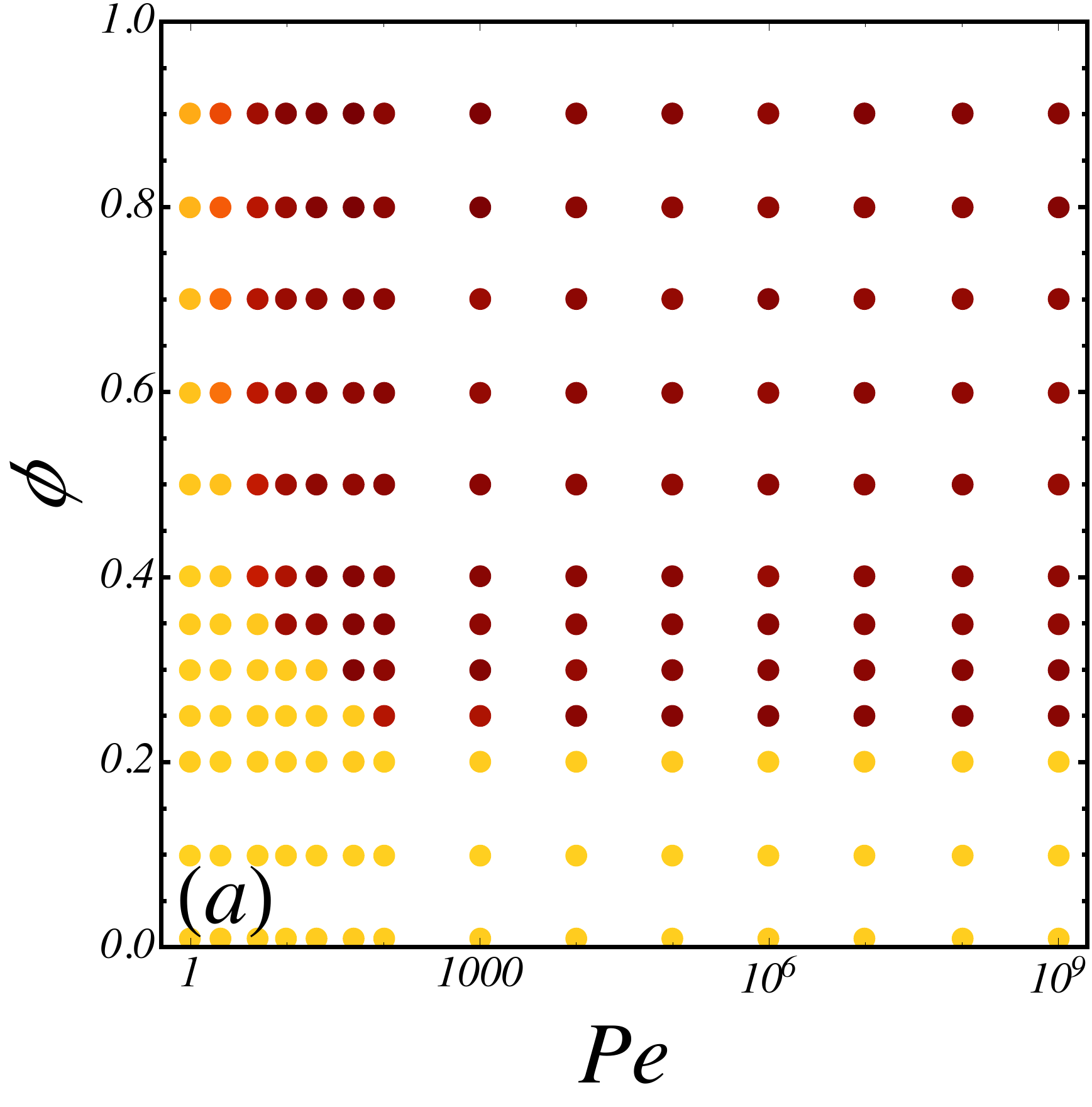}
    \includegraphics[height = .35\columnwidth]{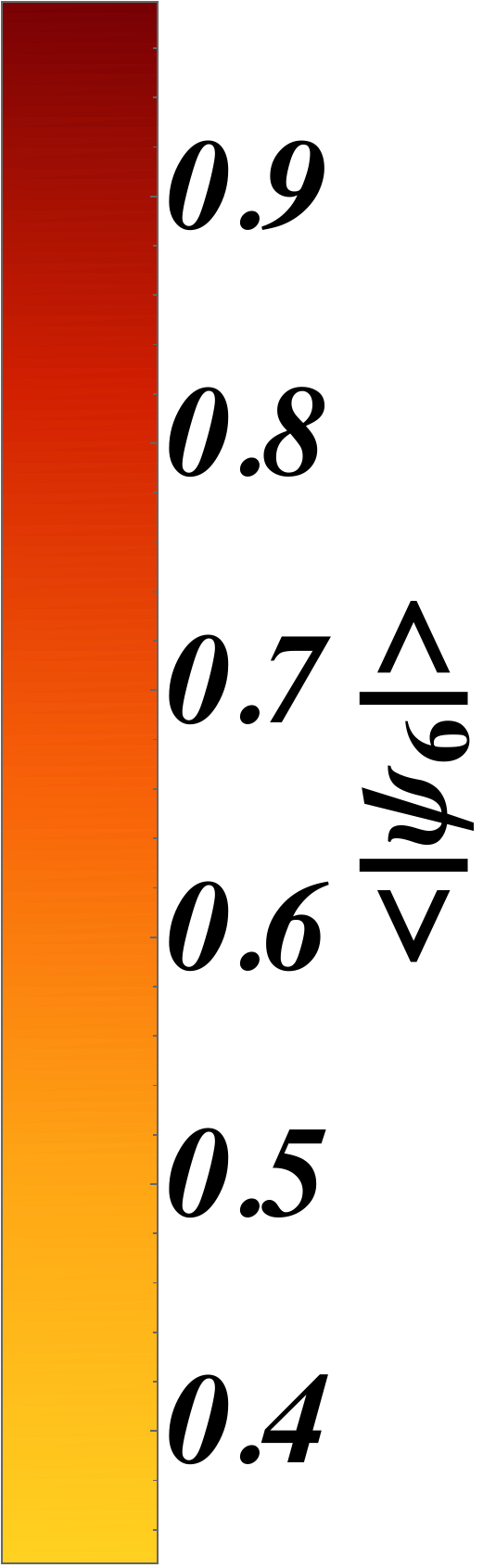}
    \includegraphics[height = .35\columnwidth]{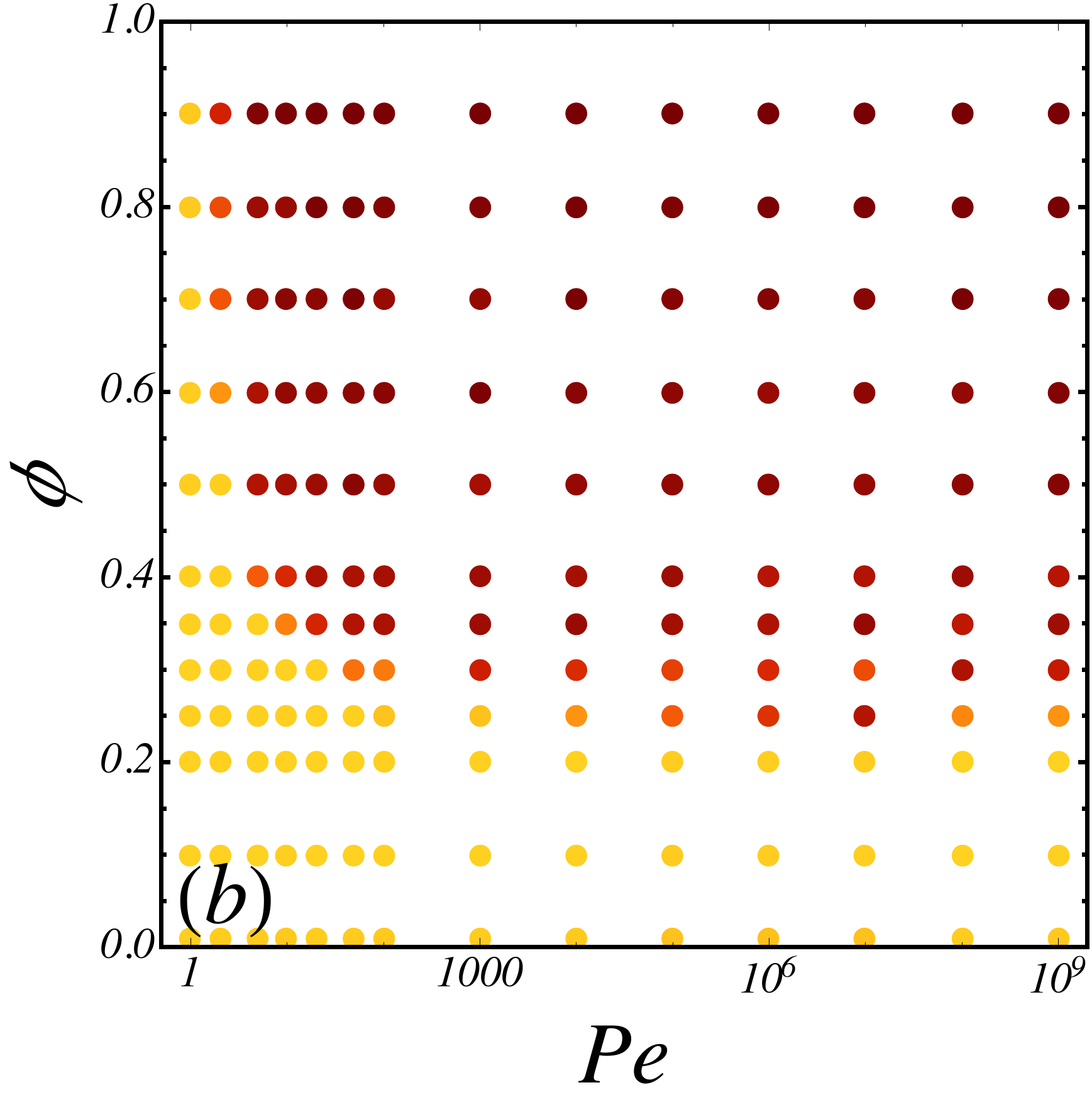}
    \includegraphics[height = .35\columnwidth]{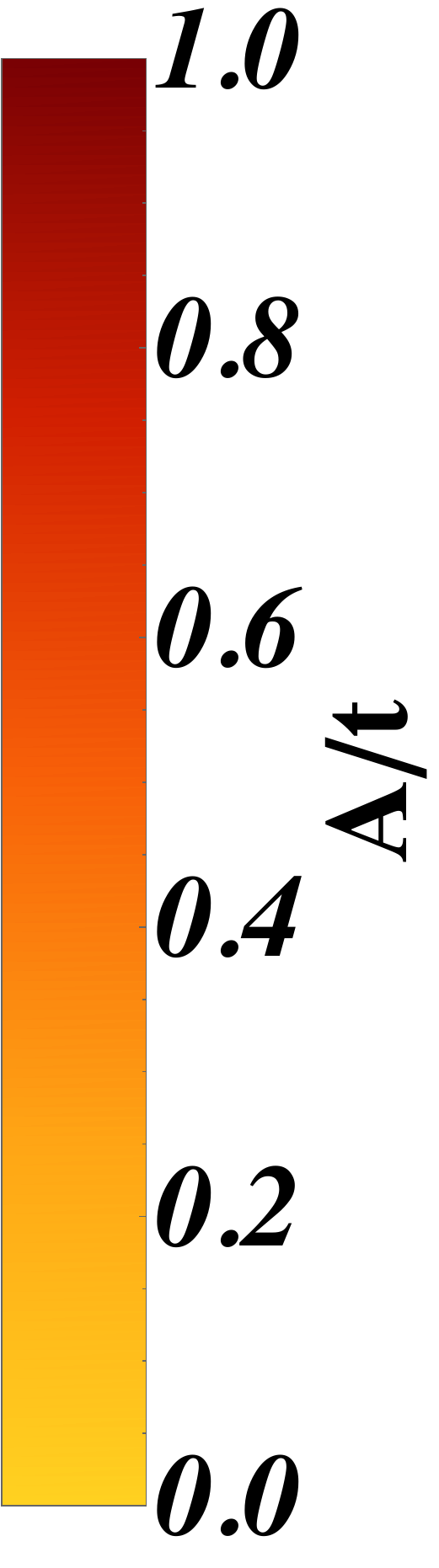}
    \caption{\textbf{Mapping out jams.}
    $(a)$  Map of the average of the modulus of the microscopic hexatic vectors, in the case of translational noise, in the $(Pe,\phi)$ plane.
    The colour code of each point is explicited in the legend bar.
   $(b)$ Corresponding map of the long-time value of the average age of live particles divided by the total simulation time.}
    \label{fig:PhaseDiagram}
\end{figure}
Throughout the main text, we map out jammed, ballistically renewed, diffusively renewed, and MIPS domains when varying the density, noise amplitudes, or relaxation rate of the polarity associated to self-propulsion.
In this subsection, we explain in more detail how these domains were identified in simulations.
Jams are identified by measuring the observables mentioned in Fig.~1 of the main text, namely, 
\begin{enumerate}
    \item[(i)] the average value $\overline{A}/t$ of the age $\{A_i = t - t^{\text{birth}}_i\}$ of the particles left inside the system divided by the total simulation time $t$, where $t^{\text{birth}}_i$ is the date of birth of the current $i$-th particle in the system. This observable approaches $1$ when no particle dies in the system, and $0$ when particles escape at a sufficiently fast rate;
    \item[(ii)] the average of the modulus of the local hexatic order parameter, $\overline{\left\| \bm{\psi_6}\right\|}$. This observable is close to $1$ when the observed phase exhibits strong local hexatic order, meaning that the nearest neighbours of each particle are organised in a regular hexagon. It is, consequently, low in isotropic fluid phases.
\end{enumerate}
Maps of these quantities used to identify the jammed region in the case of translational noise only are shown in Fig.~\ref{fig:PhaseDiagram}.
One observes a very sharp contrast between jammed and non-jammed phases, even at high packing fractions.
Note that the age of particles typically converges slower than the hexatic order parameter at densities close to the jamming density: this is due to the fact that the nucleation of jams typically takes longer at lower densities.
\begin{figure}
    \centering
    \includegraphics[width = 0.32\textwidth]{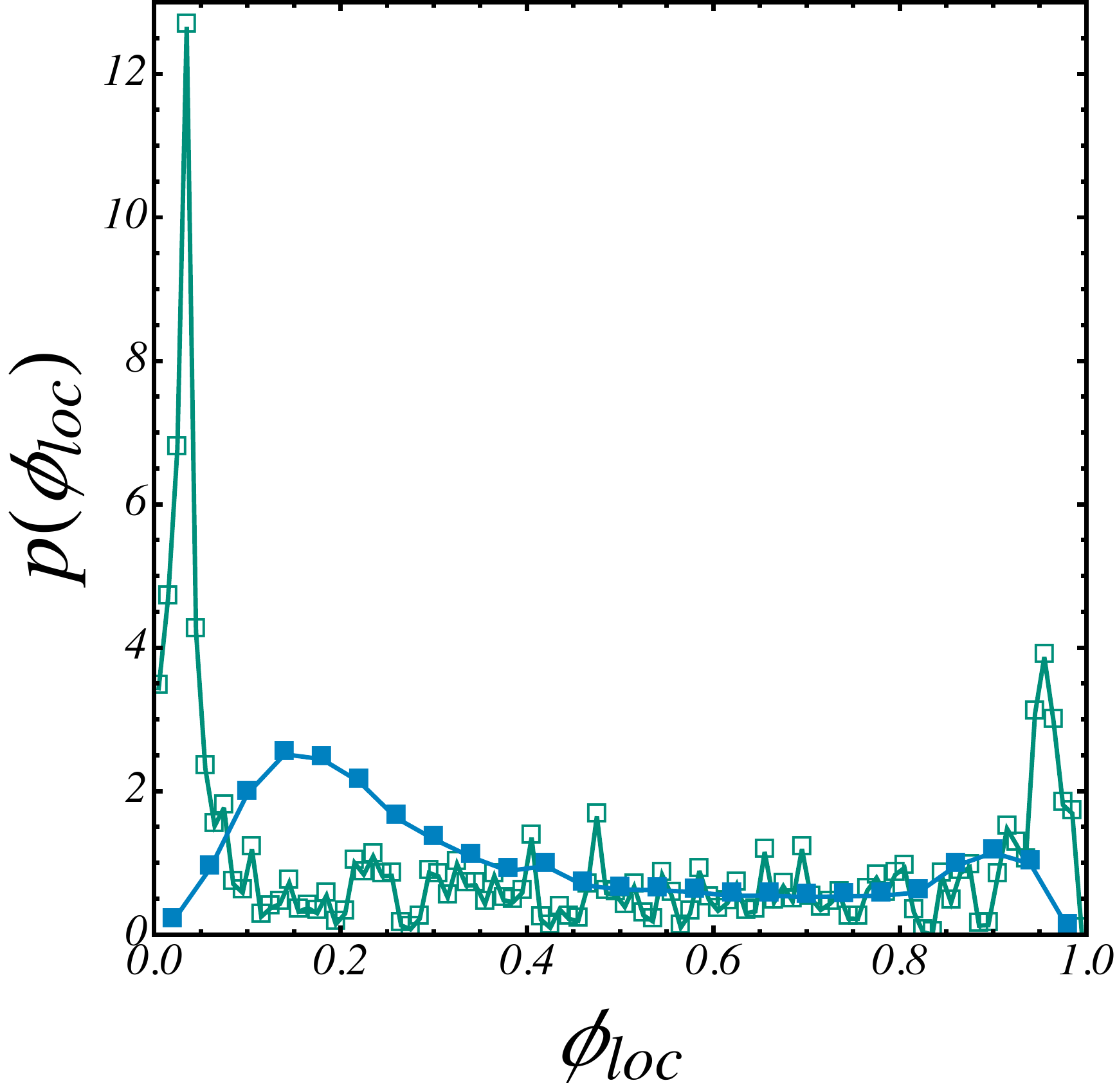}
    \caption{\textbf{Example of measured coarse-grained density distributions.} 
    Curves obtained for $N = 2048, \phi = 0.4, 1/Pe = 0, 1/Pe_r = 1, \Omega_r = 0.01$ (blue filled squares), and $\Omega_r = 0.2$ (teal empty squares). They were obtained with about $10^2$ configurations.}
    \label{fig:localdensity}
\end{figure}

In the main text, we also map out regions in which the system is phase separated, and measure the corresponding densities.
In order to do so, having chosen a set of parameters, we run long simulations ($t \sim 10^4 \geq 10^8 dt$) and save configurations with a spacing between snapshots long enough that they can be considered to be uncorrelated, provided that they are not jammed ($\Delta t \sim 10^2$).
We also start $n_{runs} \sim 10$ independent runs of each of these simulations when measuring coarse-grained densities. 
In each of these configurations, we cut the simulation box into a set of smaller square boxes with sidelength $\ell$, and save the value $\phi_{loc}$ of the measured local packing fraction in that smaller box.
Stocking the $L^2/\ell^2$ values over $n_{snap}$ snapshots and $n_{runs}$ runs, we measure a histogram of the local densities, two examples of which are shown in Fig.~\ref{fig:localdensity}.
In phase-separated cases, this distribution is bimodal, and typically becomes very peaked as the system becomes a jam-vacuum coexistence.
The maxima of that distribution can be used to estimate the values $\phi_{low}$ and $\phi_{high}$ of the two coexisting densities.
Whenever presented in the main text, natural fluctuations of these densities are estimated by measuring the standard deviation of the list of $n_{runs}$ densities $\phi_{low}$ each obtained from a single run but $n_{snap}$ snapshots.

\section{Finite Size Effects on Jamming\label{app:FiniteSizeEffectsJamming}}

\begin{figure}
    \centering
    \includegraphics[width = 0.48\columnwidth]{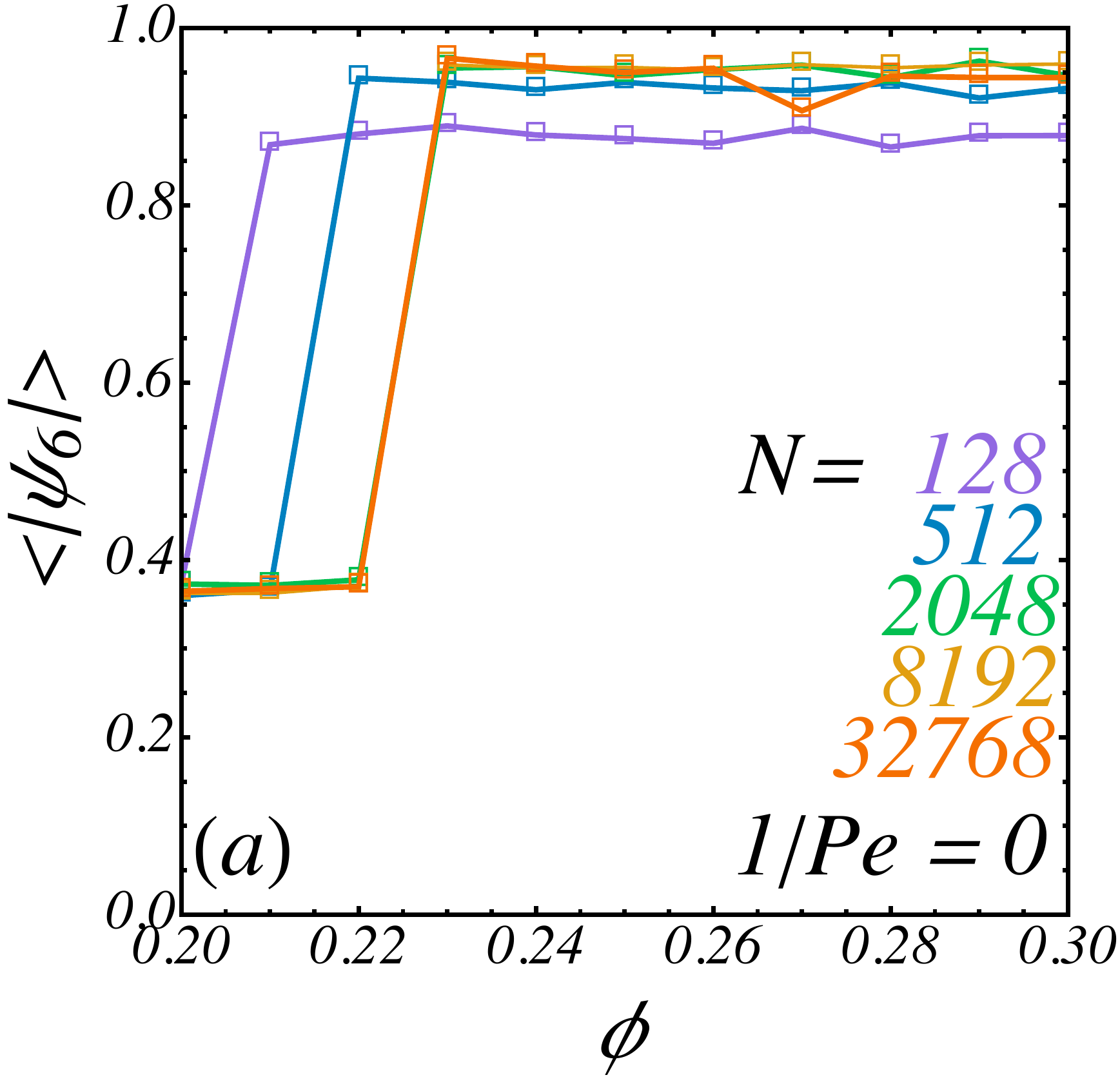}
    \includegraphics[width = 0.48\columnwidth]{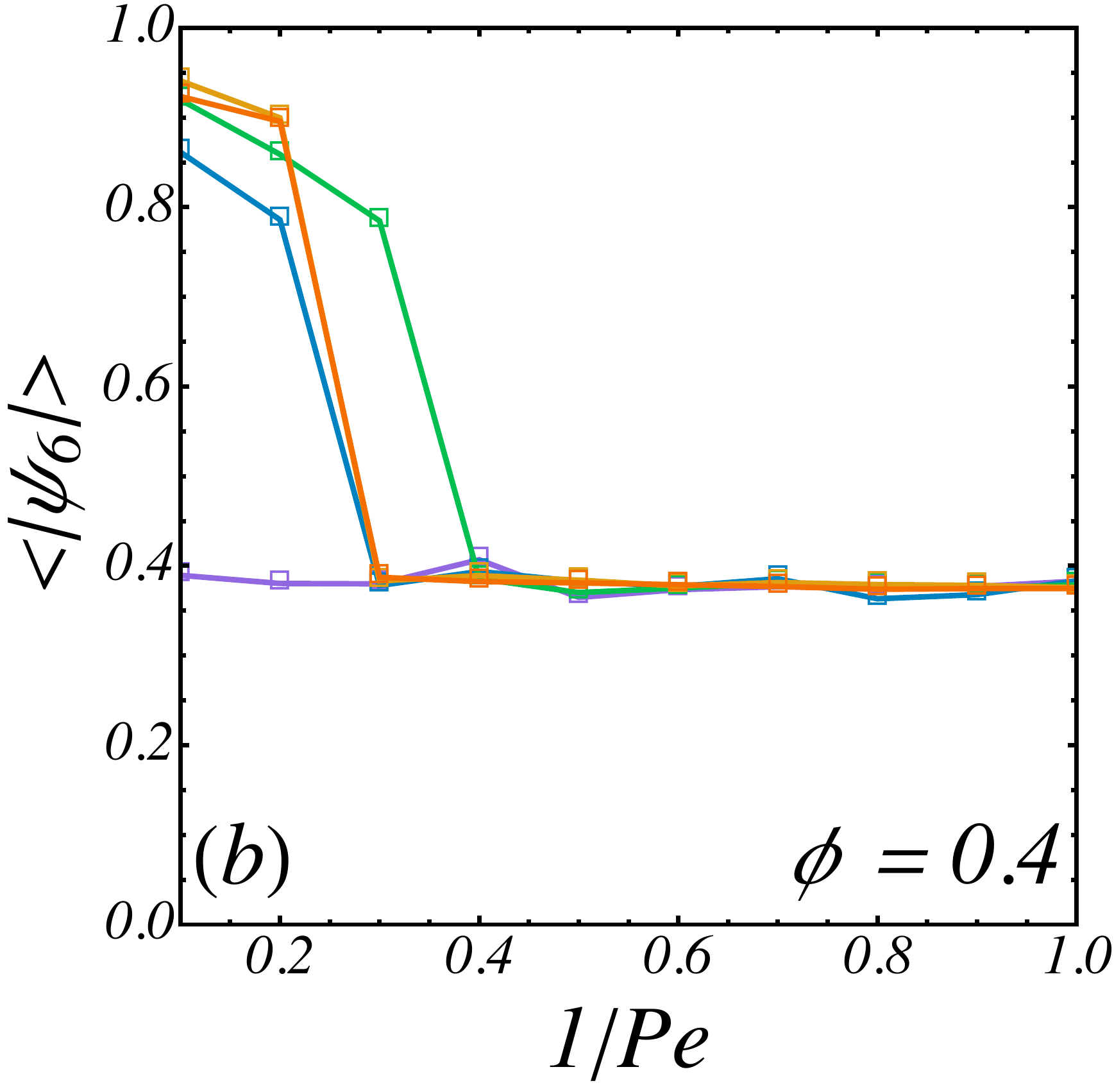}
    \caption{\textbf{Size Effects on Jamming of HBPs.}
    Average of the modulus of the local hexatic order against $(a)$ the density at zero noise and $(b)$ the dimensionless noise amplitude $1/Pe$ at $\phi = 0.4$, in one realisation of the steady state each, across several system sizes between $N = 512$ and $N = 32768$.
    }
    \label{fig:SizeJammingHBPs}
\end{figure}

In the main text, we report occurrences of jams for a large set of different parameters but, for simplicity, focusing only on one system size ($N=2048$) as much as possible.
In this appendix, we present a few results on the finite size effects on the jamming transition.
First, in Fig.~\ref{fig:SizeJammingHBPs}, we focus on HBPs.
In panel $(a)$, we show a good indicator of the jamming transition, the average value of the local hexatic order parameter, against the density, when setting the noise to zero and varying the system size.
We show that the system jams over a density that seems to converge towards the jamming density described in the main text, $\phi \approx 0.23$.
In panel $(b)$, we repeat the same analysis, but this time setting the density above the jamming value ($\phi = 0.4 > \phi_J$) and varying the dimensionless noise amplitude $1/Pe$.
We show that, up to some fluctuations, a jamming transition is observed when $N \geq 512$, at a noise amplitude that seems to converge towards a value $1/Pe \approx 0.2 -- 0.3$.
Note that the smallest system here never jams at any shown value of the noise: very small jams are seemingly less stable against perturbations.

\begin{figure}
    \centering
    \includegraphics[width = 0.48\columnwidth]{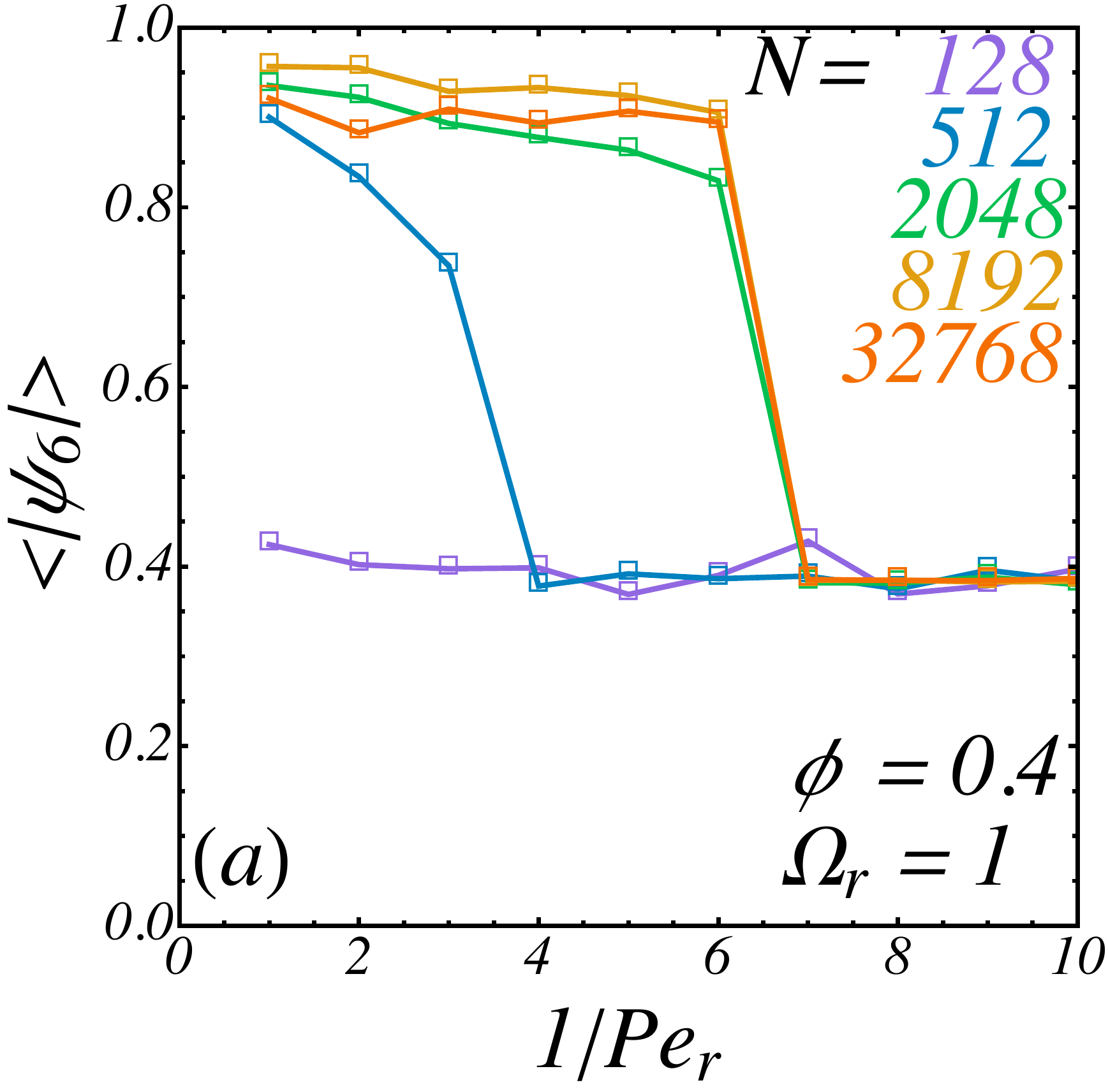}
    \includegraphics[width = 0.48\columnwidth]{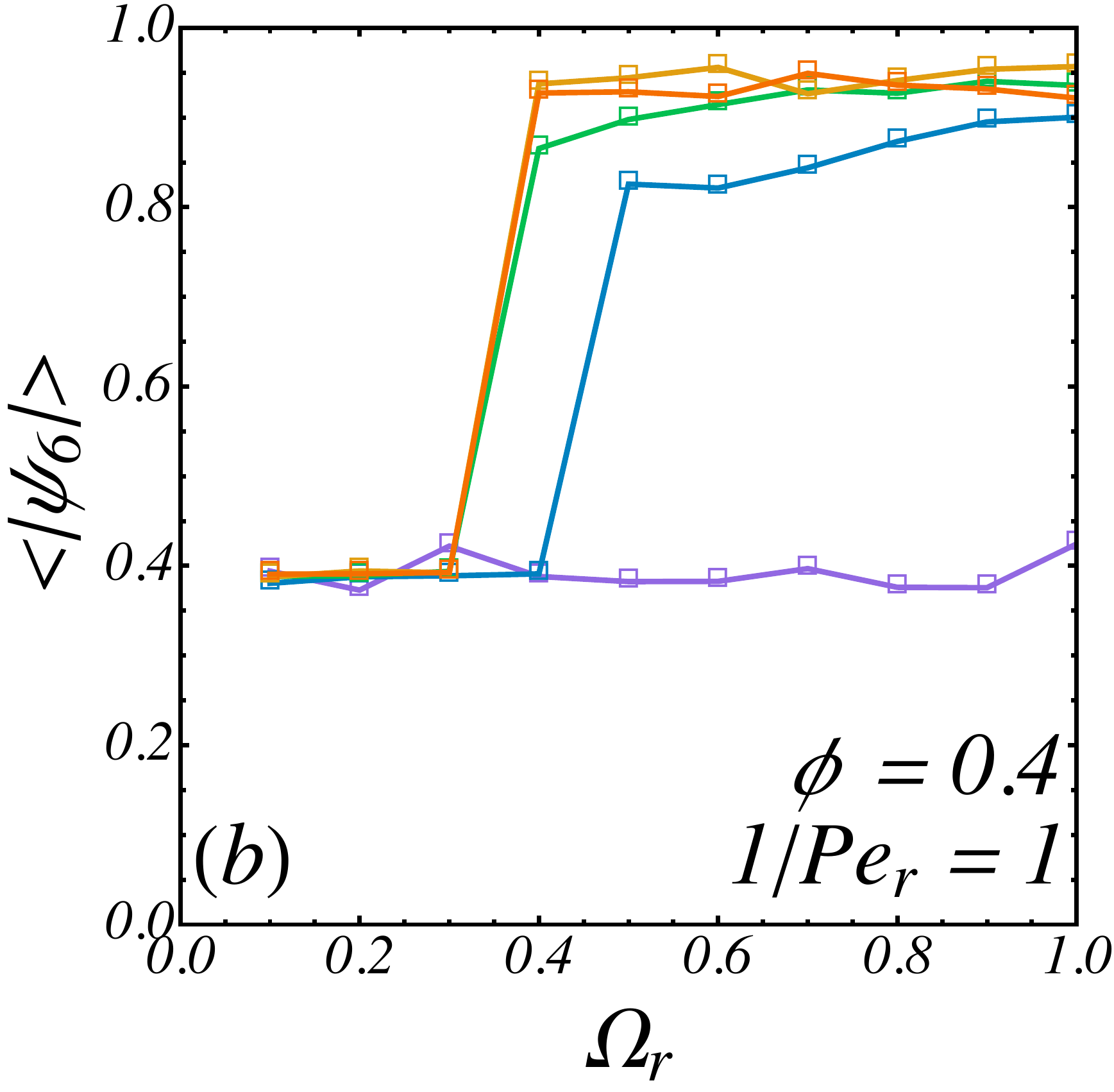}
    \caption{\textbf{Size Effects on Jamming of HABPs.}
    Average of the modulus of the local hexatic order against $(a)$ the dimensionless rotational noise amplitude $1/Pe_r$, and $(b)$ the dimensionless relaxation rate $\Omega_r$ at $\phi = 0.4$, in one realisation of the steady state each, across several system sizes between $N = 512$ and $N = 32768$.
    }
    \label{fig:SizeJammingHABPs}
\end{figure}

We now turn our attention to HABPs, in which one can tweak two other parameters: the relaxation rate at a fixed rotational noise, and the rotational noise at a fixed relaxation rate.
We show the results in Fig.~\ref{fig:SizeJammingHABPs}.
First, in panel $(a)$, we vary the dimensionless rotational noise amplitude $1/Pe_r$ at a fixed relaxation rate $\Omega_r = 1$.
Like in HBPs, small jams are less stable against perturbations, so that the $N=128$ line is unjammed by values of the noise not attained here, and the $N=512$ line displays unjamming at a value of the noise that is significantly smaller than the one observed in larger systems.
For $N > 512$ however, the system is jammed until a value of the noise that seems to converge to a finite value $1/Pe_r \approx 6$.
Finally, in panel $(b)$, we vary the dimensionless relaxation rate $\Omega_r$ at a fixed rotational noise amplitude, $1/Pe_r = 1$.
We show once again that $N=128$ is too small for jams to be stable against perturbations, but that past $N \geq 512$, jams are observed when relaxation is fast enough, beyond a value that seems to converge to $\Omega_r \approx 0.3$.

All in all, we showed that jamming occurs in a domain of parameters that seem to converge to well-defined values in the $N \to \infty$ limit.
At all sizes tested in this paper, the transition seems discontinuous.
Additionally, it seems that jams are rather unstable against perturbations in the small $N$ limit.

\section{MIPS clusters as seeds for jams}

\subsection{ABP phase diagram\label{app:ABPMIPS}}

\begin{figure}
    \centering
    \includegraphics[width=0.96\columnwidth]{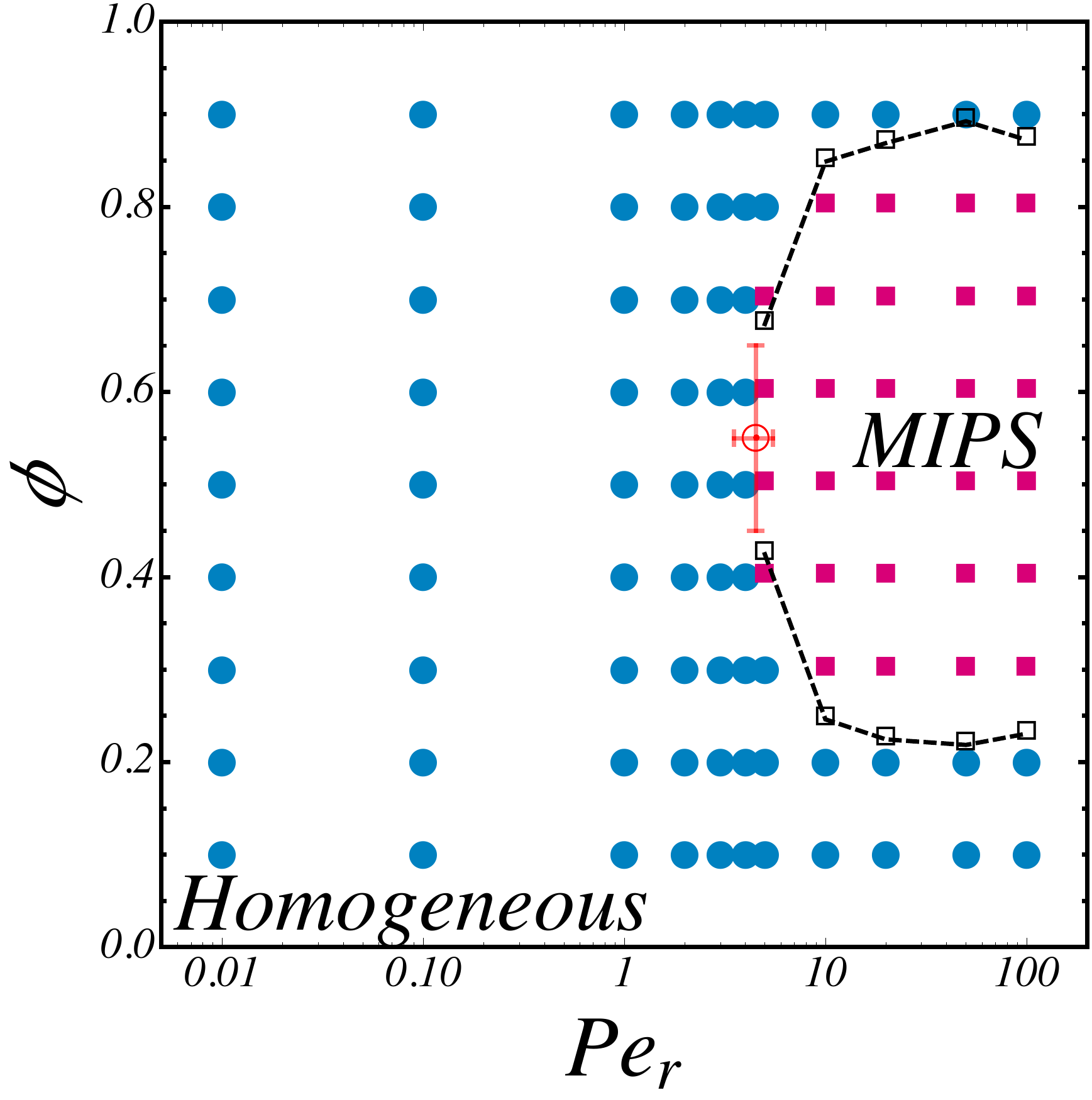}
    \caption{\textbf{Phase diagram of ABPs.}
    We map uniform states (blue disks), and bimodal density distributions (magenta squares), used to locate the two branches of phase separation (black squares, with $95\%$ confidence intervals). They meet at a critical point (red circle).
    In this figure, $N = 2048$.}
    \label{fig:MIPSPhaseDiagram}
\end{figure}

In the main text, we discuss the link between the jamming packing fraction $\phi_J$ of Homing Active Brownian Particles, and the lower-branch value of Motility-Induced Phase Separation in Active Brownian Particles in the limit $Pe_r \to \infty$.
To back up the comparison between these two phenomena, we set the relaxation rate $\Omega_r$ and translational noise amplitude $Pe^{-1} \propto D_0$ strictly to zero, thus defining a model of ABPs, and we simulate the system for various values of the packing fraction $\phi$ and rotational Péclet.
The results are summarised in Fig.~\ref{fig:MIPSPhaseDiagram}, which was obtained for $N=2048$.
We find a MIPS domain at high enough values of the rotational Péclet number (low enough values of the noise), using the method described in Sec.~\ref{sec:PhDConstruction}, we construct the distribution of the local, coarse-grained packing fraction.
When possible, we locate the phase separation branches by computing the coarse-grained density distribution, obtained by segmenting the simulation box into smaller boxes with side-length $\ell = 10a$ and counting the particles inside each box, across several (about $10^2$) independent snapshots of the system for each set of parameters.
As expected from previous studies~\cite{Solon2015d,Nie2020}, this domain is U-shaped, stemming from a critical point that we find around $(Pe_{r,C},\phi_C) \approx (4.5, 0.55)$, and lower branch that saturates at a finite density.

\subsection{Transient regime\label{app:JammingvsMIPS}}

\begin{figure*}
    \centering
    \includegraphics[width=0.135\textwidth]{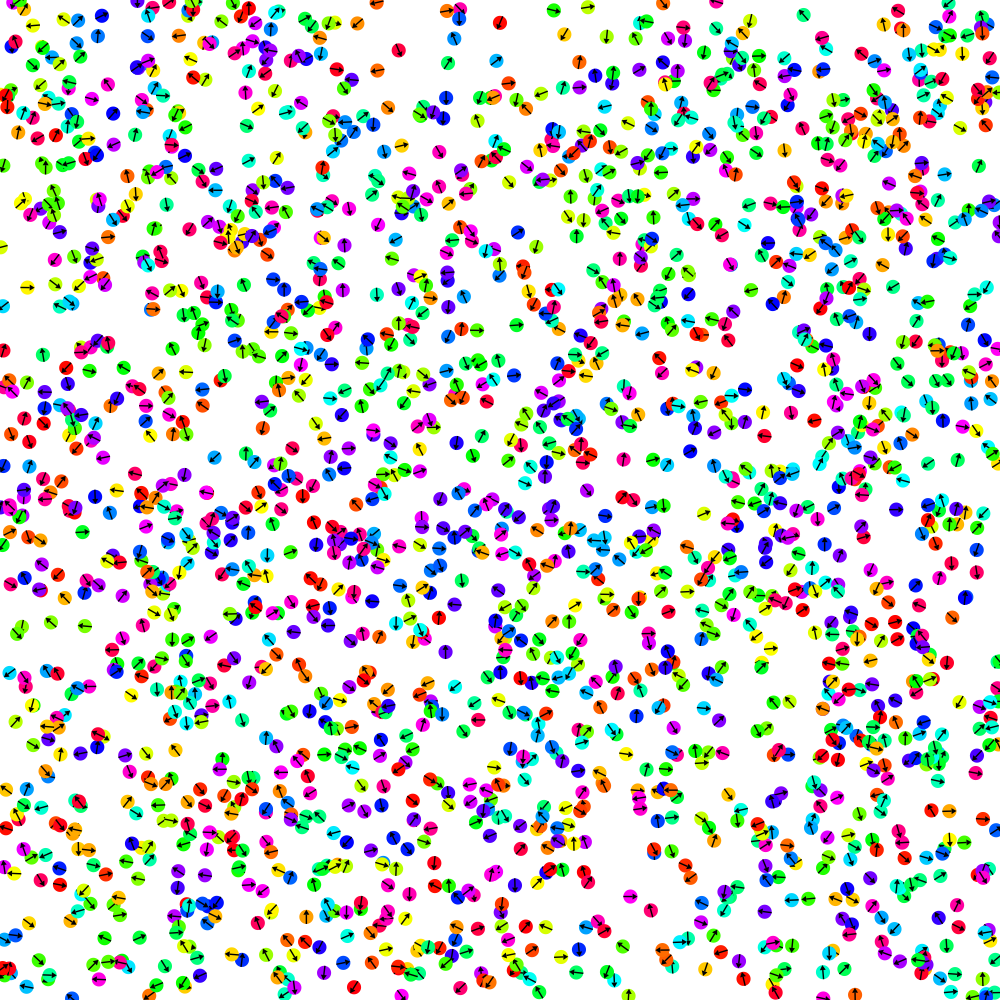}
    \includegraphics[width=0.135\textwidth]{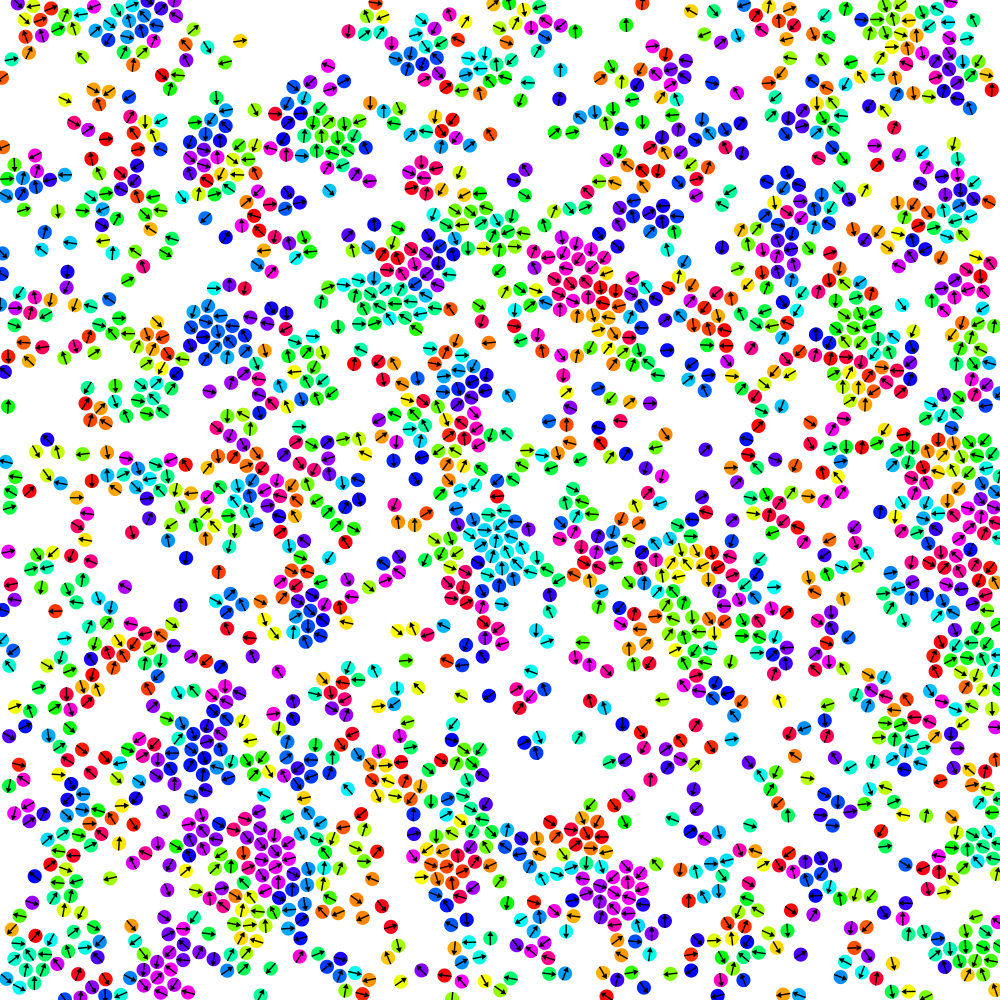}
    \includegraphics[width=0.135\textwidth]{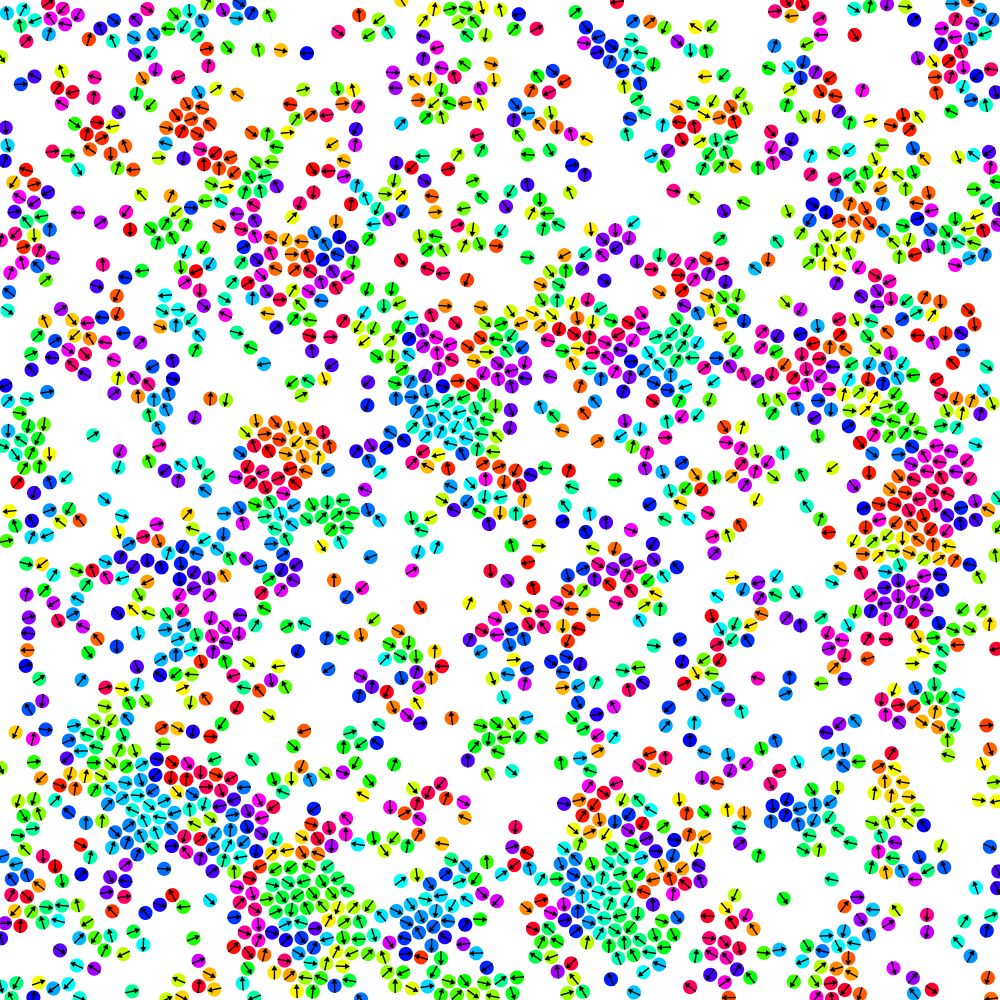}
    \includegraphics[width=0.135\textwidth]{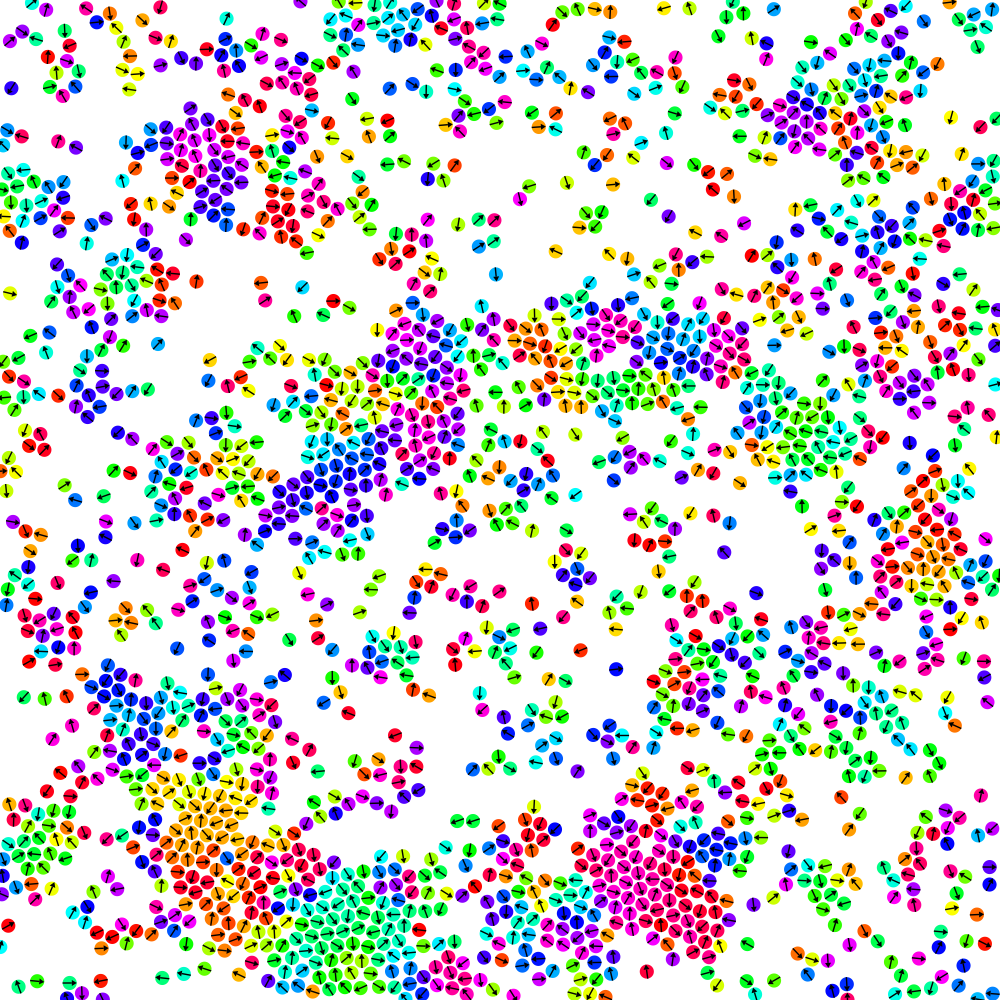}
    \includegraphics[width=0.135\textwidth]{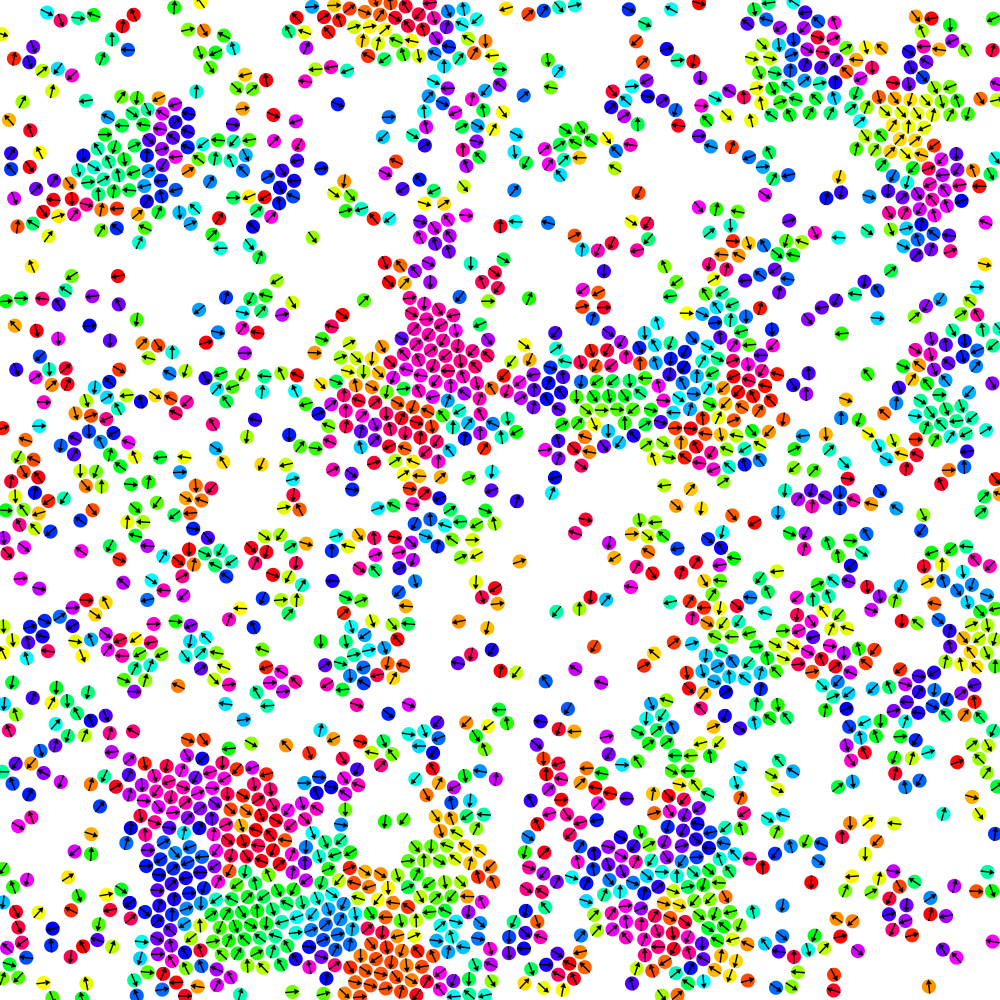}
    \includegraphics[width=0.135\textwidth]{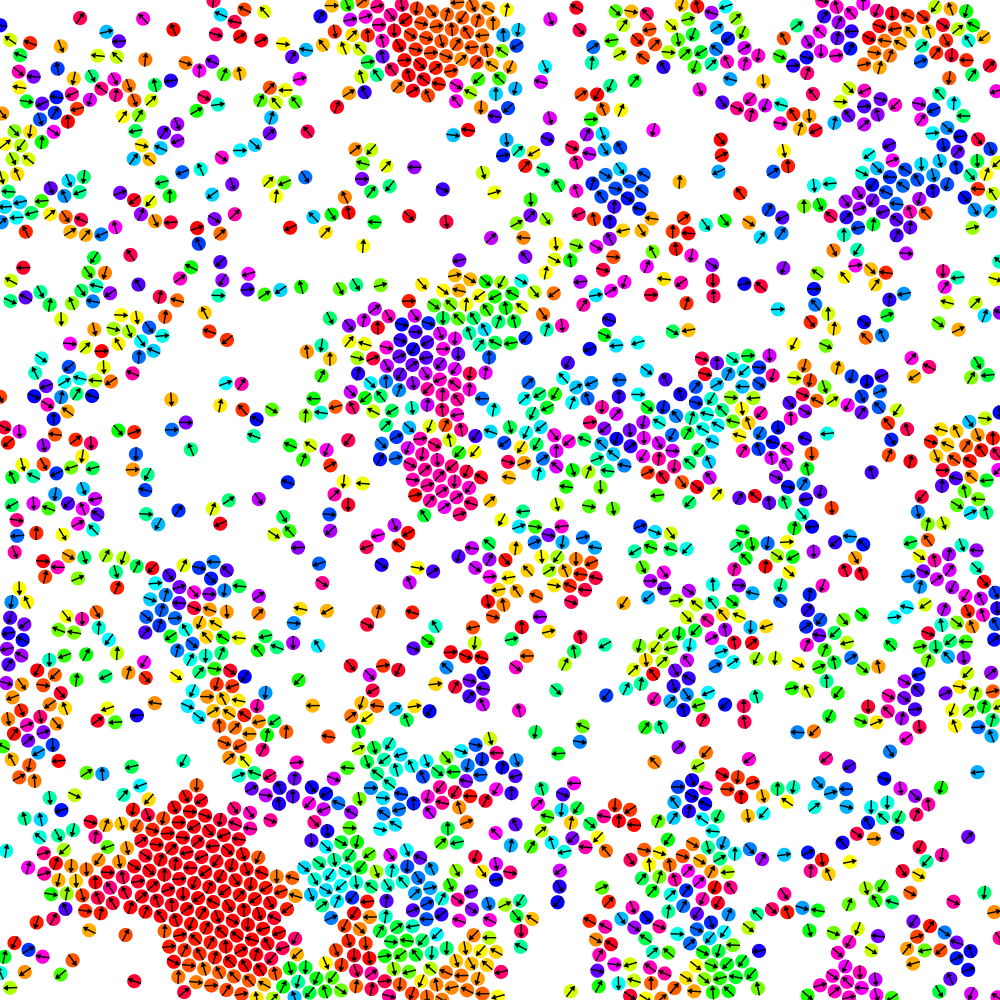}
    \includegraphics[width=0.135\textwidth]{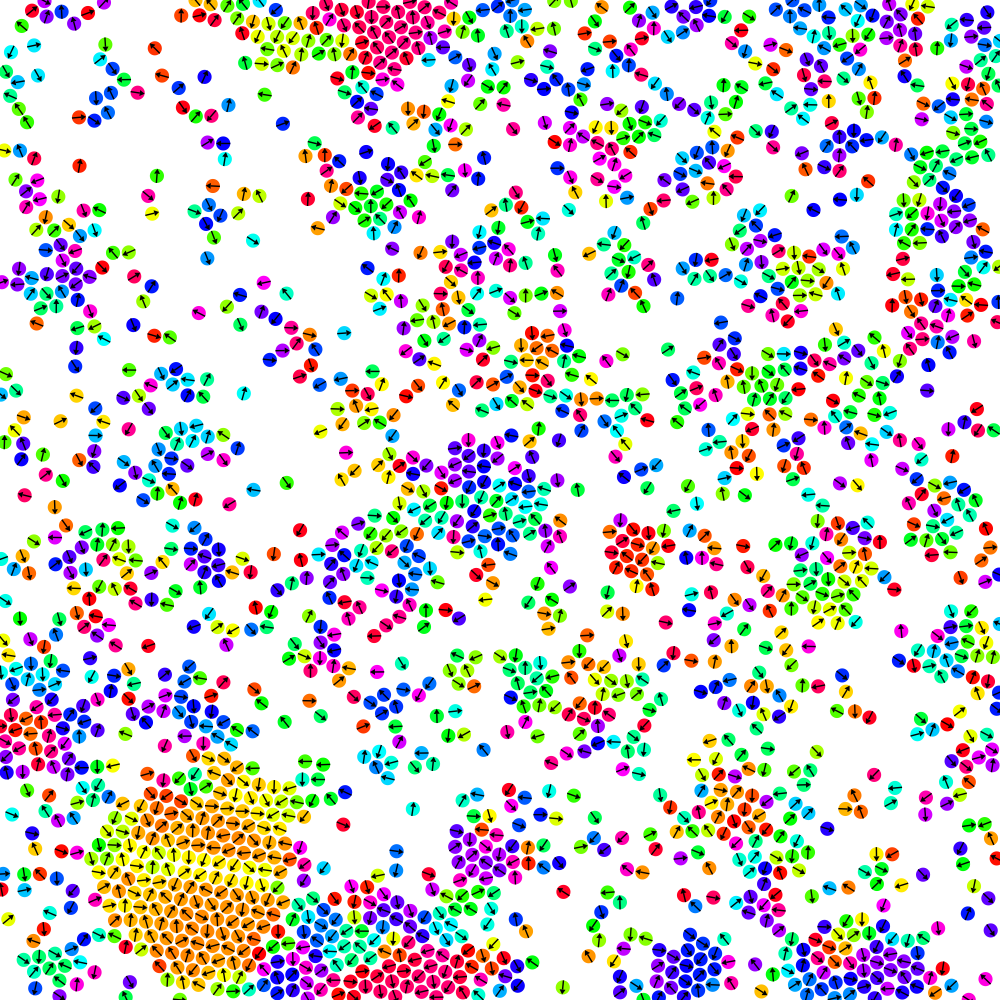} \\
    \includegraphics[width=0.135\textwidth]{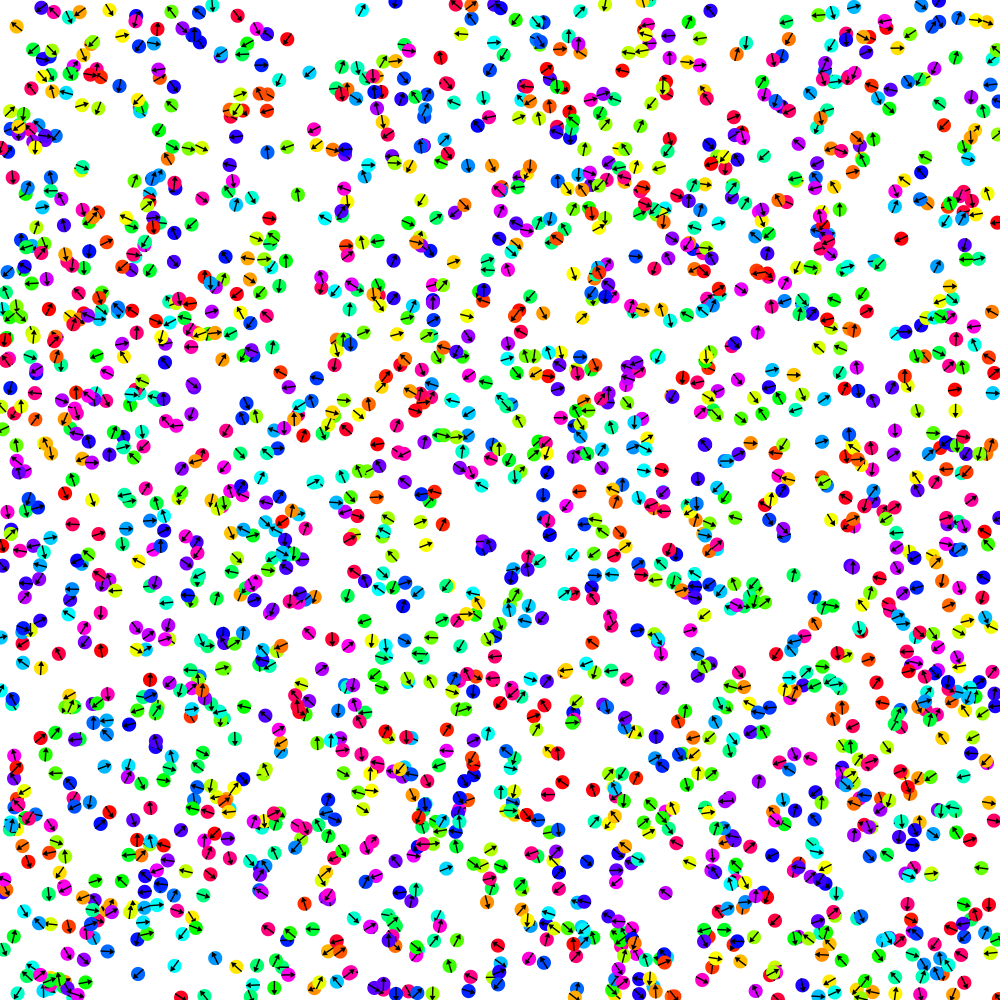}
    \includegraphics[width=0.135\textwidth]{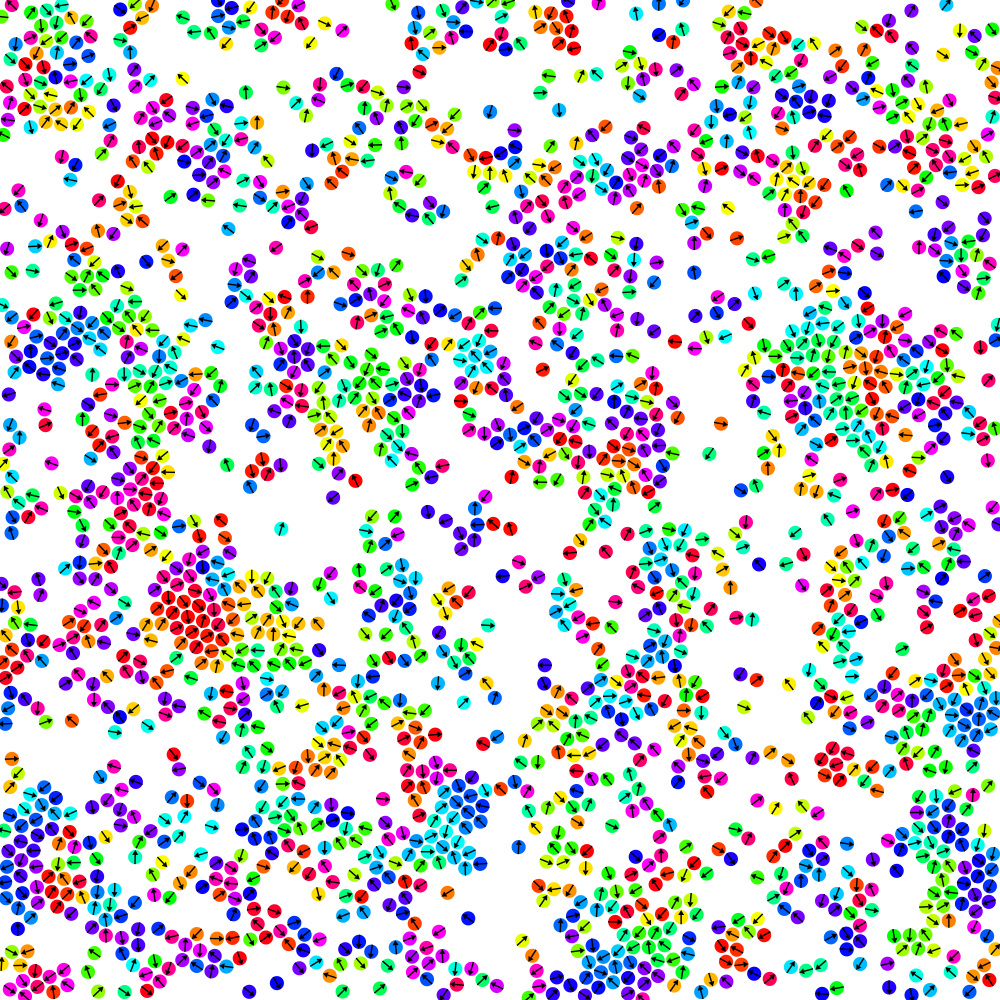}
    \includegraphics[width=0.135\textwidth]{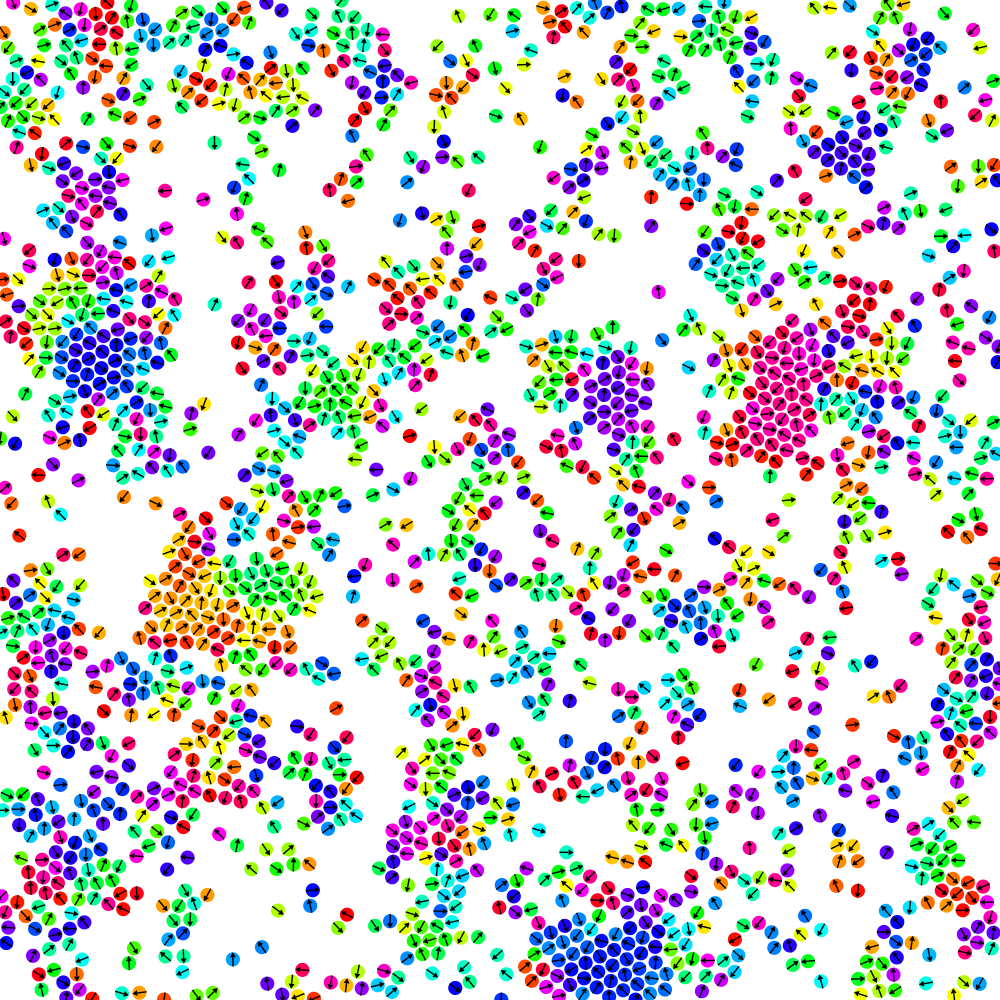}
    \includegraphics[width=0.135\textwidth]{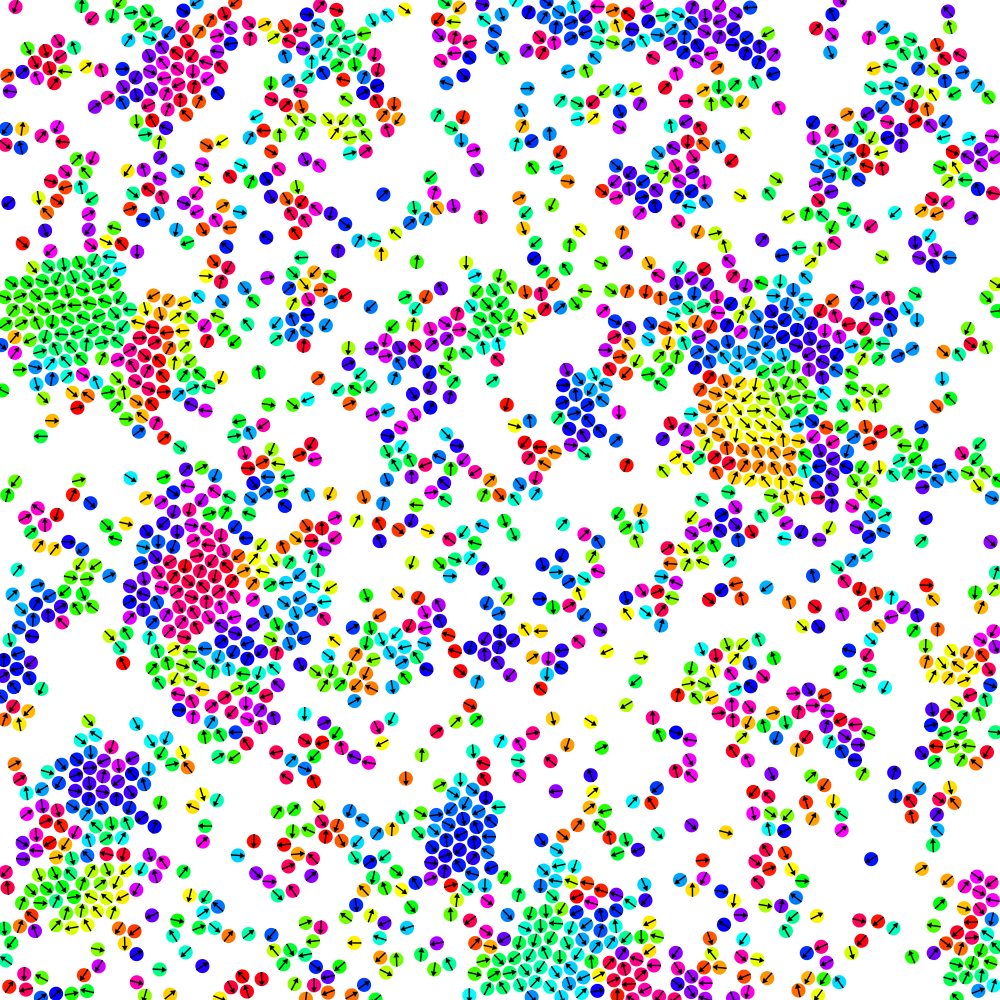}
    \includegraphics[width=0.135\textwidth]{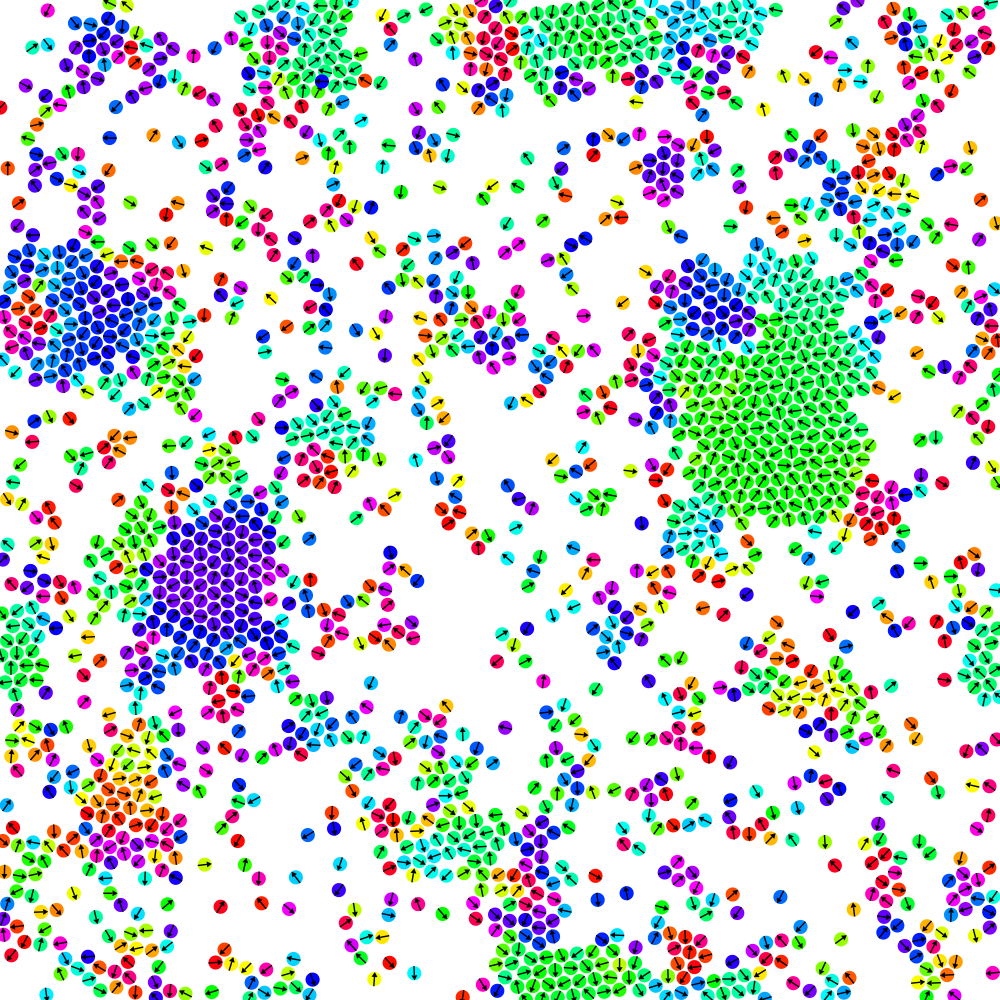}
    \includegraphics[width=0.135\textwidth]{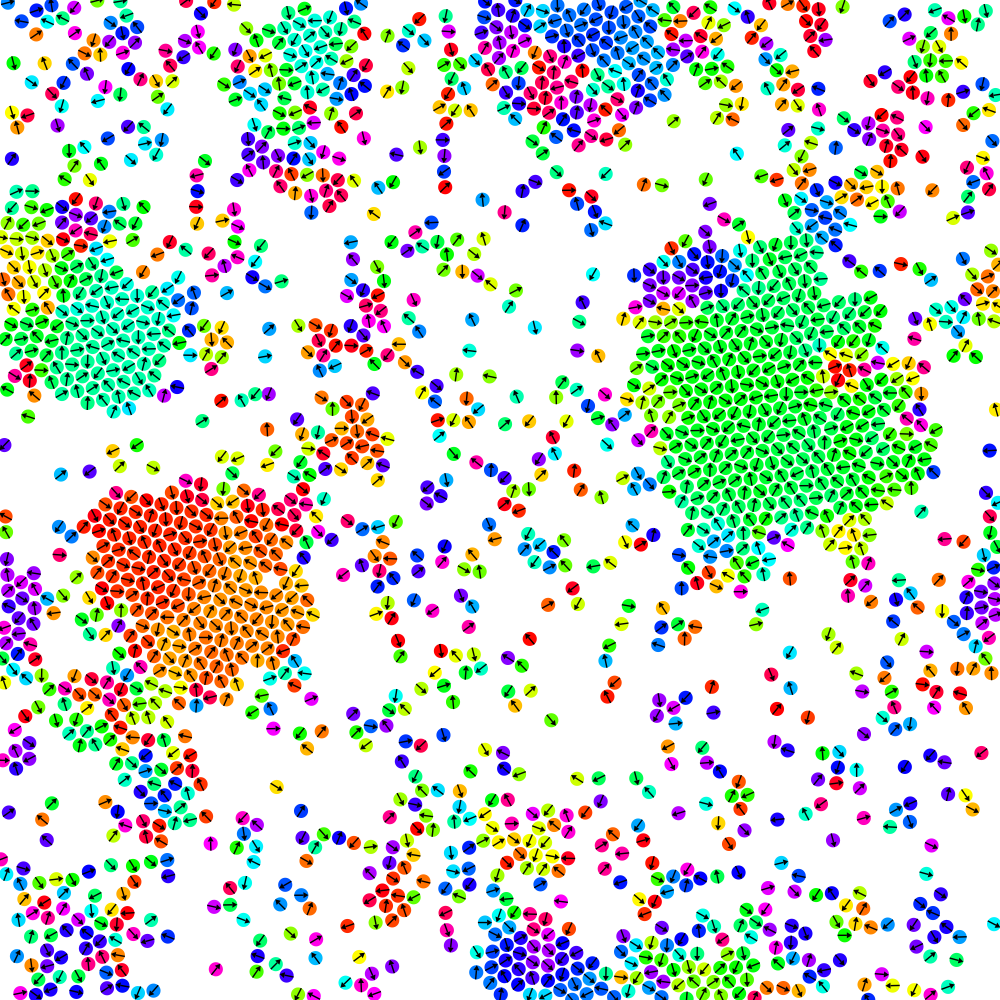}
    \includegraphics[width=0.135\textwidth]{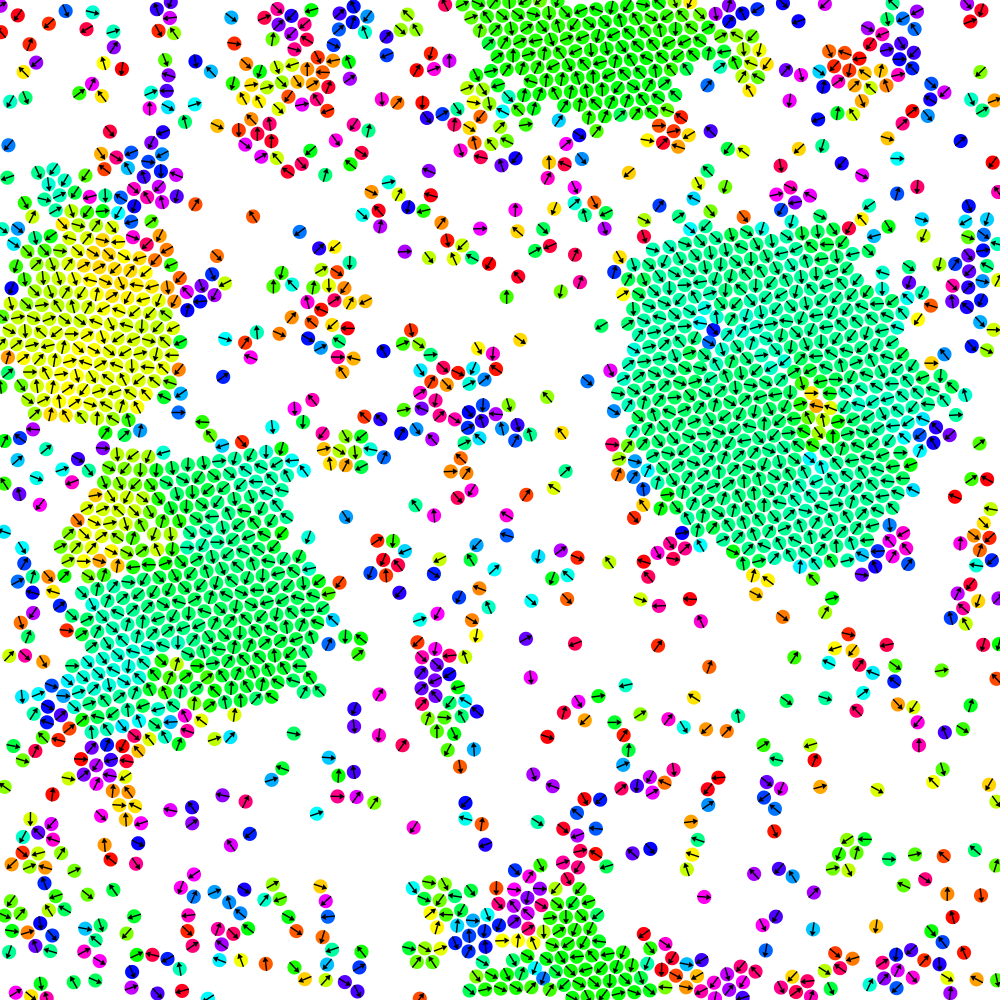} \\
    \includegraphics[width=0.135\textwidth]{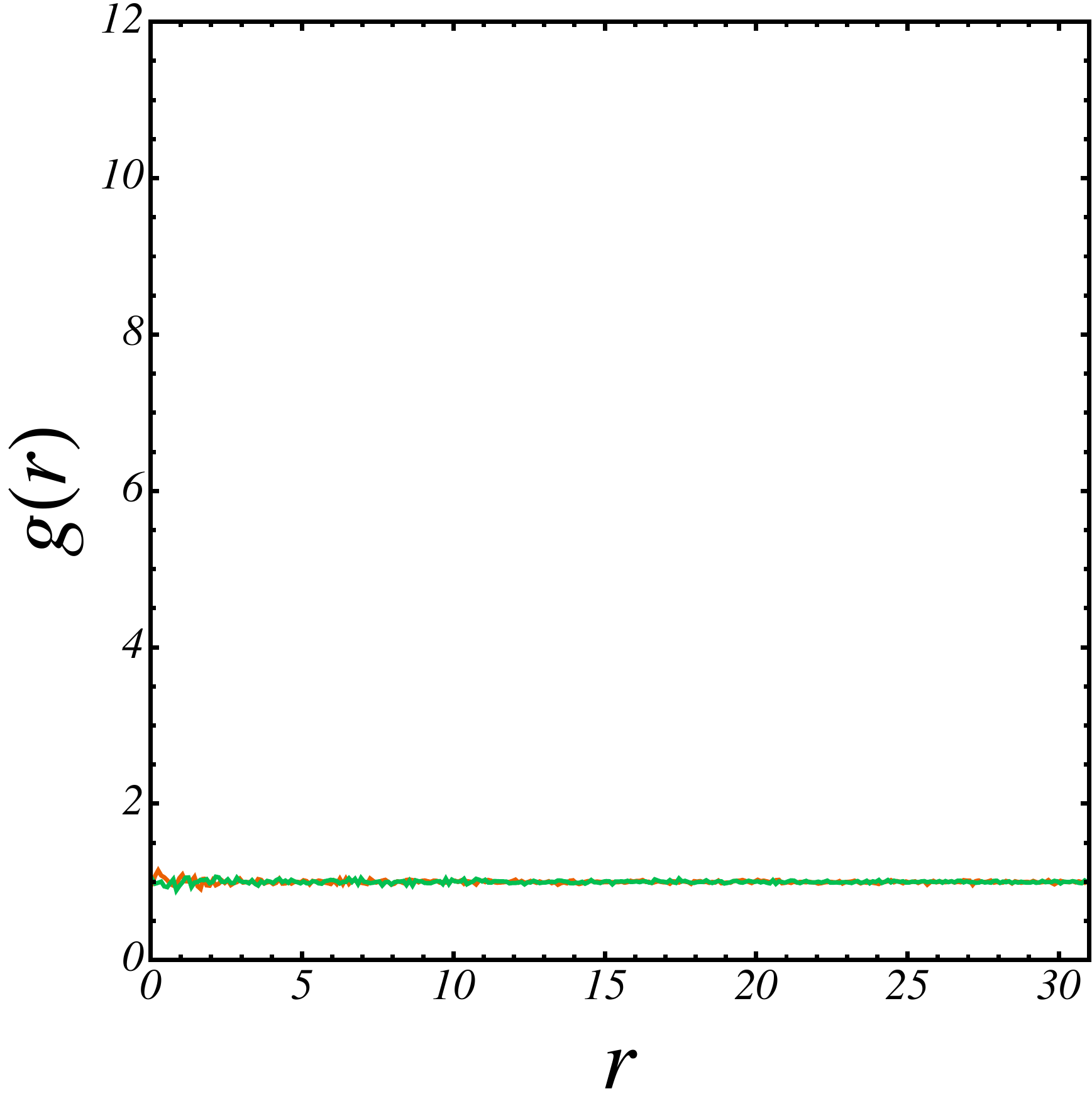}
    \includegraphics[width=0.135\textwidth]{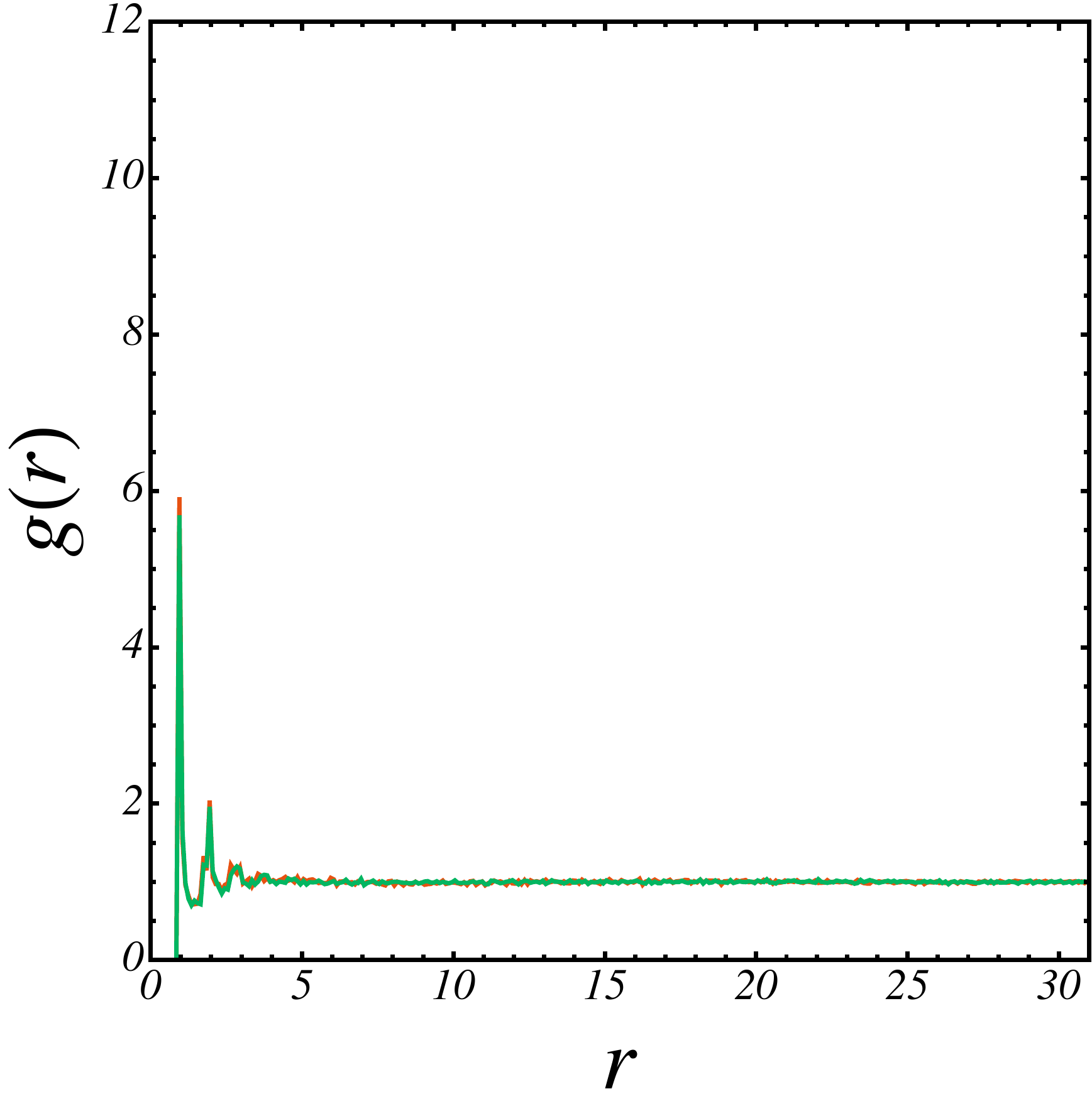}
    \includegraphics[width=0.135\textwidth]{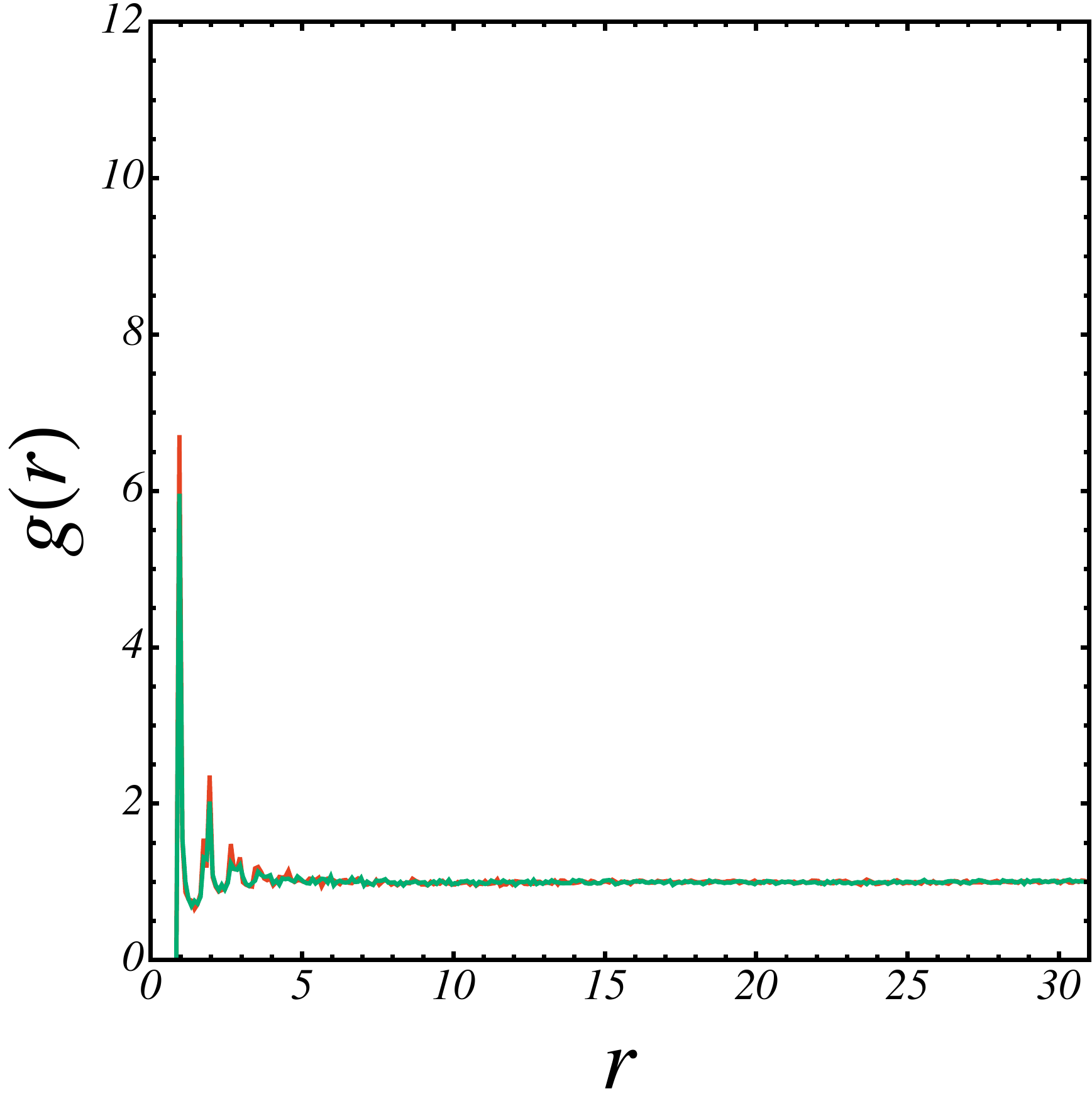}
    \includegraphics[width=0.135\textwidth]{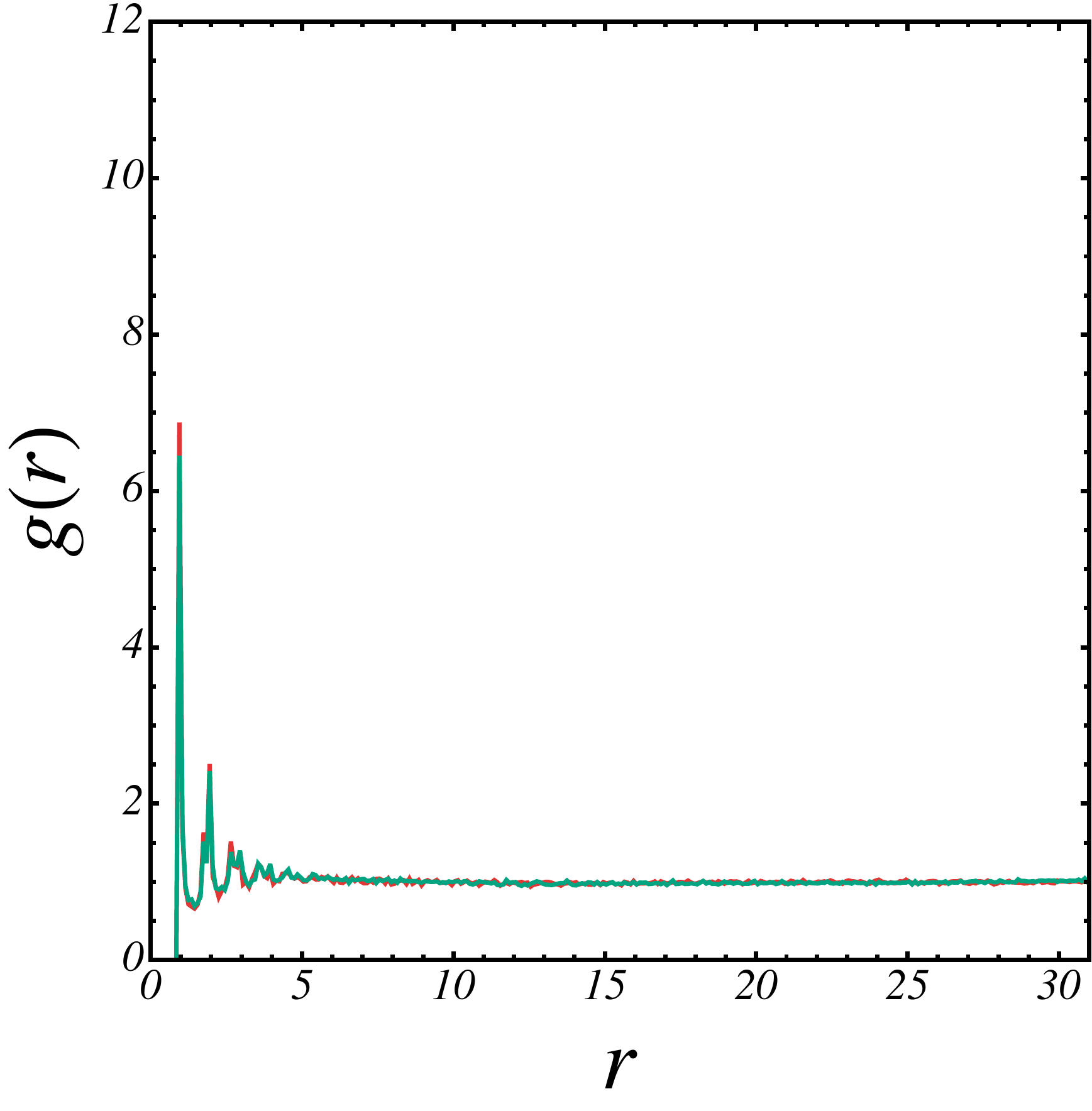}
    \includegraphics[width=0.135\textwidth]{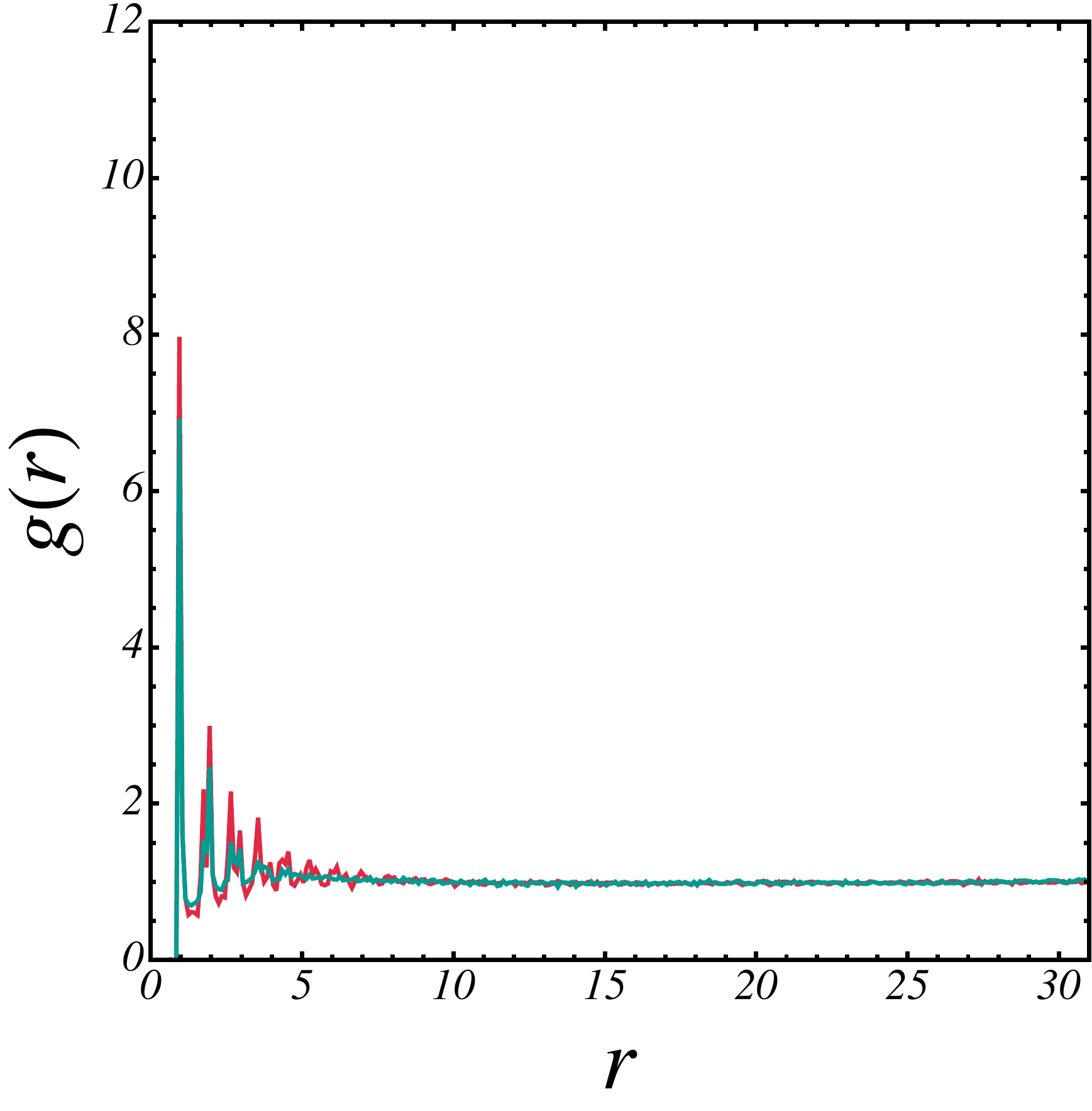}
    \includegraphics[width=0.135\textwidth]{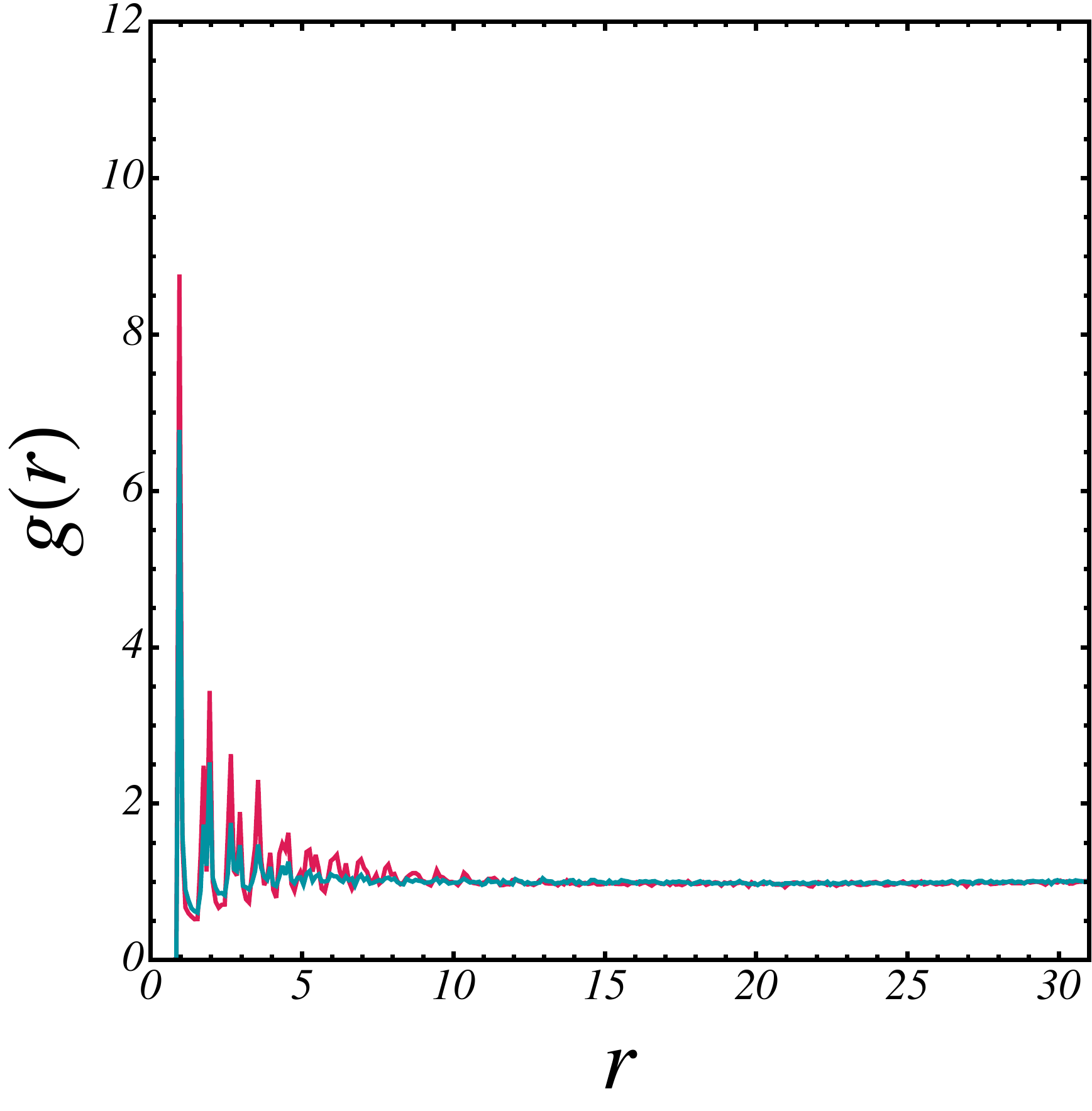}
    \includegraphics[width=0.135\textwidth]{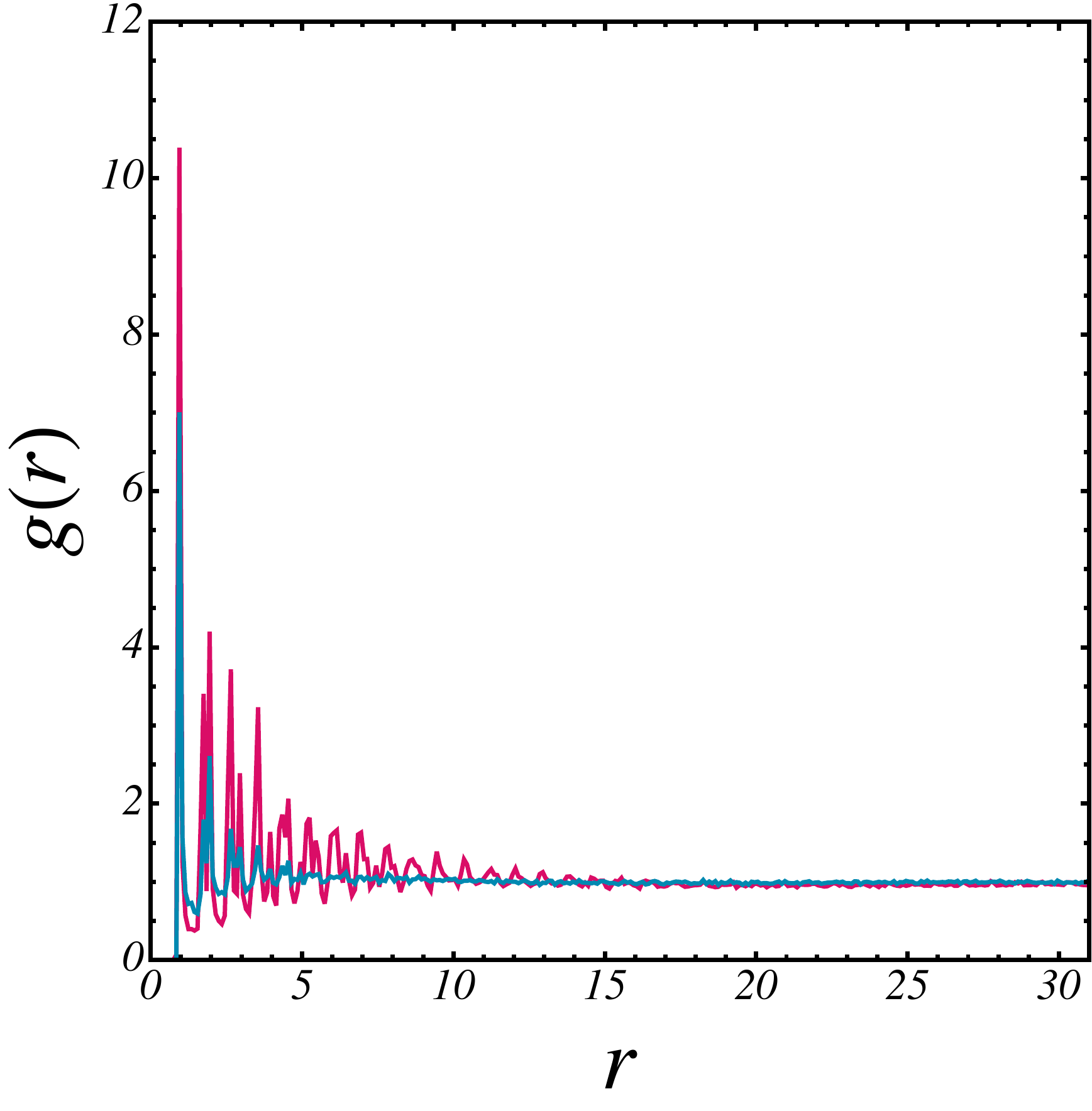} \\
    \includegraphics[width=0.135\textwidth]{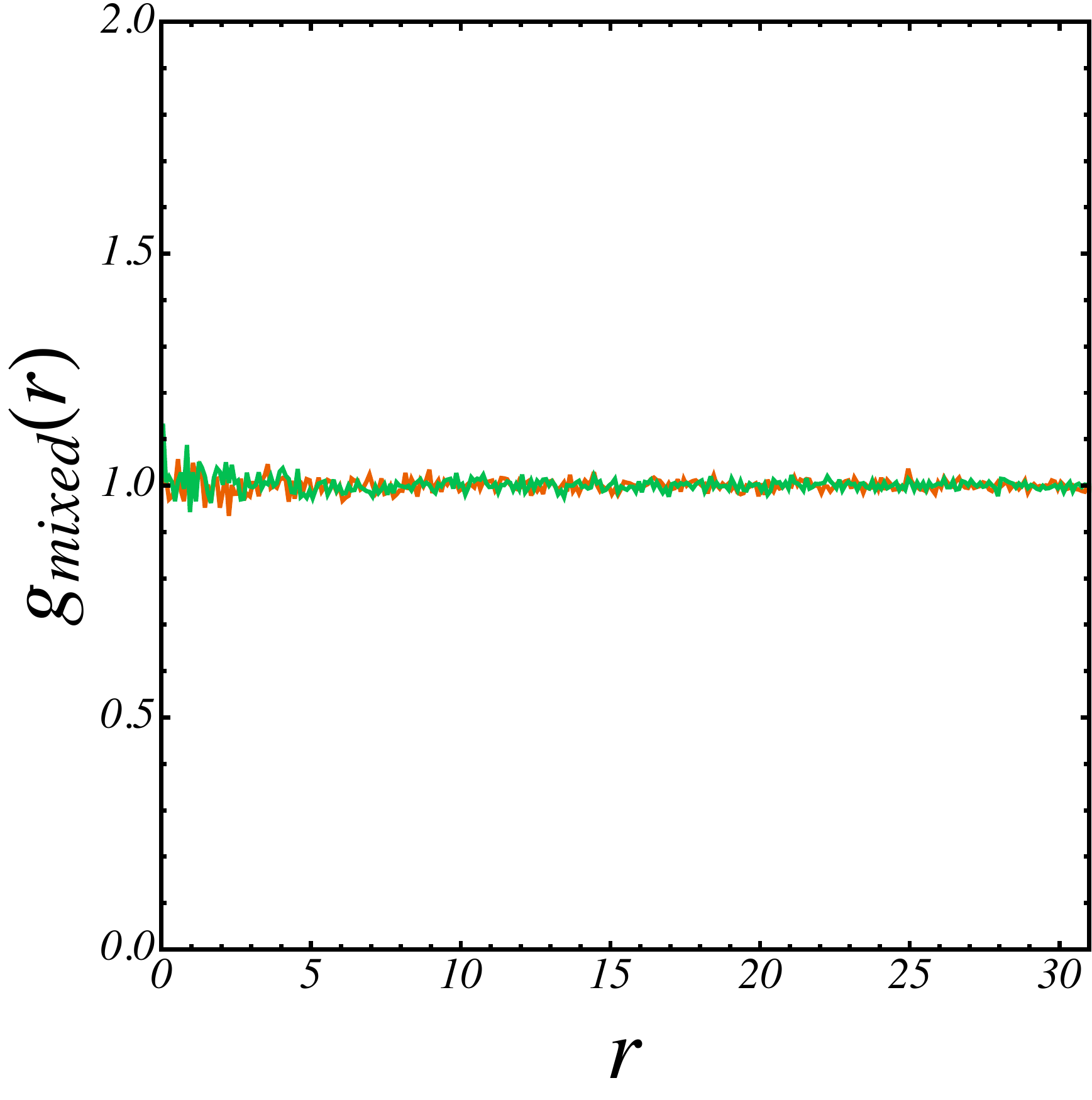}
    \includegraphics[width=0.135\textwidth]{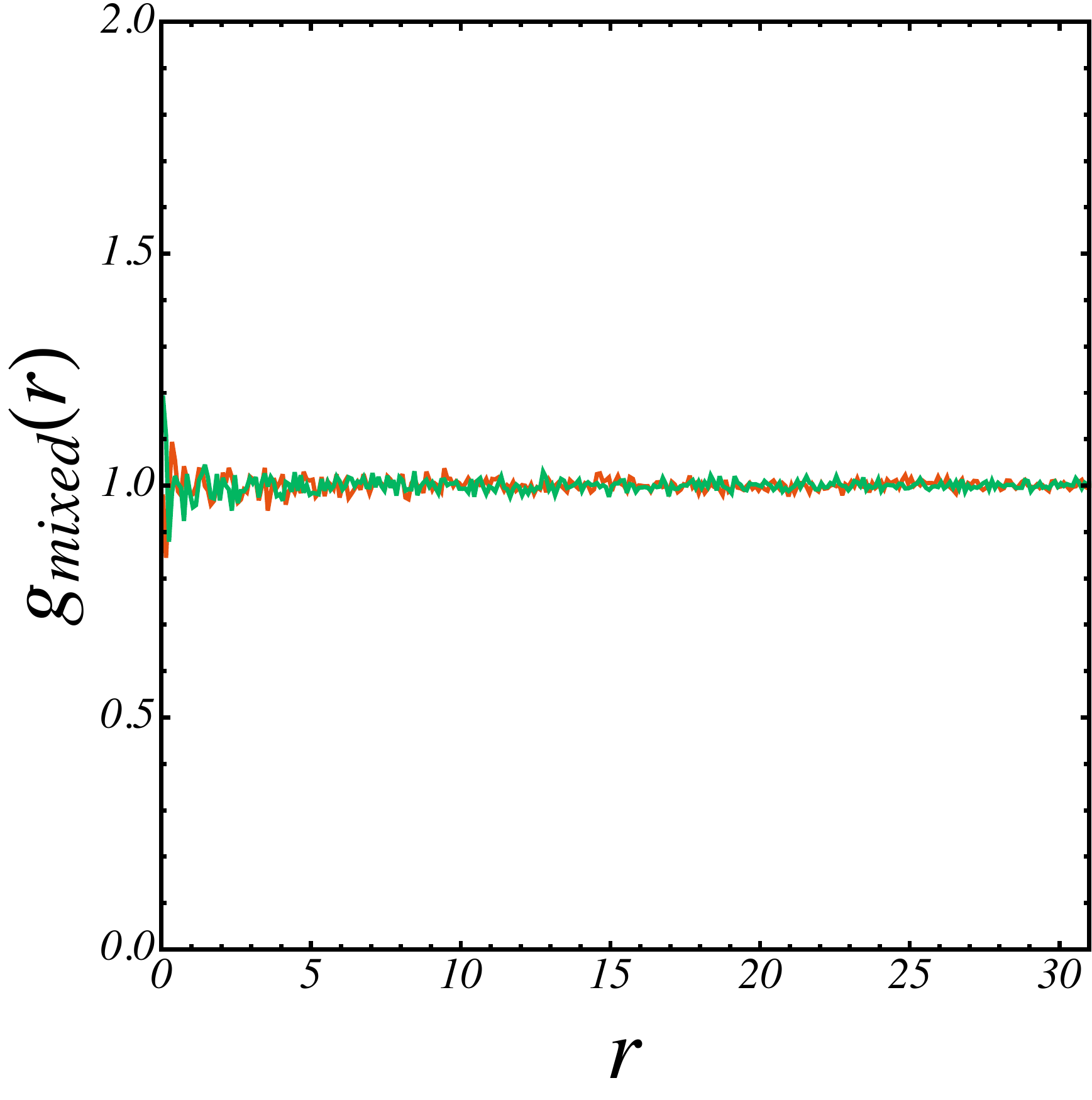}
    \includegraphics[width=0.135\textwidth]{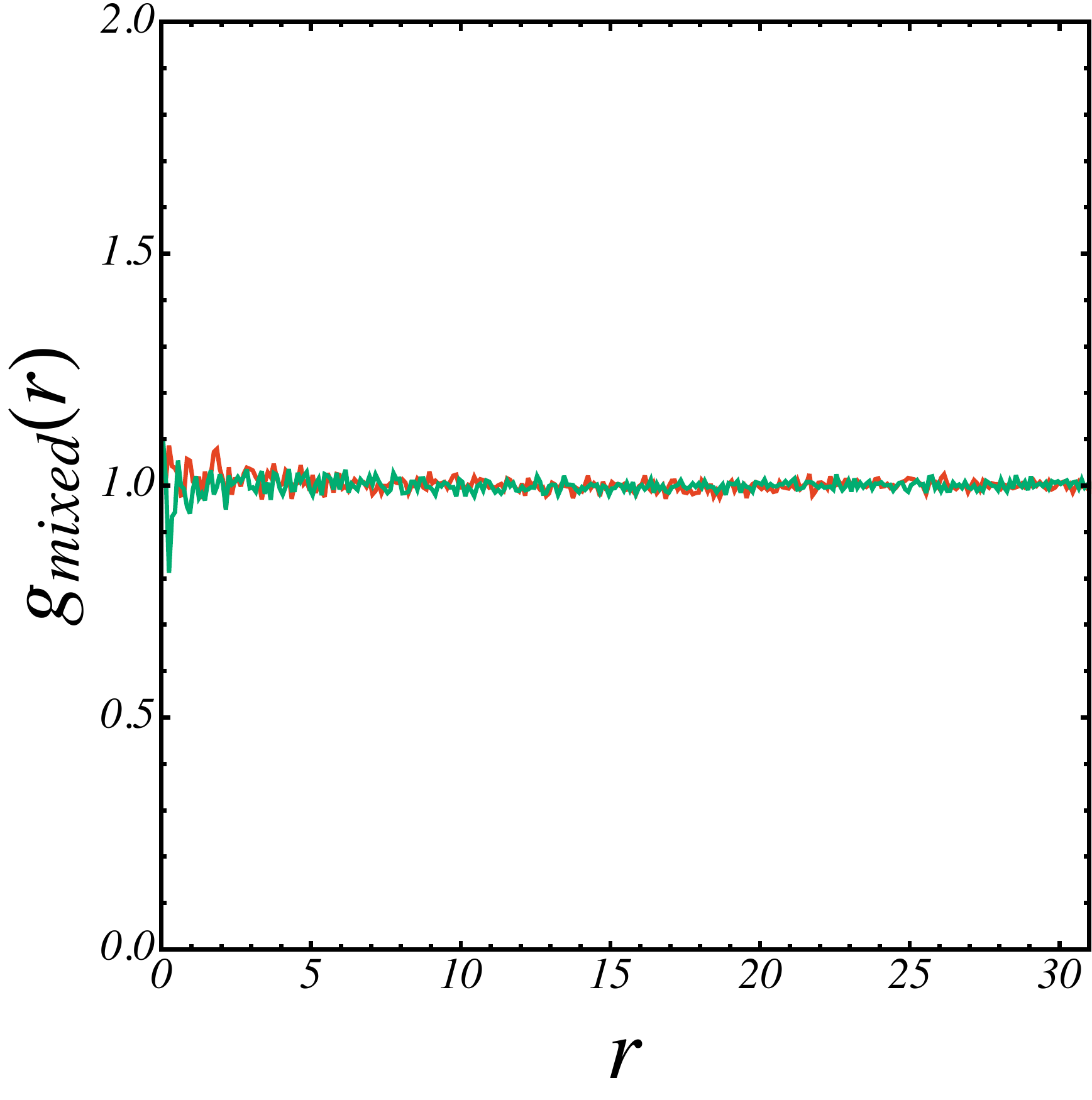}
    \includegraphics[width=0.135\textwidth]{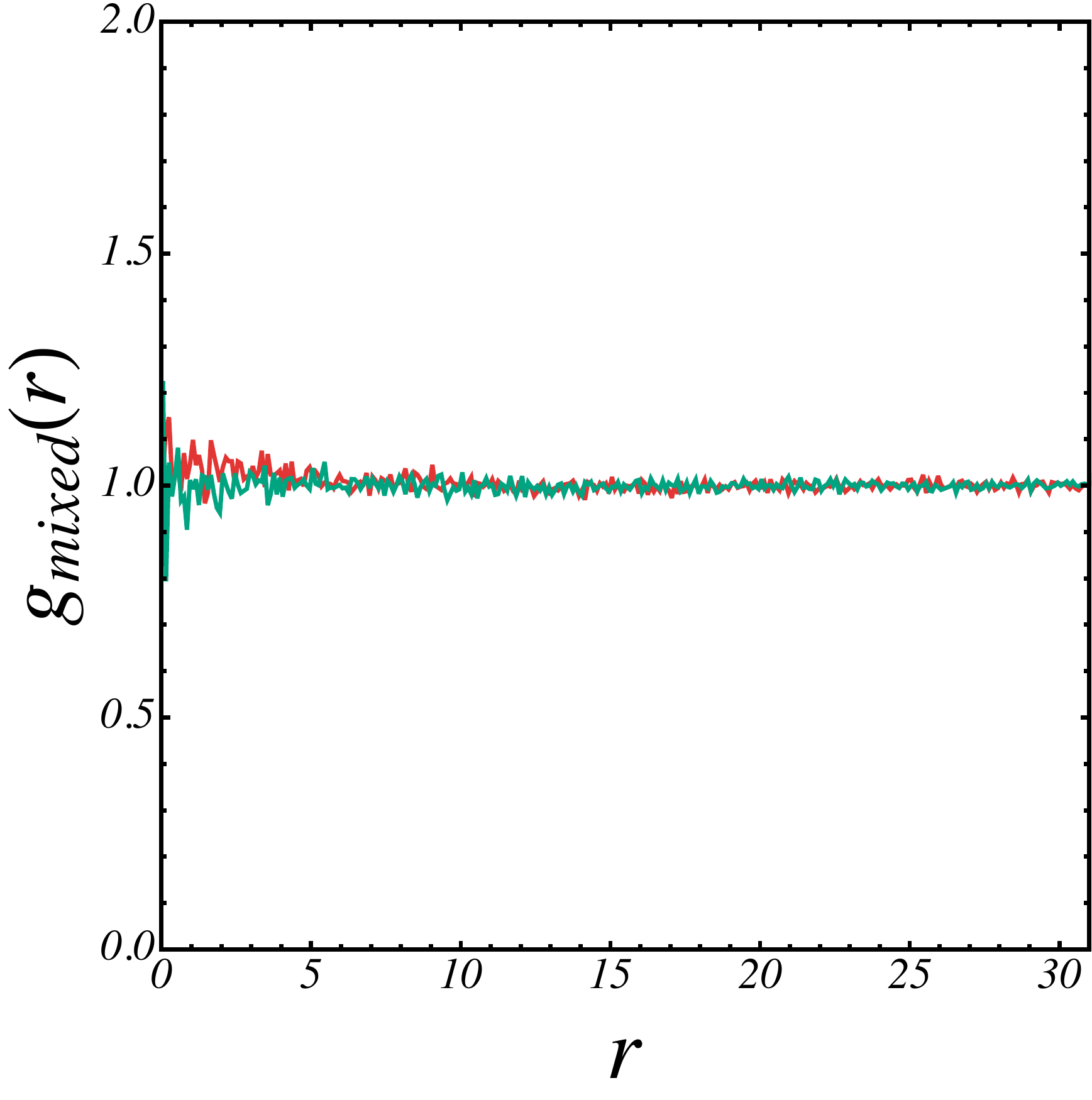}
    \includegraphics[width=0.135\textwidth]{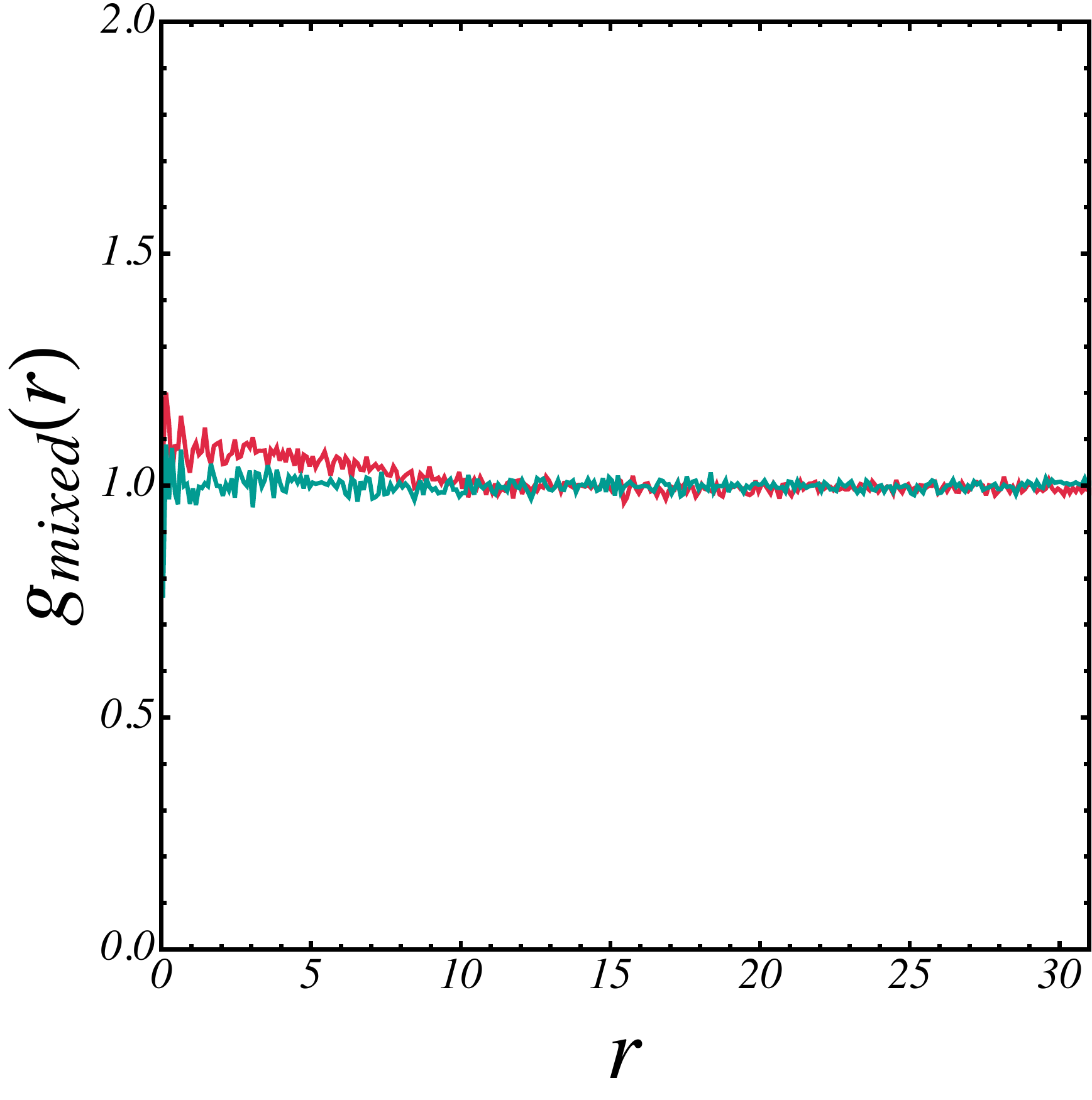}
    \includegraphics[width=0.135\textwidth]{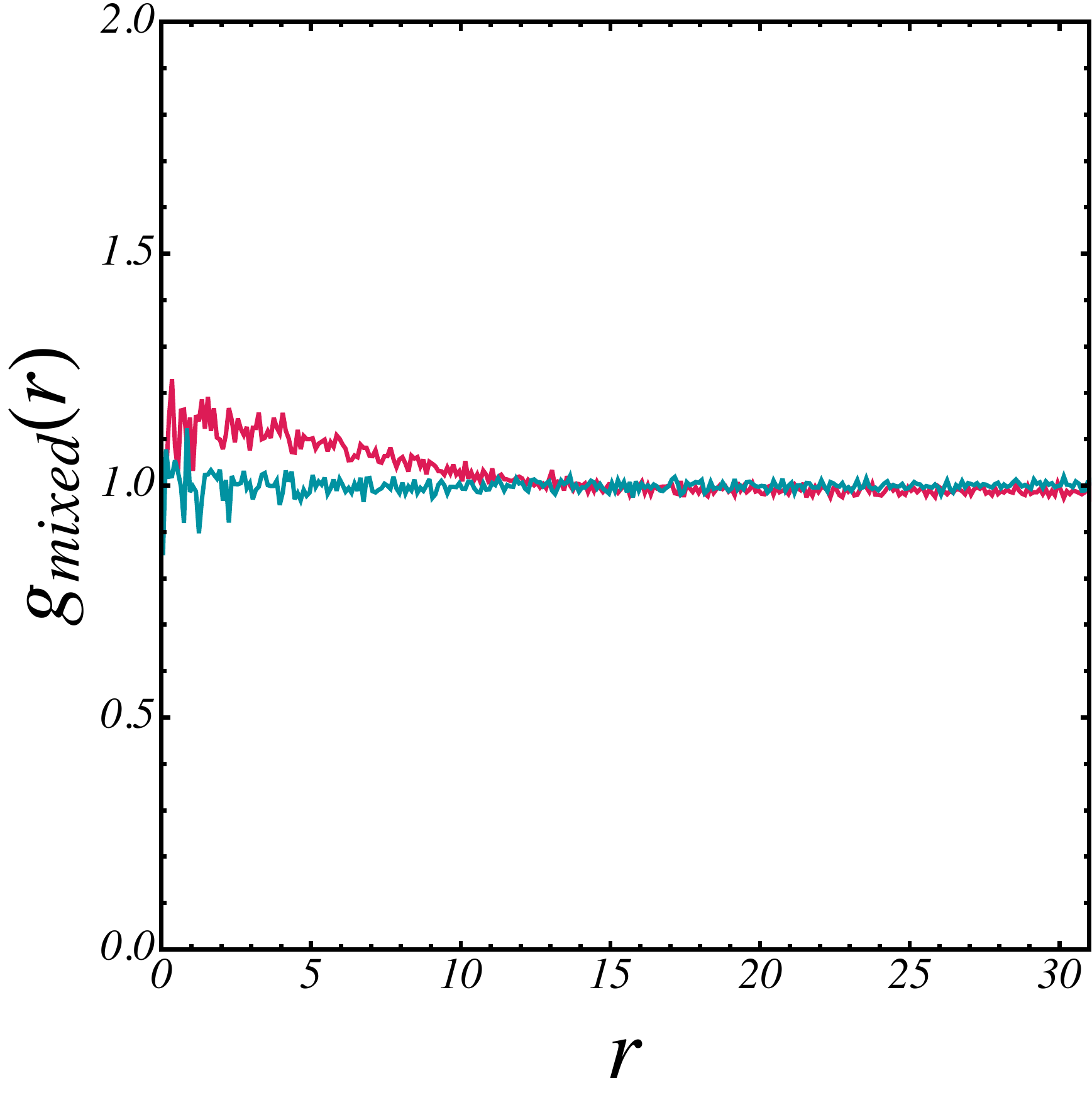}
    \includegraphics[width=0.135\textwidth]{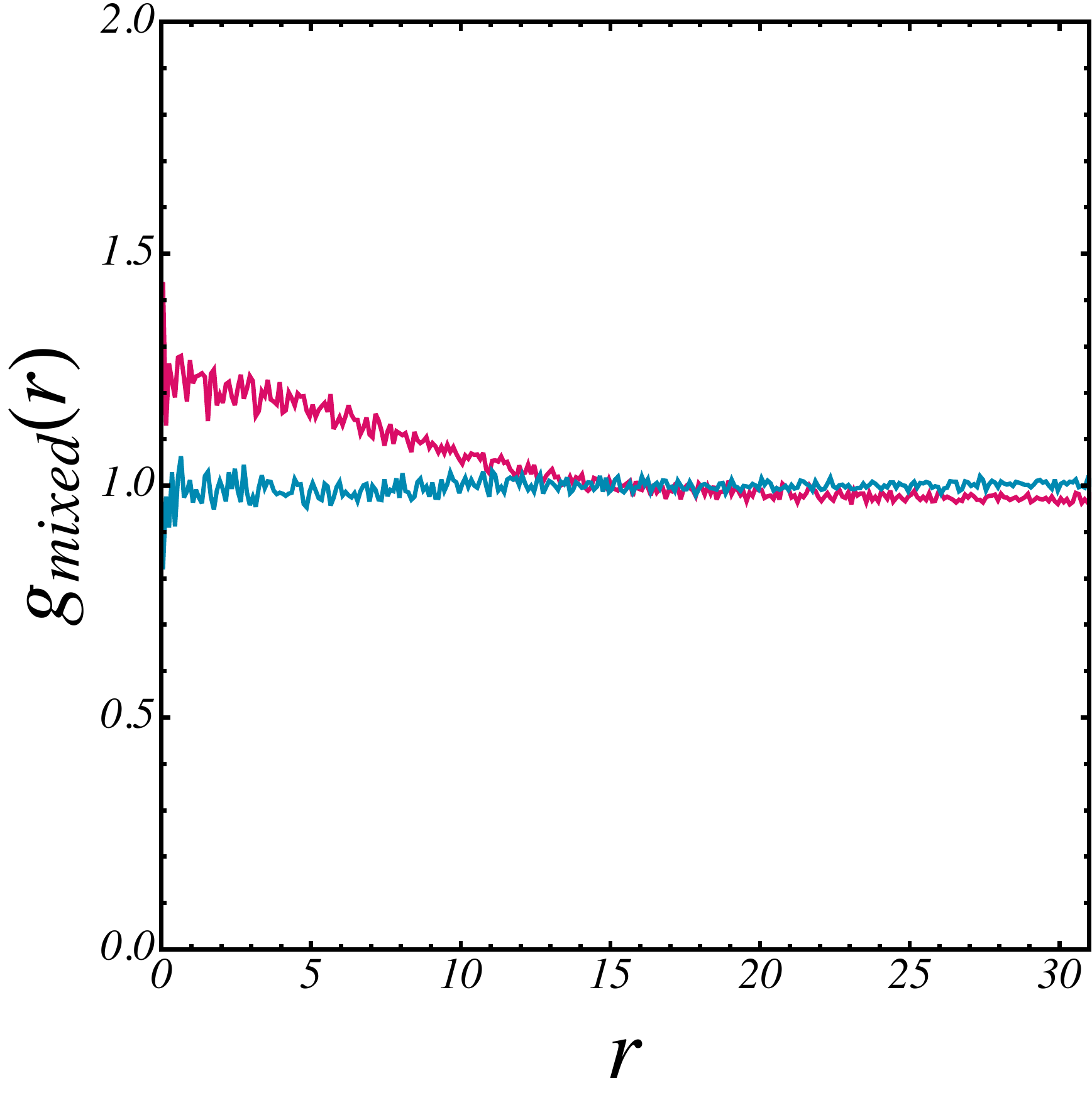}
    \caption{\textbf{Formation of jams.}
    Top row: snapshots from the ABP part of \textit{Comparison.mp4 --}, corresponding to times $T = 0, 1, 2, 3, 4, 5, 6$ seconds in the video.
    Second row: corresponding HABP snapshots.
    Third row: corresponding $g(r)$ of ABPs (green to blue) and HABPs (orange to red).
    Bottom row: corresponding $g_{mixed}(r)$.
    }
    \label{fig:moviedetails}
\end{figure*}
In the main text, we propose a mechanism for the formation of jams that goes as follows: at large relaxation rates, and starting from a high-enough density, HABPs are essentially very persistent ABPs.
For such particles, there is a finite density over which motility-induced phase separation is observed at zero noise, as reported by a large number of numerical and analytical studies~\cite{Redner2013,Solon2015d,VanDamme2019,Nie2020}.
From a microscopic, kinetic theory perspective, the formation of clusters that leads to MIPS is simply due to the fact that a few ABPs meeting in a head-on collision can get stuck together for rather long times~\cite{Redner2013,Nie2020}.
This mechanism has no reason to disappear for high-relaxation HABPs: they are still very persistent self-propelled particles.
Therefore, above the MIPS density, one expects HABPs to start forming finite clusters of a denser fluid.
Then, the role of targets kicks in: if a cluster becomes large enough, a fraction of the particles in the system will have targets lying directly inside the cluster.
Even if by some collision these particles manage to move along the side of the cluster, their self-propulsion will persistently point towards the inside of the cluster.
Their evaporation is therefore strongly suppressed, even at finite, small values of the rotational noise.
Furthermore, particles that manage to reach targets outside of clusters are removed from the system and replaced by particles whose targets can this time end up in jams.
In the end, above the noiseless MIPS lower critical density, seeds born from the same mechanism as MIPS will start growing and, if they get large enough, will eventually adsorb all particles from the surrounding gas.
This picture is supported by the coincidence of the jamming and lower-branch MIPS densities in both $2d$ and $3d$, and by the co-accumulation of particles and targets in the same regions in steady-state, shown in the main text.

A final proof is given by the short-time dynamics of a system that eventually jams, compared to usual ABPs.
In \textit{Comparison.mp4 --}, we show that, starting from a uniform density, an ABP and an HABP system are indistinguishable at short times as they start forming clusters.
This can be checked more quantitatively by plotting density-density correlation functions at different times that lie in the transient regime, like in Fig.~\ref{fig:moviedetails}.
In that figure, the top two rows are snapshots from \textit{Comparison.mp4 --}, from the ABP (top) and HABP (bottom) simulations, evenly spaced by 1 second (30 frames, or 15 simulation time units).
The third and fourth row respectively show the corresponding $g(r)$ and $g_{mixed}(r)$ of both simulations.
These curves show that the transient regime of ABPs and HABPs are indistinguishable not only visually, but even looking at pair correlation functions, up to the fifth column of Fig.~\ref{fig:moviedetails} (time code $00:04$ of the movie).
Starting from this point, in the HABP simulation, one notices an accumulation of targets close to particles, as well as a more peaked, crystal-like structure of the $g(r)$.

\subsection{3d results}

\begin{figure*}
    \centering
    \includegraphics[height = 0.30\textwidth]{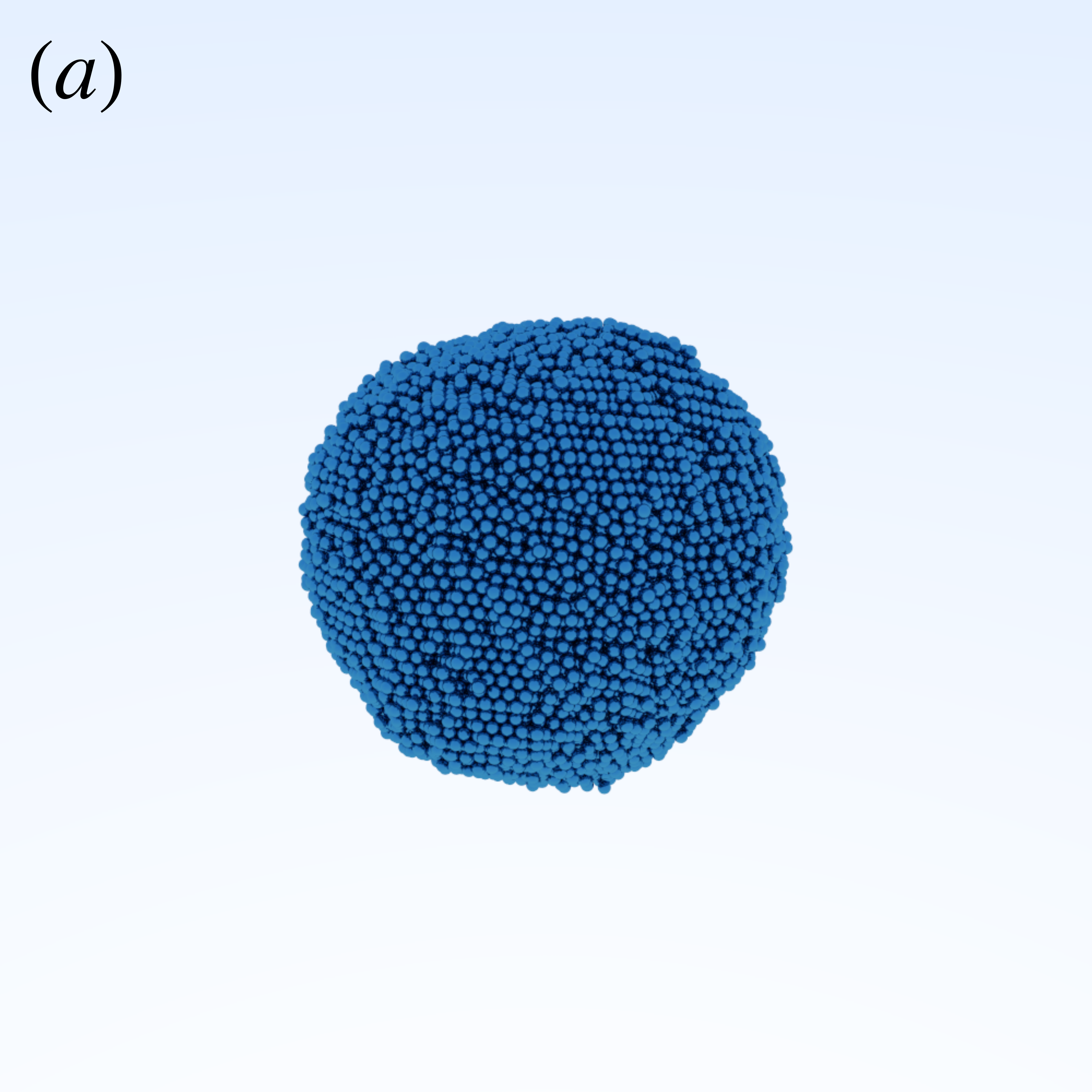}
    \includegraphics[height = 0.30\textwidth]{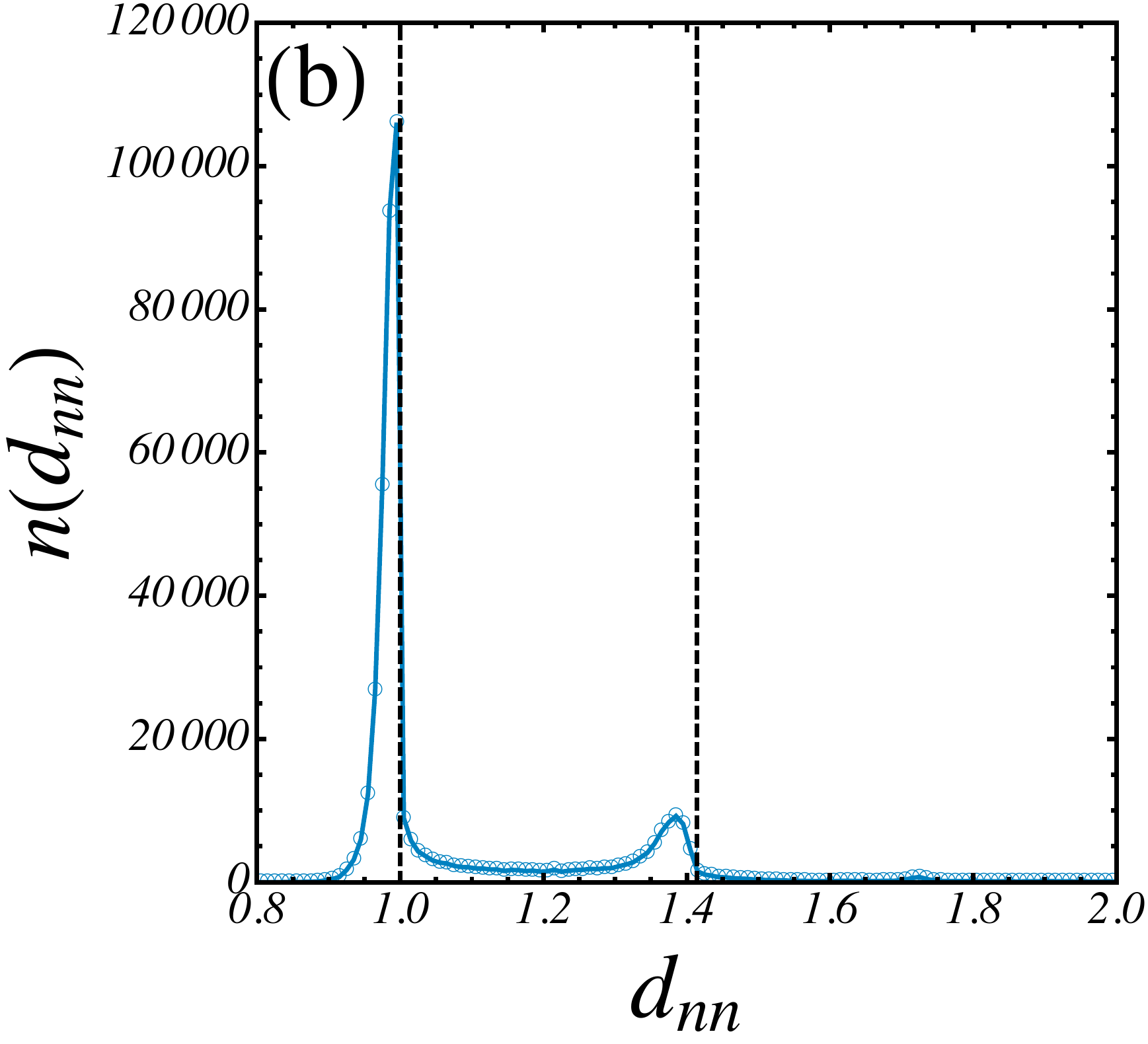}
    \includegraphics[height = 0.30\textwidth]{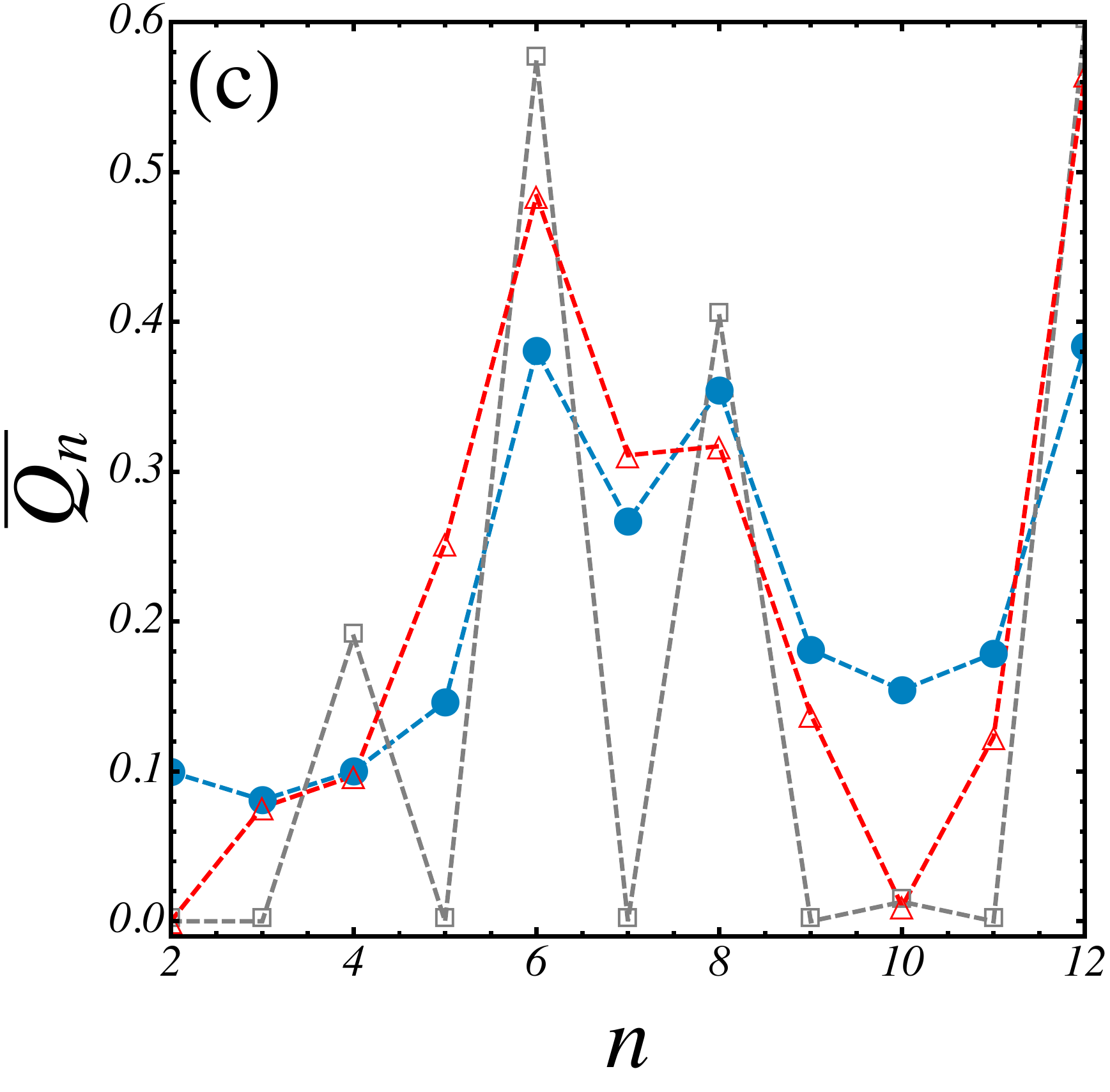}
    \caption{\textbf{3d jamming}
    $(a)$ Snapshot of a $3d$ jam obtained at $\phi = 0.30$ and for $N = 32768$ particles.
    $(b)$ Corresponding histogram of the distances $d_{nn}$ to Voronoi neighbours, in units of $a$ the particle diameter. 
    Black vertical lines indicate the special values $d_{nn} = a$ (particles are tangent) and $d_{nn} = \sqrt{2} a$ (expected peak of the next-nearest-neighbour histogram)
    $(c)$ Corresponding bond-orientational order parameters, averaged over particles within the jam (filled blue disks). 
    For comparison, we plot the expected values for ideal, infinite FCC crystals (empty grey squares) and HCP crystals (empty red triangles).
    Dashed lines are just guides for the eye and have no physical meaning.
    }
    \label{fig:3dresults}
\end{figure*}
To complement the clues presented in the main text that suggest a link between the onset of MIPS and jam formation from small MIPS clusters, we conduct a few simulations of the $3d$ version of the model, following the simulation design presented in Sec.~\ref{app:3dext}.
Without carrying out a full simulation campaign, we simply seek to confirm the approximate equality between the jamming packing fractiong $\phi_J$ of zero-noise, fast-relaxing HABPs, and the lower branch packing fraction $\phi_{low}$ of MIPS for zero-noise $3d$ ABPs.
Note that in $3d$, the packing fraction is defined as the volume fraction
\begin{align}
    \phi \equiv N \frac{\pi a^3}{6 L^3},
\end{align}
with $a$ the particle diameter (set to $1$ in simulation units) and $L$ the sidelength of the periodic, cubic simulation box.

We simulate, on the one hand, systems of zero-noise HABPs with quasi-instantaneous relaxation rate $\Omega_r = 100$ at various densities for $N = 32768$ particles, and seek the onset of jamming, if any.
We do observe jams for $\phi_J \geq 0.29$, and example of which is represented in Fig.~\ref{fig:3dresults}$(a)$.
Note that these jams are, like in $2d$ very close to spherical close packing: evaluating the density of the jam by approximating its volume via the convex hull, we find a density inside the jam of the order of $72\%$, between the values of body-centered cubic (BCC, $\phi_{BCC} = \pi \sqrt{3}/8 \approx 0.68$) and face-centered cubic (FCC, $\phi_{FCC} = \pi / 3 \sqrt{2} \approx 0.74$) or hexagonal compact ($HCP$, $\phi_{HCP} \approx 0.74$) crystals, and significantly higher than random close packing (RCP, $\phi_{RCP} \approx 0.64$)~\cite{Jaeger1992}.
Looking at the surface of jams, like in Fig.~\ref{fig:3dresults}$(a)$, one sees both triangular-lattice parts corresponding and square-lattice parts.
This is also illustrated in the flyover of \textit{2048\_jamoverview.gif}, which corresponds to a smaller jam ($N = 2048$ particles at $\phi = 0.30$).
These observations are highly suggestive of the presence of FCC or HCP structure.
Indeed, both structures feature a triangular lattices in one crystallographic direction, but square and (looser) rectangular lattices in other directions.
To further confirm this picture, in Fig.~\ref{fig:3dresults} $(b)$ we plot the histogram of distances to the Voronoi neighbours of each particle within a jam.
In a pure FCC crystal or HCP, one would expect only one peak at $d_{nn} = a$ for all 12 nearest neighbours, and one usually observes a leak of next-nearest neighbours into nearest-neighbours at distances comparable to (but smaller than) $\sqrt{2} a$ due to topological instabilities of the Voronoi mesh construction~\cite{Kumar2005}.
This is exactly the picture shown in Fig.~\ref{fig:3dresults} $(b)$.
It is interesting to note that there is no significant feature at distance $2 a/\sqrt{3}$, which would correspond to 6 out of 14 first Voronoi neighbours in a BCC structure: among usual crystalline structures, the jam seems to be overwhelmingly FCC or HCP.
Furthermore, in order to recognise these structures and, sometimes, discriminate between them, one often computes the bond-orientational order parameters (BOOPs) $\overline{Q_n}$ for $n$ going from $n = 2$ to $12$, which are the $3d$ equivalents of the $\psi_n$ order parameters in $2d$~\cite{Steinhardt1983,Mickel2013}.
We show in the Fig.~\ref{fig:3dresults}$(c)$ the values $\overline{Q_n}$ averaged over a single jam (filled blue disks), using Voronoi neighbours in the calculation of the $Q_n$.
They are plotted together with the values expected in a perfect FCC (empty grey squares) and a perfect HCP (empty red triangles) crystal.
While the measured values are not as sharp as one would expect in a perfect infinite crystal with perfect nearest-neighbour detection, they do display significant $n = 6$, $n = 8$, and $n=12$ orders as expected from FCC and HCP predictions.
Although a more detailed study would be necessary to distinguish between FCC and HCP structures, the observation of significant $n=5$ and $n=7$ orders point to an HCP structure.
Recent research has shown via principal component analysis that linear combinations of BOOPs can more accurately discriminate between fluid, and different crystalline structures~\cite{VanDamme2020}.
In particular, linear combinations that boil down to $y_1 \approx \overline{Q_4} + \overline{Q_6} + \overline{Q_8} + \overline{Q_{10}}+ \overline{Q_{12}}$, $y_2 \approx \overline{Q_4} - \overline{Q_6} + 2\overline{Q_{10}} - \overline{Q_{12}}$, and $y_3 \approx \overline{Q_4} - 2\overline{Q_{8}} + \overline{Q_{12}}$, are the most determining indicators when trying to distinguish crystalline structures.
The first one, $y_1$ is strongly positive in crystalline structures but small in fluid structures, the second one, $y_2$ is strongly positive in crystals with tetrahedral local structures like BCC but negative in crystals with hexagonal structures like FCC or HCP, and the third one, $y_3$, compares the weight of $4-$ and $8-$fold symmetries.
Using these approximate proxies for the principal components of Ref.~\cite{VanDamme2020} on our data, we find $(y_1,y_2,y_3) \approx (1.4, -0.35, 0.22)$, to be compared to $(y_1,y_2,y_3)_{FCC} \approx (1.8, -0.96, 0.40)$ in a perfect FCC structure and $(y_1,y_2,y_3)_{HCP} \approx (1.5, -0.93, -1.1)$ in a perfect HCP structure.
This simple approximation shows that the observed jam has a large $y_1$, a signature of crystalline order, a negative $y_2$ suggestive of local hexagonal order, and a slightly positive $y_3$ that gives a tiny bit more weight to the hypothesis of FCC structure compared to HCP.
All in all, it seems that running an HABP simulation with high relaxation rates and slightly above the jamming density leads to dense, crystalline packings that are significantly denser than RCP.
\begin{figure}
    \centering
    \includegraphics[width = 0.96\columnwidth]{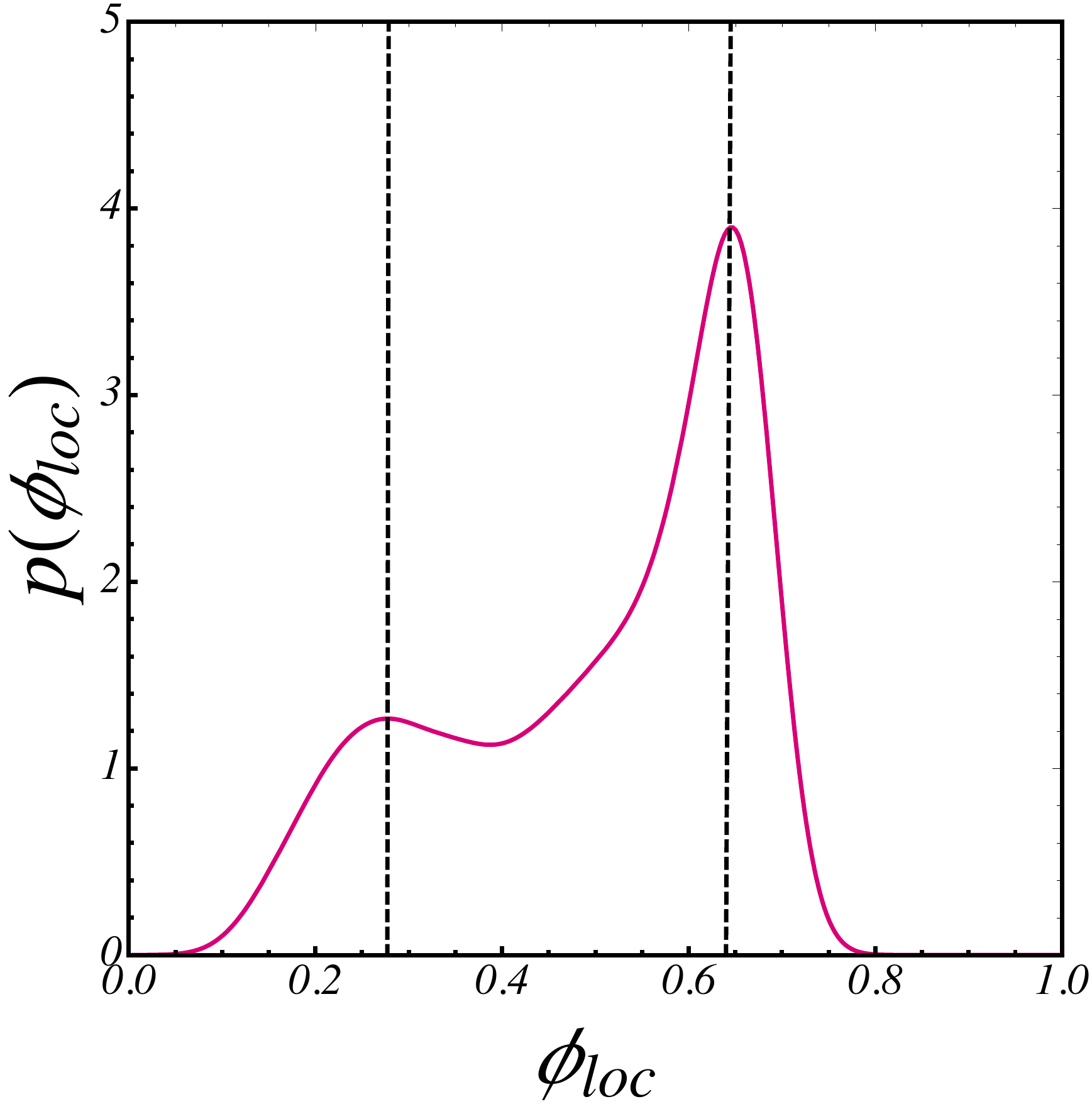}
    \caption{\textbf{3d MIPS in ABPs.}
    Plot of the local packing fraction obtained in a system of $N = 32768$ ABPs at zero noise, at an overall packing fraction $\phi = 0.50$.
    The black vertical lines show the estimated maxima: one at $\phi \approx 0.28$ and the other one at $\phi \approx 0.65$.
    }
    \label{fig:3dMIPS}
\end{figure}

To check the value of the lower branch of MIPS in $3d$, we run a long simulation of a system of $N = 32768$ ABPs at an overall packing fraction $\phi = 0.50$ and measure the local packing fraction in $8a\times 8a \times 8a$ cubic boxes across $n_{snapshots} \sim 20$ snapshots, leading to $n_{boxes} \sim 10^3$. The resulting distribution of local packing fractions is plotted in Fig.~\ref{fig:3dMIPS}.
The lower-branch density lies around $\phi_{low} \sim 0.28$, which is consistent with a recent study of MIPS in repulsive ABPs in $3d$~\cite{VanDamme2019}.

Comparing the two threshold densities (the lower branch of MIPS and the jamming density of HABPs), it seems that, like in $2d$, their $3d$ values are approximately equal to each other.
This observation is an additional clue pointing to a mechanism for jam formation that relies on seeds caused by MIPS.
Furthermore, note that the density of the higher-branch of MIPS, at $\phi_{high} \approx 0.65$ (which also agrees with Van Damme \textit{et al.}~\cite{VanDamme2019}), is significantly lower than the density measured in $3d$ jams: it seems that the addition of targets significantly compresses dense clusters, in a way reminiscent of the effect of fixed obstacles on MIPS~\cite{Reichhardt2014}.
This effect was likely present in $2d$ as well, but less noticeable due to the proximity between the higher branch of $2d$ MIPS and the $2d$ close-packing density.

\section{Optimality near jamming. \label{app:MoreDensityLines}}

\begin{figure}
    \centering
    \includegraphics[width = 0.63\columnwidth]{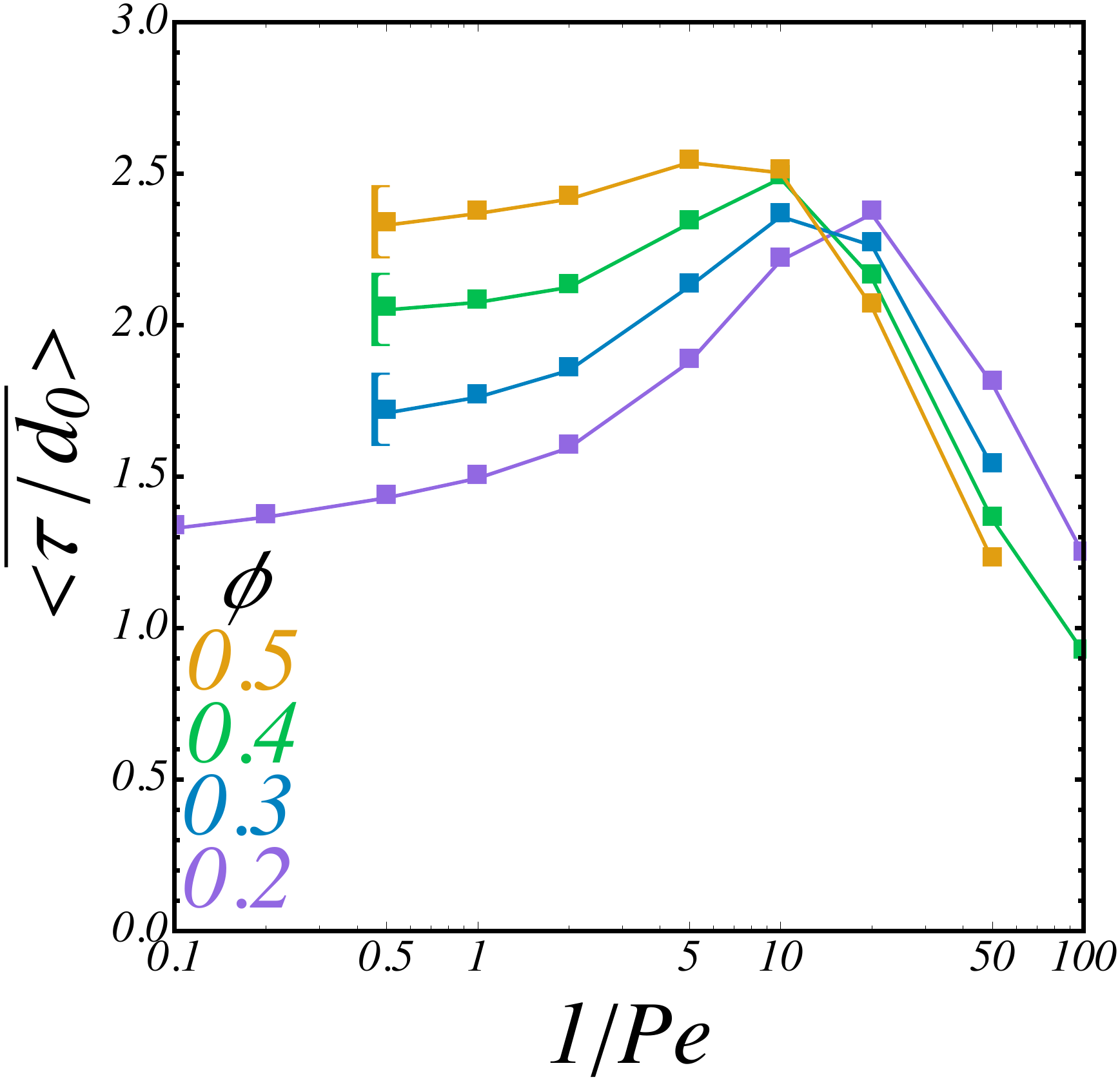}
    \caption{\textbf{Optimality near jamming: extra density lines.}
    Inverse effective travel speed $\langle \overline{\tau / d_0} \rangle$ against the inverse Péclet number $1/Pe$ in HBPs, for a few more densities than in the main text (see colour code in the inset).
    For all curves, $N = 2048$.}
    \label{fig:MoreDensitiesHBPs}
\end{figure}

In the main text, we claim that arrivals at any given density above the jamming density are fastest at the lowest possible value of the noise that unjams the system.
This is supported by curves showing the evolution of arrival times against a noise amplitude at one density above jamming.
In this appendix, we show that this result is independent of the density, in the simple example of HBPs (Model I).
In Fig.~\ref{fig:MoreDensitiesHBPs}, we show the variation of the average inverse speed $\langle\overline{\tau / d_0}\rangle$ (averaged over both realisations and initial distances) against the dimensionless noise amplitude $1/Pe$, averaging over all initial distances and about $10^6$ arrivals per point, in a system of $N = 2048$ HBPs and for $\phi = 0.2, 0.3, 0.4, 0.5$.
At all densities, the trend is the one described in the main text: travel at a constant density is faster near jamming, and remains the fastest option until travel becomes diffusive.
Furthermore, as discussed in the main text, as the system size increases, the maximum of arrival times is pushed to higher and higher noise amplitudes, so that only the optimum near jamming is relevant in the large system size limit.
Also note that increasing the density always increases the travel time in the ballistic regime, as already noted in the main text.

\section{Size scalings of arrival times.\label{app:sizescalings}}

\begin{figure}
    \centering
    \includegraphics[width = 0.48\columnwidth]{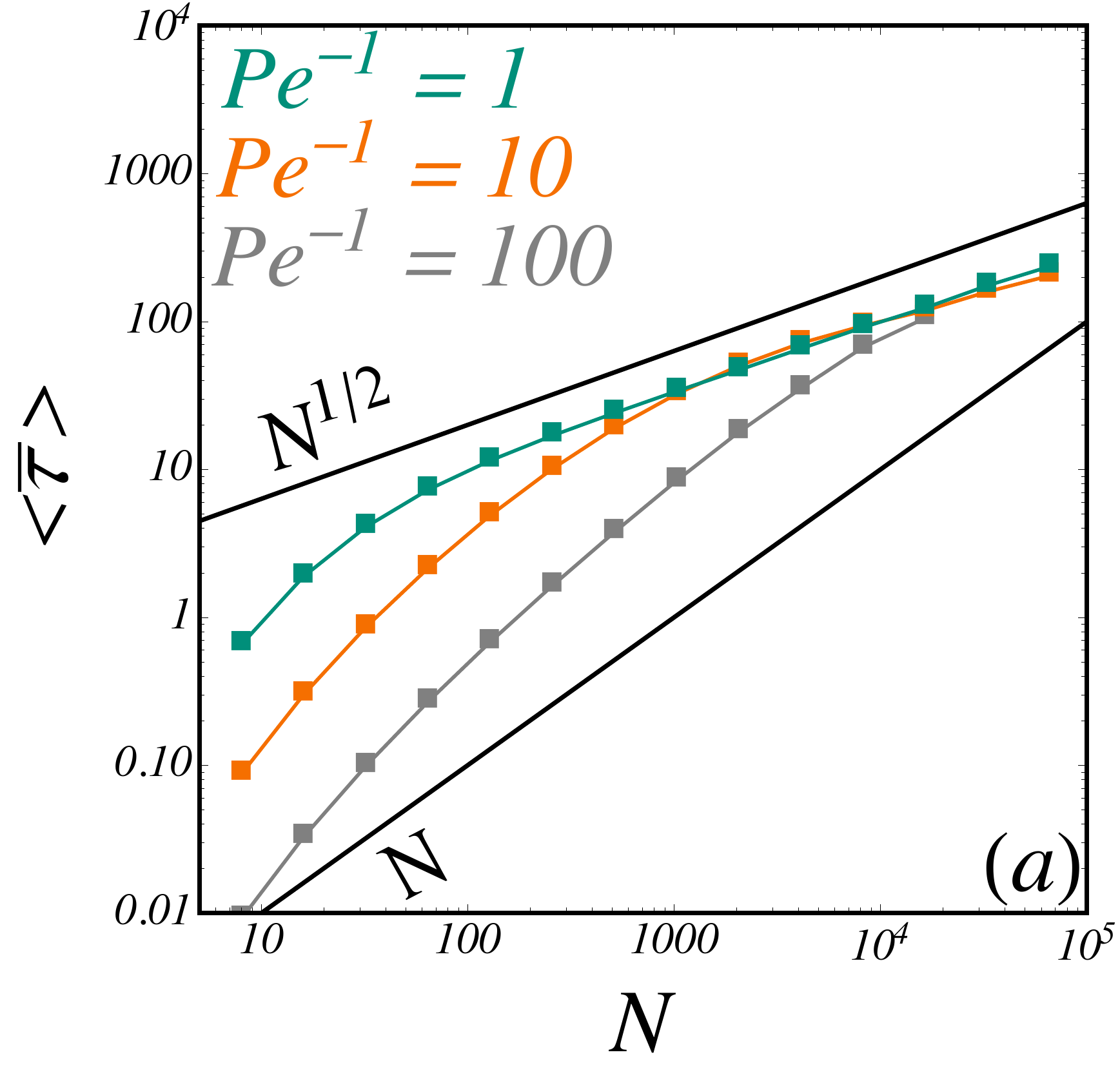}
    \includegraphics[width = 0.48\columnwidth]{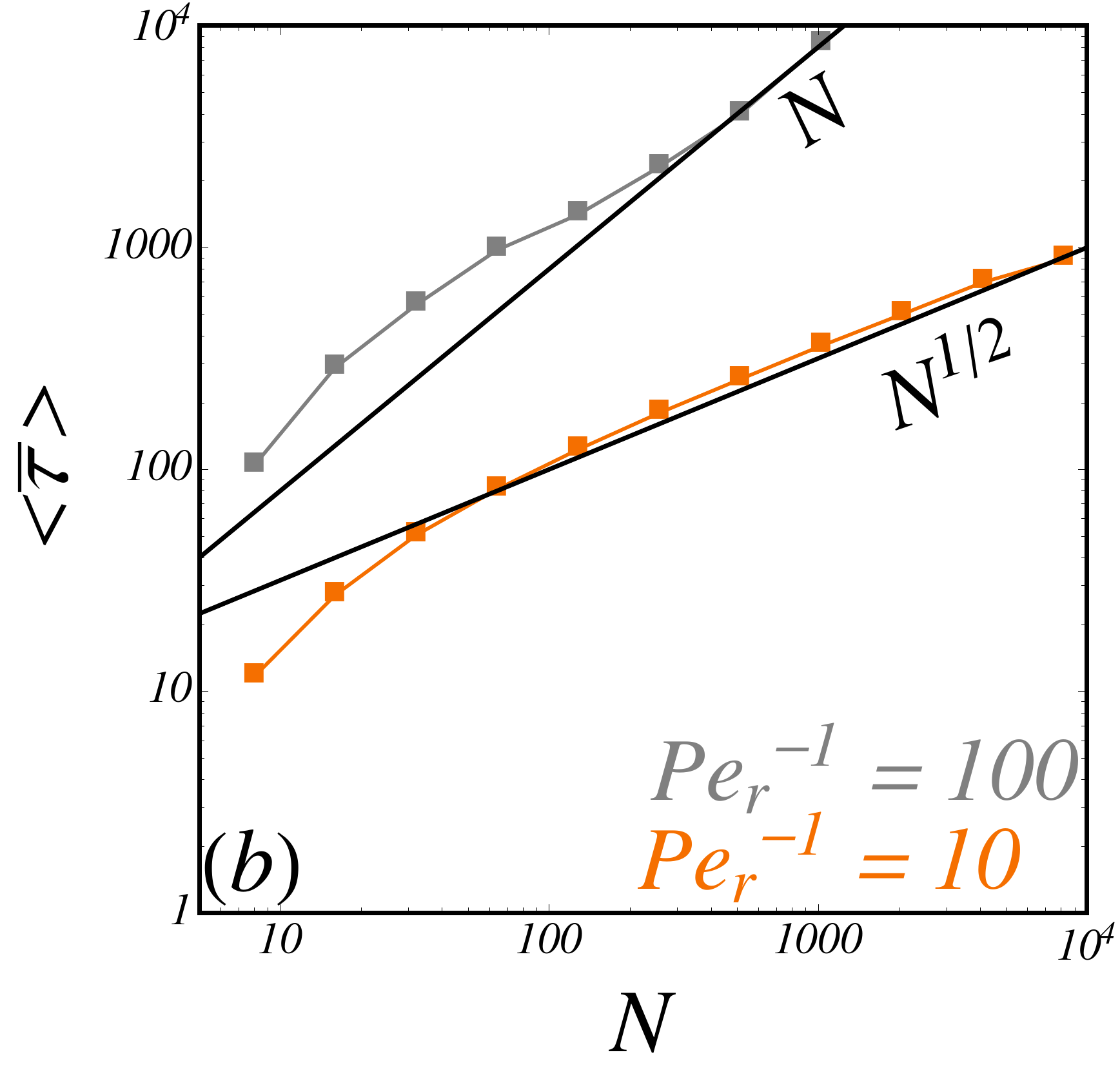}
    \caption{\textbf{Size scalings of arrival times.} $(a)$ Average arrival time $\langle\overline{\tau}\rangle$ against number of particles in Homing Brownian Particles at $\phi = 0.4$ and a few $Pe$. 
    $(b)$ Equivalent figure for Homing Active Brownian Particles at $\phi = 0.4$ and $\Omega_r = 1.0$.
    In both panels, solid black lines indicate the diffusive ($N \propto L^2$) and ballistic ($N^{1/2} \propto L$) scalings.}
    \label{fig:sizescalings}
\end{figure}

In the main text, we describe renewal as being ``ballistic'' or ``diffusive'' based on the behaviour of the arrival time of particles against the initial distance to their targets but for a single system size.
We also checked that, if one keeps the density and noise amplitude constant but changes the system size, the arrival time follows simple scalings with the number of particles.
This is shown in Fig.~\ref{fig:sizescalings}.
In panel $(a)$, we show size scalings of arrival times for Homing Brownian Particles ($Pe_r, \Omega_r \to \infty$) for a few values of $Pe$ at $\phi = 0.4$.
In large enough systems, the arrival time eventually scales ballistically, \textit{i.e.} $\langle\overline{\tau}\rangle \propto L \propto N^{1/2}$.
However, in smaller systems, the scaling typically becomes closer to a diffusive one, \textit{i.e.} $\langle\overline{\tau}\rangle \propto L^2 \propto N$.
Furthermore, as suggested in the main text, we confirm that the crossover between the ballistic and the diffusive regimes is pushed to larger and larger values of noise as the system size is increased.
Therefore, in the limit $N\to\infty$, arrivals are on average ballistic, with an effective speed that decreases as the noise amplitude increases.

In panel $(b)$, we show similar scalings for Homing Brownian Particles with zero translational noise ($Pe \to \infty$), this time setting $\phi = 0.4$ and $\Omega_r = 1.0$.
Next to the melting line, the arrivals are ballistic, and they become diffusive at larger noise amplitudes and large system sizes.

\section{Additional cuts in the full HABP phase diagram\label{app:OmegarPhi}}

\begin{figure}
    \centering
    \includegraphics[width=0.29\columnwidth]{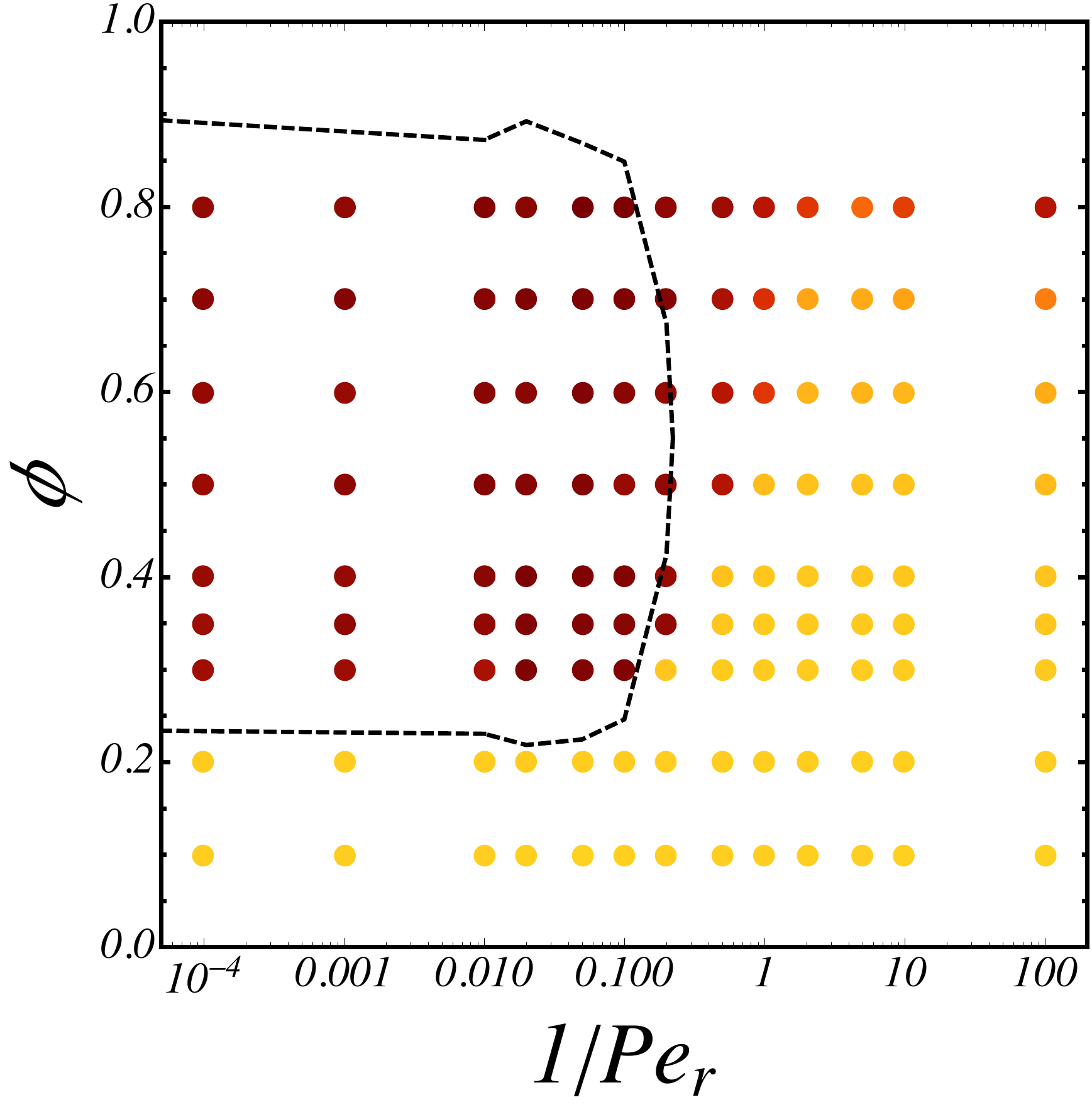}
    \includegraphics[width=0.29\columnwidth]{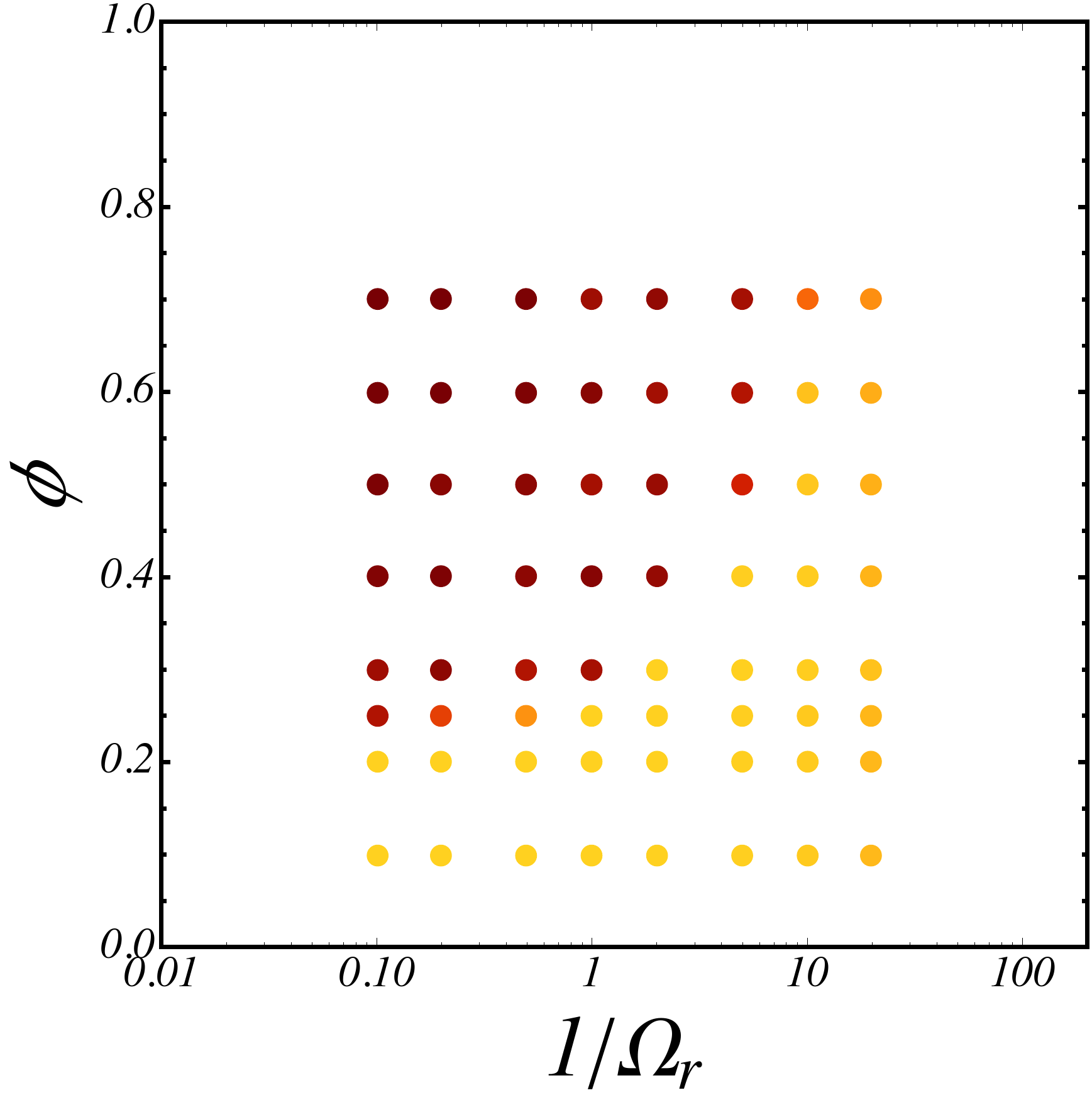}
    \includegraphics[width=0.29\columnwidth]{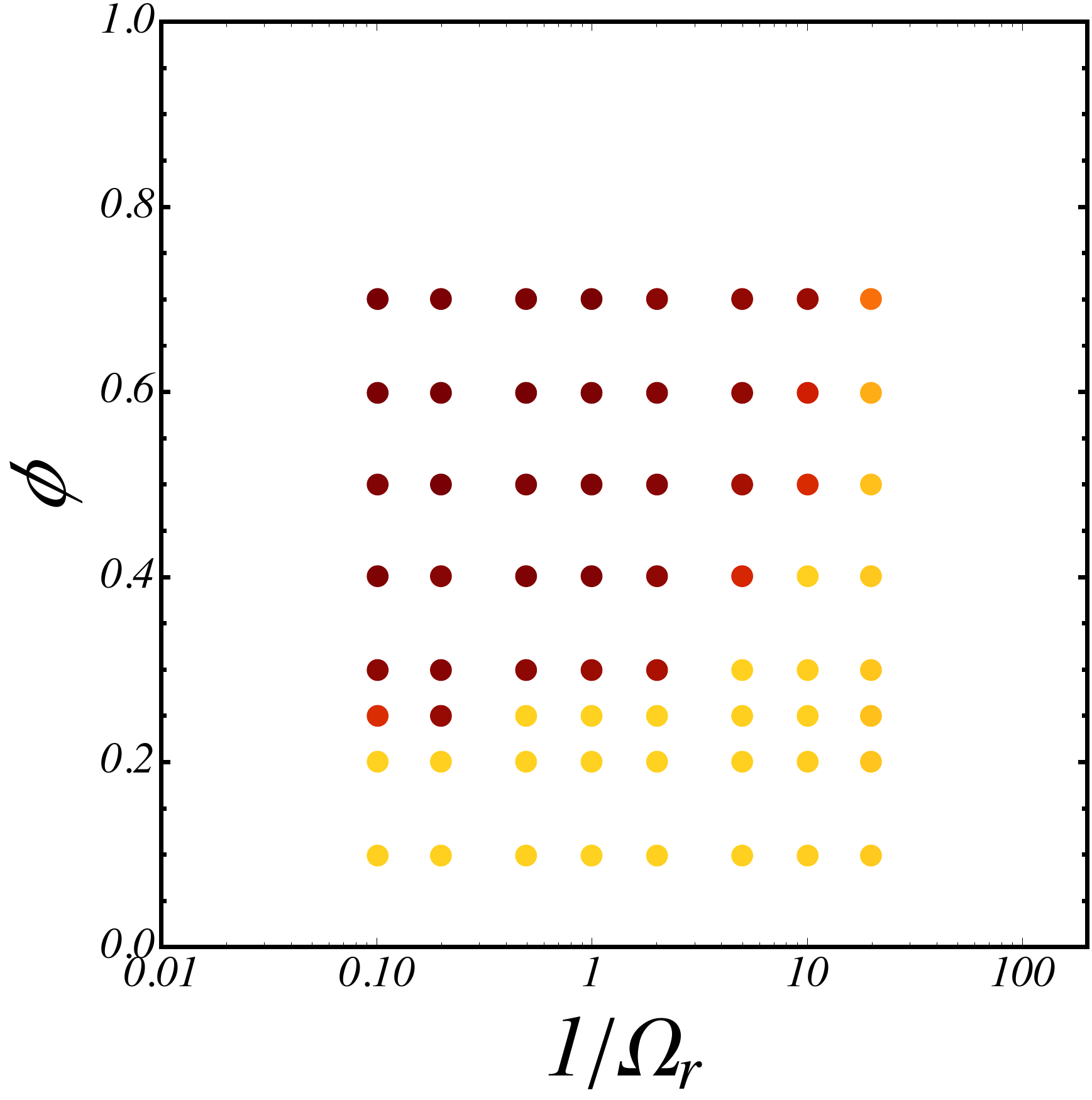}
    \includegraphics[height = .29\columnwidth]{AppFigures/Fig_JamID/SIPsi6mapbar.pdf}
    \caption{\textbf{Extra cuts into the HABP phase diagram.}
    From left to right: cut at $\Omega_r = 0.2$ in the noise, density plane, cut at $Pe_r = 2$ in the relaxation-density plane, and cut at $Pe_r = 5$ in the relaxation-density plane.
    Here symbols are coloured according to the mean modulus of $\psi_6$.
    In the first map, non-jammed phases are all homogeneous in space (no phase separation, and the zero-relaxation MIPS domain is shown as a dashed line).
    In the second map, the noise is too large for MIPS even for ABPs, and every non-jammed point is homogeneous.
    In the third map, the noise is slightly lower than the evaluated MIPS critical noise, so that high density fluctuations persist even far from jams.}
    \label{fig:canbemadeintomaps}
\end{figure}

In the main text, we present a sketch of the full $3d$ phase diagram of HABPs in the relaxation rate, rotational noise, density volume.
To back this phase diagram, we show a few more cuts in Fig.~\ref{fig:canbemadeintomaps}.
The first cut is taken at a relaxation rate lying very close to the jamming value on the jammed side, namely $\Omega_r = 0.2$.
Even that close to the jamming value, we only observe jammed and uniform states, but no remnant of MIPS (the shape of which at zero relaxation is drawn as a dashed black line).
This suggests that the MIPS-jam plane is almost at constant relaxation rate in the $3d$ phase diagram.
The second and third cuts are two finite-noise cuts, lying on each side of the estimated critical value of noise for MIPS, $Pe_r = 2$ and $Pe_r = 5$.

\section{Trajectories of individual particles and ballistic-diffusive crossover\label{app:IndividualTrajectories}}

\begin{figure*}
    \centering
    \includegraphics[width=0.9\textwidth]{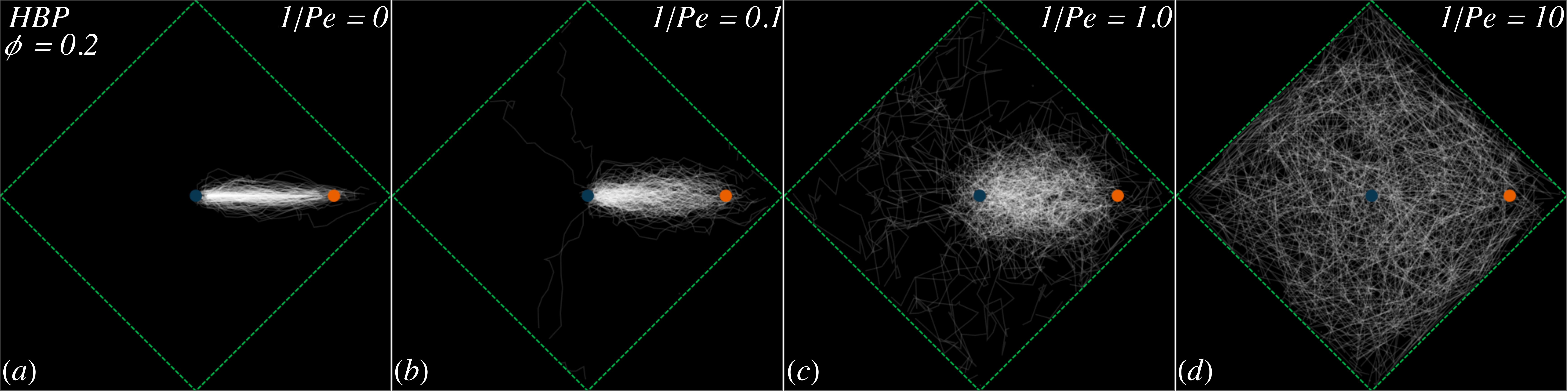}\\
    \includegraphics[width=0.9\textwidth]{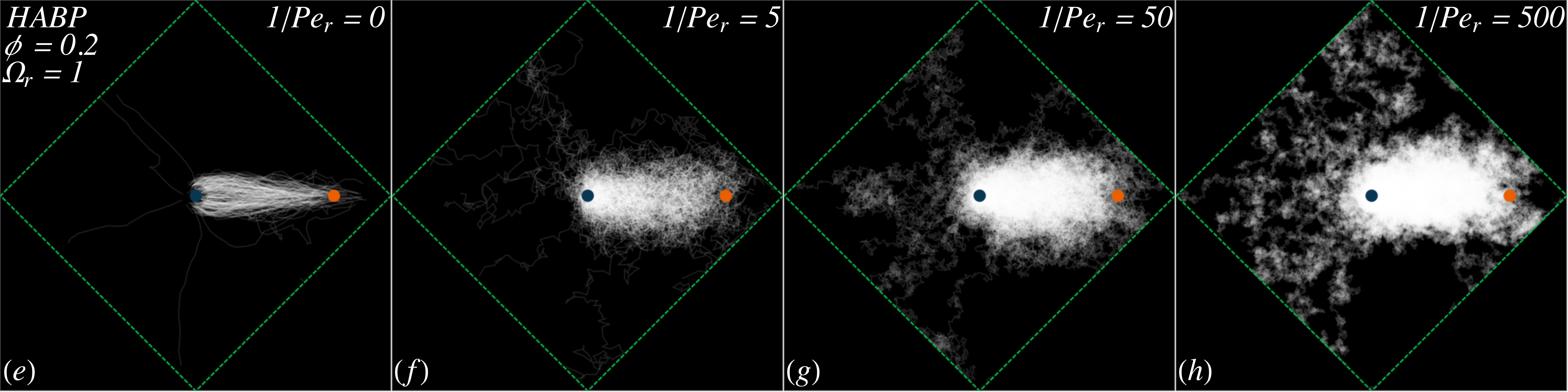}\\
    \includegraphics[width=0.9\textwidth]{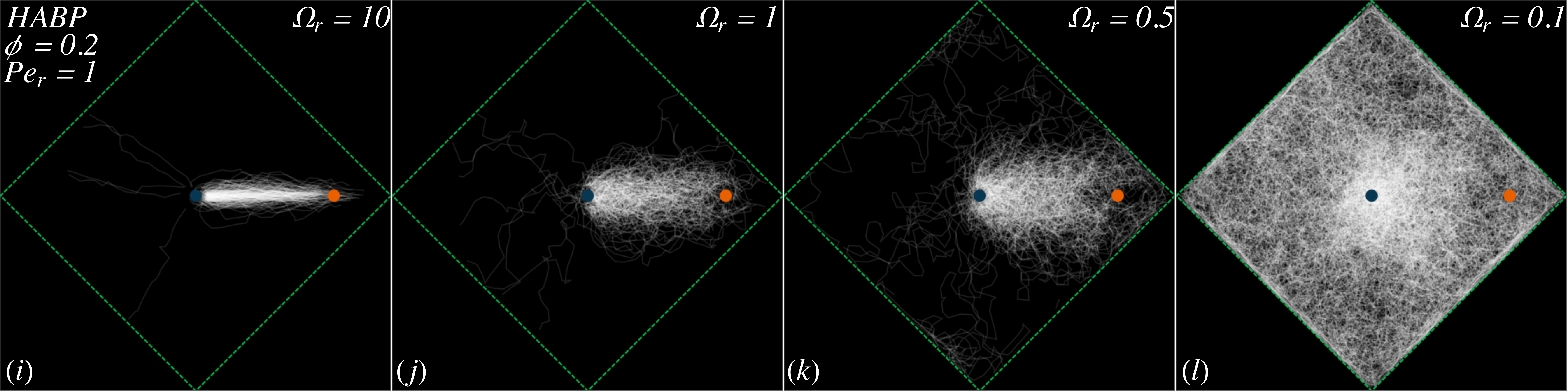}
    \caption{\textbf{Sample trajectories.}
    All panels were obtained in a small system ($N = 128$ particles) at a density below the jamming one ($\phi = 0.2$).
    In each panel, we show $256$ individual trajectories, recentered on their place of birth (blue disk), and rotated so that their target lies to the right of this point on a horizontal line.
    An orange dot shows the position of the most probable target position, at a distance $L/2$, while a green dashed line shows the bounds of the periodic simulation box.
    Each trajectory is plotted in white with a finite transparency.
    $(a)-(d)$ HBP trajectories for a few values of $Pe$.
    $(e)-(h)$ HABP trajectories at a fixed $\Omega_r = 1$ for a few values of $Pe_r$.
    $(i)-(l)$ HABP trajectories at a fixed $Pe_r = 1$ for a few values of $\Omega_r$.}
    \label{fig:Trajectories}
\end{figure*}
In the main text, we describe ballistic-diffusive crossovers for target reaching in a model of Brownian Homing Particles when tuning translational noise, and in a model of Homing Active Brownian Particles when tuning either rotational noise or the orientational relaxation rate towards the target.
In order to illustrate this crossover, we plotted typical trajectories in Fig.~2 of the main text.
These panels were obtained by simulating full systems of $N = 128$ particles at a finite density, and recording their trajectories at a rate of one point every $\Delta t = 1$ in simulation units.
We then selected only trajectories such that the initial distance to the target lies in a small interval: in the main text, the choice is $d_0 \in \left[L/2 - 3a/2; L/2 - a/2\right]$, so that targets lie close to the mode of the $d_0$ distribution but do not lie in the corners of the periodic box when recentering on the particle's birth place.
We then plotted $256$ individual trajectories, placing all the initial positions at the origin of the plot and rotating the trajectory so that the target lies on the $x$ axis and at positive values of $x$.
Each trajectory is plotted in white with a finite transparency: brighter regions encode for places where particles spent more time either within one realisation (for instance because of diffusive motion, that typically visits the same place very often), or across different realisations (because different particles follow similar paths).

For completeness, in Fig.~\ref{fig:Trajectories} of this document, we produce a similar representation, with similar parameters, but this time not filtering initial distances to the target.
In that case, the orange dot represents the mode of the distribution of targets, but targets can actually lie anywhere right of the blue disk on the horizontal axis, following the probability distribution of Fig.~\ref{fig:distancedistrib}.

In Fig.~\ref{fig:Trajectories} $(a)-(d)$, we show trajectories of Homing Brownian Particles (that always point towards their target) but with a growing translational noise amplitude.
At zero noise (panel $(a)$), every particle essentially travels in a straight line, in spite of collisions with other particles: target reaching is ballistic.
When increasing the noise $(b)-(d)$, trajectories spread more and more outside of the straight line towards the target, and become more and more similar to unbiased Brownian walks with a step size that increases with the noise level.
Eventually, at high noise (panel $(d)$), trajectories cover the whole simulation box uniformly, a signature of diffusive exploration.

In Fig.~\ref{fig:Trajectories} $(e)-(h)$, we show trajectories of Homing Active Brownian Particles at a fixed value of the relaxation rate ($\Omega_r = 1.0$) but for a few values of the rotational noise amplitude.
At zero noise (panel $(e)$), the particles travel to the target almost following a straight line, but with some slight initial curvature due to the finite relaxation rate.
As the noise increases  (panels $(f)-(h)$), the trajectories spread further and further in the direction perpendicular to the initial particle-to-target vector (vertically in this representation), and look more and more diffusive but this time with an effective diffusion coefficient that decreases as the noise increases.
As a result, the effective step size of the walks decreases as the noise increases: while trajectories significantly stray away from the straight path to the target, they do so extremely slowly.
Like for translational noise, trajectories typically cover an area comparable to that of the whole box at very large noise amplitudes (panel $(h)$).
Note that, in that case, the most visited region is a roughly elliptic region symmetric around the straight path to the target, which is reminiscent of usual $1d$ Brownian bridge~\cite{Majumdar2015,Szavits-Nossan2015}: even for purely Brownian walks, plotting trajectories of walks conditional to their arrival at some point of space is equivalent to drawing biased random walks.
This shape is observed in panel $(h)$ but not $(d)$ because in the case of rotational noise, the effective spatial diffusion constant decreases as the rotational noise is tuned up: therefore, even in small systems, at large enough rotational noise one can observe patterns usually expected only in the limit of very large initial distance to the target.

Finally, in Fig.~\ref{fig:Trajectories} $(i)-(l)$, we show trajectories of Homing Active Brownian Particles at a fixed value of the rotational noise ($Pe_r = 1.0$) but for a few values of the relaxation rate.
At very large relaxation rates (panel $(i)$), one recovers a picture similar to that of panel $(a)$, with particles reaching their targets in a straight line as the fast relaxation rate vastly overpowers the rotational noise.
As the relaxation of self-propulsion towards the target becomes weaker than rotational noise (panels $(j)-(l)$), the typical trajectory of particles strays away from a straight line to the target, first simply by broadening laterally, and eventually covers the whole box (panel $(l)$).
\normalem
\vspace{-15mm}
\bibliography{PostDoc-DovLevine}

\begin{thebibliography}{63}%
\makeatletter
\providecommand \@ifxundefined [1]{%
 \@ifx{#1\undefined}
}%
\providecommand \@ifnum [1]{%
 \ifnum #1\expandafter \@firstoftwo
 \else \expandafter \@secondoftwo
 \fi
}%
\providecommand \@ifx [1]{%
 \ifx #1\expandafter \@firstoftwo
 \else \expandafter \@secondoftwo
 \fi
}%
\providecommand \natexlab [1]{#1}%
\providecommand \enquote  [1]{``#1''}%
\providecommand \bibnamefont  [1]{#1}%
\providecommand \bibfnamefont [1]{#1}%
\providecommand \citenamefont [1]{#1}%
\providecommand \href@noop [0]{\@secondoftwo}%
\providecommand \href [0]{\begingroup \@sanitize@url \@href}%
\providecommand \@href[1]{\@@startlink{#1}\@@href}%
\providecommand \@@href[1]{\endgroup#1\@@endlink}%
\providecommand \@sanitize@url [0]{\catcode `\\12\catcode `\$12\catcode
  `\&12\catcode `\#12\catcode `\^12\catcode `\_12\catcode `\%12\relax}%
\providecommand \@@startlink[1]{}%
\providecommand \@@endlink[0]{}%
\providecommand \url  [0]{\begingroup\@sanitize@url \@url }%
\providecommand \@url [1]{\endgroup\@href {#1}{\urlprefix }}%
\providecommand \urlprefix  [0]{URL }%
\providecommand \Eprint [0]{\href }%
\providecommand \doibase [0]{http://dx.doi.org/}%
\providecommand \selectlanguage [0]{\@gobble}%
\providecommand \bibinfo  [0]{\@secondoftwo}%
\providecommand \bibfield  [0]{\@secondoftwo}%
\providecommand \translation [1]{[#1]}%
\providecommand \BibitemOpen [0]{}%
\providecommand \bibitemStop [0]{}%
\providecommand \bibitemNoStop [0]{.\EOS\space}%
\providecommand \EOS [0]{\spacefactor3000\relax}%
\providecommand \BibitemShut  [1]{\csname bibitem#1\endcsname}%
\let\auto@bib@innerbib\@empty
\bibitem [{\citenamefont {Karlin}\ and\ \citenamefont
  {Peres}(2017)}]{KarlinPeres}%
  \BibitemOpen
  \bibfield  {author} {\bibinfo {author} {\bibfnamefont {A.~R.}\ \bibnamefont
  {Karlin}}\ and\ \bibinfo {author} {\bibfnamefont {Y.}~\bibnamefont {Peres}},\
  }\href@noop {} {\emph {\bibinfo {title} {{Game Theory , Alive}}}}\ (\bibinfo
  {publisher} {American Mathematical Society},\ \bibinfo {year}
  {2017})\BibitemShut {NoStop}%
\bibitem [{\citenamefont {Helbing}(2001)}]{Helbing2001}%
  \BibitemOpen
  \bibfield  {author} {\bibinfo {author} {\bibfnamefont {D.}~\bibnamefont
  {Helbing}},\ }\href@noop {} {\bibfield  {journal} {\bibinfo  {journal}
  {Reviews of Modern Physics}\ }\textbf {\bibinfo {volume} {73}},\ \bibinfo
  {pages} {1067} (\bibinfo {year} {2001})}\BibitemShut {NoStop}%
\bibitem [{\citenamefont {Orosz}\ \emph {et~al.}(2010)\citenamefont {Orosz},
  \citenamefont {{Eddie Wilson}},\ and\ \citenamefont {Stefan}}]{Orosz2010}%
  \BibitemOpen
  \bibfield  {author} {\bibinfo {author} {\bibfnamefont {G.}~\bibnamefont
  {Orosz}}, \bibinfo {author} {\bibfnamefont {R.}~\bibnamefont {{Eddie
  Wilson}}}, \ and\ \bibinfo {author} {\bibfnamefont {G.}~\bibnamefont
  {Stefan}},\ }\href@noop {} {\bibfield  {journal} {\bibinfo  {journal}
  {Philosophical Transactions of the Royal Society A: Mathematical, Physical
  and Engineering Sciences}\ }\textbf {\bibinfo {volume} {368}},\ \bibinfo
  {pages} {4455} (\bibinfo {year} {2010})}\BibitemShut {NoStop}%
\bibitem [{\citenamefont {Nakayama}\ \emph {et~al.}(2016)\citenamefont
  {Nakayama}, \citenamefont {Kikuchi}, \citenamefont {Shibata}, \citenamefont
  {Sugiyama}, \citenamefont {Tadaki},\ and\ \citenamefont
  {Yukawa}}]{Nakayama2016}%
  \BibitemOpen
  \bibfield  {author} {\bibinfo {author} {\bibfnamefont {A.}~\bibnamefont
  {Nakayama}}, \bibinfo {author} {\bibfnamefont {M.}~\bibnamefont {Kikuchi}},
  \bibinfo {author} {\bibfnamefont {A.}~\bibnamefont {Shibata}}, \bibinfo
  {author} {\bibfnamefont {Y.}~\bibnamefont {Sugiyama}}, \bibinfo {author}
  {\bibfnamefont {S.~I.}\ \bibnamefont {Tadaki}}, \ and\ \bibinfo {author}
  {\bibfnamefont {S.}~\bibnamefont {Yukawa}},\ }\href@noop {} {\bibfield
  {journal} {\bibinfo  {journal} {New Journal of Physics}\ }\textbf {\bibinfo
  {volume} {18}},\ \bibinfo {pages} {043040} (\bibinfo {year}
  {2016})}\BibitemShut {NoStop}%
\bibitem [{\citenamefont {Aoyama}\ \emph {et~al.}(2020)\citenamefont {Aoyama},
  \citenamefont {Yoshioka}, \citenamefont {Shimokawa},\ and\ \citenamefont
  {Morita}}]{Aoyama2020}%
  \BibitemOpen
  \bibfield  {author} {\bibinfo {author} {\bibfnamefont {E.}~\bibnamefont
  {Aoyama}}, \bibinfo {author} {\bibfnamefont {K.}~\bibnamefont {Yoshioka}},
  \bibinfo {author} {\bibfnamefont {S.}~\bibnamefont {Shimokawa}}, \ and\
  \bibinfo {author} {\bibfnamefont {H.}~\bibnamefont {Morita}},\ }\href@noop {}
  {\bibfield  {journal} {\bibinfo  {journal} {Asian Transport Studies}\
  }\textbf {\bibinfo {volume} {6}},\ \bibinfo {pages} {100015} (\bibinfo {year}
  {2020})}\BibitemShut {NoStop}%
\bibitem [{\citenamefont {Benhamou}(2004)}]{Benhamou2004}%
  \BibitemOpen
  \bibfield  {author} {\bibinfo {author} {\bibfnamefont {S.}~\bibnamefont
  {Benhamou}},\ }\href@noop {} {\bibfield  {journal} {\bibinfo  {journal}
  {Journal of Theoretical Biology}\ }\textbf {\bibinfo {volume} {229}},\
  \bibinfo {pages} {209} (\bibinfo {year} {2004})}\BibitemShut {NoStop}%
\bibitem [{\citenamefont {Benhamou}(2014)}]{Benhamou2014}%
  \BibitemOpen
  \bibfield  {author} {\bibinfo {author} {\bibfnamefont {S.}~\bibnamefont
  {Benhamou}},\ }\href@noop {} {\bibfield  {journal} {\bibinfo  {journal}
  {Ecology Letters}\ }\textbf {\bibinfo {volume} {17}},\ \bibinfo {pages} {261}
  (\bibinfo {year} {2014})}\BibitemShut {NoStop}%
\bibitem [{\citenamefont {Moussa{\"{i}}d}\ \emph {et~al.}(2009)\citenamefont
  {Moussa{\"{i}}d}, \citenamefont {Helbing}, \citenamefont {Garnier},
  \citenamefont {Johansson}, \citenamefont {Combe},\ and\ \citenamefont
  {Theraulaz}}]{Moussaid2009}%
  \BibitemOpen
  \bibfield  {author} {\bibinfo {author} {\bibfnamefont {M.}~\bibnamefont
  {Moussa{\"{i}}d}}, \bibinfo {author} {\bibfnamefont {D.}~\bibnamefont
  {Helbing}}, \bibinfo {author} {\bibfnamefont {S.}~\bibnamefont {Garnier}},
  \bibinfo {author} {\bibfnamefont {A.}~\bibnamefont {Johansson}}, \bibinfo
  {author} {\bibfnamefont {M.}~\bibnamefont {Combe}}, \ and\ \bibinfo {author}
  {\bibfnamefont {G.}~\bibnamefont {Theraulaz}},\ }\href@noop {} {\bibfield
  {journal} {\bibinfo  {journal} {Proceedings of the Royal Society B:
  Biological Sciences}\ }\textbf {\bibinfo {volume} {276}},\ \bibinfo {pages}
  {2755} (\bibinfo {year} {2009})}\BibitemShut {NoStop}%
\bibitem [{\citenamefont {Moussa{\"{i}}d}\ \emph {et~al.}(2012)\citenamefont
  {Moussa{\"{i}}d}, \citenamefont {Guillot}, \citenamefont {Moreau},
  \citenamefont {Fehrenbach}, \citenamefont {Chabiron}, \citenamefont
  {Lemercier}, \citenamefont {Pettr{\'{e}}}, \citenamefont {Appert-Rolland},
  \citenamefont {Degond},\ and\ \citenamefont {Theraulaz}}]{Moussaid2012}%
  \BibitemOpen
  \bibfield  {author} {\bibinfo {author} {\bibfnamefont {M.}~\bibnamefont
  {Moussa{\"{i}}d}}, \bibinfo {author} {\bibfnamefont {E.~G.}\ \bibnamefont
  {Guillot}}, \bibinfo {author} {\bibfnamefont {M.}~\bibnamefont {Moreau}},
  \bibinfo {author} {\bibfnamefont {J.}~\bibnamefont {Fehrenbach}}, \bibinfo
  {author} {\bibfnamefont {O.}~\bibnamefont {Chabiron}}, \bibinfo {author}
  {\bibfnamefont {S.}~\bibnamefont {Lemercier}}, \bibinfo {author}
  {\bibfnamefont {J.}~\bibnamefont {Pettr{\'{e}}}}, \bibinfo {author}
  {\bibfnamefont {C.}~\bibnamefont {Appert-Rolland}}, \bibinfo {author}
  {\bibfnamefont {P.}~\bibnamefont {Degond}}, \ and\ \bibinfo {author}
  {\bibfnamefont {G.}~\bibnamefont {Theraulaz}},\ }\href@noop {} {\bibfield
  {journal} {\bibinfo  {journal} {PLoS Computational Biology}\ }\textbf
  {\bibinfo {volume} {8}},\ \bibinfo {pages} {e1002442} (\bibinfo {year}
  {2012})}\BibitemShut {NoStop}%
\bibitem [{\citenamefont {Yajima}\ \emph {et~al.}(2020)\citenamefont {Yajima},
  \citenamefont {Yoshii},\ and\ \citenamefont {Sumino}}]{Yajima2020}%
  \BibitemOpen
  \bibfield  {author} {\bibinfo {author} {\bibfnamefont {S.}~\bibnamefont
  {Yajima}}, \bibinfo {author} {\bibfnamefont {K.}~\bibnamefont {Yoshii}}, \
  and\ \bibinfo {author} {\bibfnamefont {Y.}~\bibnamefont {Sumino}},\
  }\href@noop {} {\bibfield  {journal} {\bibinfo  {journal} {Journal of the
  Physical Society of Japan}\ }\textbf {\bibinfo {volume} {89}},\ \bibinfo
  {pages} {074003} (\bibinfo {year} {2020})}\BibitemShut {NoStop}%
\bibitem [{\citenamefont {Nash}\ \emph {et~al.}(2010)\citenamefont {Nash},
  \citenamefont {Daniel}, \citenamefont {Koenig},\ and\ \citenamefont
  {Feiner}}]{Nash2010}%
  \BibitemOpen
  \bibfield  {author} {\bibinfo {author} {\bibfnamefont {A.}~\bibnamefont
  {Nash}}, \bibinfo {author} {\bibfnamefont {K.}~\bibnamefont {Daniel}},
  \bibinfo {author} {\bibfnamefont {S.}~\bibnamefont {Koenig}}, \ and\ \bibinfo
  {author} {\bibfnamefont {A.}~\bibnamefont {Feiner}},\ }\href@noop {}
  {\bibfield  {journal} {\bibinfo  {journal} {Journal of Artificial
  Intelligence Research}\ }\textbf {\bibinfo {volume} {39}},\ \bibinfo {pages}
  {533} (\bibinfo {year} {2010})}\BibitemShut {NoStop}%
\bibitem [{\citenamefont {Standley}\ and\ \citenamefont
  {Korf}(2011)}]{Standley2011}%
  \BibitemOpen
  \bibfield  {author} {\bibinfo {author} {\bibfnamefont {T.}~\bibnamefont
  {Standley}}\ and\ \bibinfo {author} {\bibfnamefont {R.}~\bibnamefont
  {Korf}},\ }\href@noop {} {\bibfield  {journal} {\bibinfo  {journal}
  {Proceedings of the Twenty-Second international joint conference on
  Artificial Intelligence}\ }\textbf {\bibinfo {volume} {1}},\ \bibinfo {pages}
  {668} (\bibinfo {year} {2011})}\BibitemShut {NoStop}%
\bibitem [{\citenamefont {Morris}\ \emph {et~al.}(2016)\citenamefont {Morris},
  \citenamefont {Pǎsǎreanu}, \citenamefont {Luckow}, \citenamefont {Malik},
  \citenamefont {Ma}, \citenamefont {{Satish Kumar}},\ and\ \citenamefont
  {Koenig}}]{Morris2016}%
  \BibitemOpen
  \bibfield  {author} {\bibinfo {author} {\bibfnamefont {R.}~\bibnamefont
  {Morris}}, \bibinfo {author} {\bibfnamefont {C.~S.}\ \bibnamefont
  {Pǎsǎreanu}}, \bibinfo {author} {\bibfnamefont {K.}~\bibnamefont {Luckow}},
  \bibinfo {author} {\bibfnamefont {W.}~\bibnamefont {Malik}}, \bibinfo
  {author} {\bibfnamefont {H.}~\bibnamefont {Ma}}, \bibinfo {author}
  {\bibfnamefont {T.~K.}\ \bibnamefont {{Satish Kumar}}}, \ and\ \bibinfo
  {author} {\bibfnamefont {S.}~\bibnamefont {Koenig}},\ }in\ \href@noop {}
  {\emph {\bibinfo {booktitle} {The Workshops of the Thirtieth AAAI Conference
  on Artificial Intelligence Planning for Hybrid Systems: Technical Report
  WS-16-12}}}\ (\bibinfo {year} {2016})\ pp.\ \bibinfo {pages}
  {608--614}\BibitemShut {NoStop}%
\bibitem [{\citenamefont {Ma}\ \emph {et~al.}(2017)\citenamefont {Ma},
  \citenamefont {Kumar}, \citenamefont {Li},\ and\ \citenamefont
  {Koenig}}]{Ma2017}%
  \BibitemOpen
  \bibfield  {author} {\bibinfo {author} {\bibfnamefont {H.}~\bibnamefont
  {Ma}}, \bibinfo {author} {\bibfnamefont {T.~K.}\ \bibnamefont {Kumar}},
  \bibinfo {author} {\bibfnamefont {J.}~\bibnamefont {Li}}, \ and\ \bibinfo
  {author} {\bibfnamefont {S.}~\bibnamefont {Koenig}},\ }\href@noop {}
  {\bibfield  {journal} {\bibinfo  {journal} {Proceedings of the International
  Joint Conference on Autonomous Agents and Multiagent Systems, AAMAS}\
  }\textbf {\bibinfo {volume} {2}},\ \bibinfo {pages} {837} (\bibinfo {year}
  {2017})}\BibitemShut {NoStop}%
\bibitem [{\citenamefont {Surynek}(2020)}]{Surynek2020}%
  \BibitemOpen
  \bibfield  {author} {\bibinfo {author} {\bibfnamefont {P.}~\bibnamefont
  {Surynek}},\ }in\ \href@noop {} {\emph {\bibinfo {booktitle} {18th Russian
  Conference on Artificial Intelligence}}}\ (\bibinfo {year} {2020})\ pp.\
  \bibinfo {pages} {85----99}\BibitemShut {NoStop}%
\bibitem [{\citenamefont {Mai}\ and\ \citenamefont
  {Mostaghim}(2020)}]{Mai2020}%
  \BibitemOpen
  \bibfield  {author} {\bibinfo {author} {\bibfnamefont {S.}~\bibnamefont
  {Mai}}\ and\ \bibinfo {author} {\bibfnamefont {S.}~\bibnamefont
  {Mostaghim}},\ }in\ \href@noop {} {\emph {\bibinfo {booktitle} {Swarm
  Intelligence -- 12th international conference, ANTS 2020}}}\ (\bibinfo {year}
  {2020})\ pp.\ \bibinfo {pages} {190----202}\BibitemShut {NoStop}%
\bibitem [{\citenamefont {Talamali}\ \emph {et~al.}(2021)\citenamefont
  {Talamali}, \citenamefont {Saha}, \citenamefont {Marshall},\ and\
  \citenamefont {Reina}}]{Talamali2021}%
  \BibitemOpen
  \bibfield  {author} {\bibinfo {author} {\bibfnamefont {M.~S.}\ \bibnamefont
  {Talamali}}, \bibinfo {author} {\bibfnamefont {A.}~\bibnamefont {Saha}},
  \bibinfo {author} {\bibfnamefont {J.~A.~R.}\ \bibnamefont {Marshall}}, \ and\
  \bibinfo {author} {\bibfnamefont {A.}~\bibnamefont {Reina}},\ }\href@noop {}
  {\bibfield  {journal} {\bibinfo  {journal} {Science Robotics}\ }\textbf
  {\bibinfo {volume} {6}},\ \bibinfo {pages} {1416} (\bibinfo {year}
  {2021})}\BibitemShut {NoStop}%
\bibitem [{\citenamefont {Biham}\ \emph {et~al.}(1992)\citenamefont {Biham},
  \citenamefont {Middleton},\ and\ \citenamefont {Levine}}]{Biham1992}%
  \BibitemOpen
  \bibfield  {author} {\bibinfo {author} {\bibfnamefont {O.}~\bibnamefont
  {Biham}}, \bibinfo {author} {\bibfnamefont {A.~A.}\ \bibnamefont
  {Middleton}}, \ and\ \bibinfo {author} {\bibfnamefont {D.}~\bibnamefont
  {Levine}},\ }\href@noop {} {\bibfield  {journal} {\bibinfo  {journal}
  {Physical Review A}\ }\textbf {\bibinfo {volume} {46}},\ \bibinfo {pages}
  {6124} (\bibinfo {year} {1992})}\BibitemShut {NoStop}%
\bibitem [{\citenamefont {Fily}\ and\ \citenamefont
  {Marchetti}(2012)}]{Fily2012}%
  \BibitemOpen
  \bibfield  {author} {\bibinfo {author} {\bibfnamefont {Y.}~\bibnamefont
  {Fily}}\ and\ \bibinfo {author} {\bibfnamefont {M.~C.}\ \bibnamefont
  {Marchetti}},\ }\href@noop {} {\bibfield  {journal} {\bibinfo  {journal}
  {Physical Review Letters}\ }\textbf {\bibinfo {volume} {108}},\ \bibinfo
  {pages} {235702} (\bibinfo {year} {2012})}\BibitemShut {NoStop}%
\bibitem [{\citenamefont {Lighthill}\ and\ \citenamefont
  {Whitham}(1955)}]{Lighthill1955}%
  \BibitemOpen
  \bibfield  {author} {\bibinfo {author} {\bibfnamefont {M.~J.}\ \bibnamefont
  {Lighthill}}\ and\ \bibinfo {author} {\bibfnamefont {G.~B.}\ \bibnamefont
  {Whitham}},\ }\href@noop {} {\bibfield  {journal} {\bibinfo  {journal}
  {Proceedings of the Royal Society of London. Series A. Mathematical and
  Physical Sciences}\ }\textbf {\bibinfo {volume} {229}},\ \bibinfo {pages}
  {317} (\bibinfo {year} {1955})}\BibitemShut {NoStop}%
\bibitem [{\citenamefont {Richards}(1956)}]{Richards1956}%
  \BibitemOpen
  \bibfield  {author} {\bibinfo {author} {\bibfnamefont {P.~I.}\ \bibnamefont
  {Richards}},\ }\href@noop {} {\bibfield  {journal} {\bibinfo  {journal}
  {Operations Research}\ }\textbf {\bibinfo {volume} {4}},\ \bibinfo {pages}
  {42} (\bibinfo {year} {1956})}\BibitemShut {NoStop}%
\bibitem [{Note1()}]{Note1}%
  \BibitemOpen
  \bibinfo {note} {These dynamics are typically Brownian~\cite
  {Redner2001,Benichou2010,Grebenkov2015,Agranov2017,Agranov2019}, more
  recently advected~\cite {MejiaMonasterio2020} or self-propelled~\cite
  {Basu2018}.}\BibitemShut {Stop}%
\bibitem [{\citenamefont {Chakraborty}\ \emph {et~al.}(2020)\citenamefont
  {Chakraborty}, \citenamefont {Bhunia},\ and\ \citenamefont
  {De}}]{Chakraborty2020}%
  \BibitemOpen
  \bibfield  {author} {\bibinfo {author} {\bibfnamefont {D.}~\bibnamefont
  {Chakraborty}}, \bibinfo {author} {\bibfnamefont {S.}~\bibnamefont {Bhunia}},
  \ and\ \bibinfo {author} {\bibfnamefont {R.}~\bibnamefont {De}},\ }\href@noop
  {} {\bibfield  {journal} {\bibinfo  {journal} {Scientific Reports}\ }\textbf
  {\bibinfo {volume} {10}},\ \bibinfo {pages} {8362} (\bibinfo {year}
  {2020})}\BibitemShut {NoStop}%
\bibitem [{\citenamefont {Helbing}\ \emph {et~al.}(2000)\citenamefont
  {Helbing}, \citenamefont {Farkas},\ and\ \citenamefont
  {Vicsek}}]{Helbing2000}%
  \BibitemOpen
  \bibfield  {author} {\bibinfo {author} {\bibfnamefont {D.}~\bibnamefont
  {Helbing}}, \bibinfo {author} {\bibfnamefont {I.}~\bibnamefont {Farkas}}, \
  and\ \bibinfo {author} {\bibfnamefont {T.}~\bibnamefont {Vicsek}},\
  }\href@noop {} {\bibfield  {journal} {\bibinfo  {journal} {Nature}\ }\textbf
  {\bibinfo {volume} {144}},\ \bibinfo {pages} {297} (\bibinfo {year}
  {2000})}\BibitemShut {NoStop}%
\bibitem [{\citenamefont {Cates}\ and\ \citenamefont
  {Tailleur}(2014)}]{Cates2014}%
  \BibitemOpen
  \bibfield  {author} {\bibinfo {author} {\bibfnamefont {M.~E.}\ \bibnamefont
  {Cates}}\ and\ \bibinfo {author} {\bibfnamefont {J.}~\bibnamefont
  {Tailleur}},\ }\href@noop {} {\bibfield  {journal} {\bibinfo  {journal}
  {Annual Review of Condensed Matter Physics}\ }\textbf {\bibinfo {volume}
  {6}},\ \bibinfo {pages} {219} (\bibinfo {year} {2014})}\BibitemShut {NoStop}%
\bibitem [{\citenamefont {Patch}\ \emph {et~al.}(2017)\citenamefont {Patch},
  \citenamefont {Yllanes},\ and\ \citenamefont {Marchetti}}]{Patch2017}%
  \BibitemOpen
  \bibfield  {author} {\bibinfo {author} {\bibfnamefont {A.}~\bibnamefont
  {Patch}}, \bibinfo {author} {\bibfnamefont {D.}~\bibnamefont {Yllanes}}, \
  and\ \bibinfo {author} {\bibfnamefont {M.~C.}\ \bibnamefont {Marchetti}},\
  }\href@noop {} {\bibfield  {journal} {\bibinfo  {journal} {Physical Review
  E}\ }\textbf {\bibinfo {volume} {95}},\ \bibinfo {pages} {012601} (\bibinfo
  {year} {2017})}\BibitemShut {NoStop}%
\bibitem [{\citenamefont {Bruss}\ and\ \citenamefont
  {Glotzer}(2018)}]{Bruss2018}%
  \BibitemOpen
  \bibfield  {author} {\bibinfo {author} {\bibfnamefont {I.~R.}\ \bibnamefont
  {Bruss}}\ and\ \bibinfo {author} {\bibfnamefont {S.~C.}\ \bibnamefont
  {Glotzer}},\ }\href@noop {} {\bibfield  {journal} {\bibinfo  {journal}
  {Physical Review E}\ }\textbf {\bibinfo {volume} {97}},\ \bibinfo {pages}
  {042609} (\bibinfo {year} {2018})}\BibitemShut {NoStop}%
\bibitem [{\citenamefont {{Van Damme}}\ \emph {et~al.}(2019)\citenamefont {{Van
  Damme}}, \citenamefont {Rodenburg}, \citenamefont {{Van Roij}},\ and\
  \citenamefont {Dijkstra}}]{VanDamme2019}%
  \BibitemOpen
  \bibfield  {author} {\bibinfo {author} {\bibfnamefont {R.}~\bibnamefont {{Van
  Damme}}}, \bibinfo {author} {\bibfnamefont {J.}~\bibnamefont {Rodenburg}},
  \bibinfo {author} {\bibfnamefont {R.}~\bibnamefont {{Van Roij}}}, \ and\
  \bibinfo {author} {\bibfnamefont {M.}~\bibnamefont {Dijkstra}},\ }\href@noop
  {} {\bibfield  {journal} {\bibinfo  {journal} {Journal of Chemical Physics}\
  }\textbf {\bibinfo {volume} {150}} (\bibinfo {year} {2019})}\BibitemShut
  {NoStop}%
\bibitem [{\citenamefont {Nie}\ \emph {et~al.}(2020)\citenamefont {Nie},
  \citenamefont {Chattoraj}, \citenamefont {Piscitelli}, \citenamefont {Doyle},
  \citenamefont {Ni},\ and\ \citenamefont {{Pica Ciamarra}}}]{Nie2020}%
  \BibitemOpen
  \bibfield  {author} {\bibinfo {author} {\bibfnamefont {P.}~\bibnamefont
  {Nie}}, \bibinfo {author} {\bibfnamefont {J.}~\bibnamefont {Chattoraj}},
  \bibinfo {author} {\bibfnamefont {A.}~\bibnamefont {Piscitelli}}, \bibinfo
  {author} {\bibfnamefont {P.}~\bibnamefont {Doyle}}, \bibinfo {author}
  {\bibfnamefont {R.}~\bibnamefont {Ni}}, \ and\ \bibinfo {author}
  {\bibfnamefont {M.}~\bibnamefont {{Pica Ciamarra}}},\ }\href@noop {}
  {\bibfield  {journal} {\bibinfo  {journal} {Physical Review Research}\
  }\textbf {\bibinfo {volume} {2}},\ \bibinfo {pages} {023010} (\bibinfo {year}
  {2020})}\BibitemShut {NoStop}%
\bibitem [{Note2()}]{Note2}%
  \BibitemOpen
  \bibinfo {note} {ABPs also have a strictly frozen phase, but at packing
  fractions considerably higher than what we observe~\cite
  {Henkes2011,Digregorio2018,VanDamme2019,Mandal2020}.}\BibitemShut {Stop}%
\bibitem [{\citenamefont {Farage}\ \emph {et~al.}(2015)\citenamefont {Farage},
  \citenamefont {Krinninger},\ and\ \citenamefont {Brader}}]{Farage2015}%
  \BibitemOpen
  \bibfield  {author} {\bibinfo {author} {\bibfnamefont {T.~F.}\ \bibnamefont
  {Farage}}, \bibinfo {author} {\bibfnamefont {P.}~\bibnamefont {Krinninger}},
  \ and\ \bibinfo {author} {\bibfnamefont {J.~M.}\ \bibnamefont {Brader}},\
  }\href@noop {} {\bibfield  {journal} {\bibinfo  {journal} {Physical Review E
  - Statistical, Nonlinear, and Soft Matter Physics}\ }\textbf {\bibinfo
  {volume} {91}},\ \bibinfo {pages} {042310} (\bibinfo {year}
  {2015})}\BibitemShut {NoStop}%
\bibitem [{\citenamefont {Solon}\ \emph {et~al.}(2016)\citenamefont {Solon},
  \citenamefont {Stenhammar}, \citenamefont {Cates}, \citenamefont {Kafri},\
  and\ \citenamefont {Tailleur}}]{Solon2016}%
  \BibitemOpen
  \bibfield  {author} {\bibinfo {author} {\bibfnamefont {A.~P.}\ \bibnamefont
  {Solon}}, \bibinfo {author} {\bibfnamefont {J.}~\bibnamefont {Stenhammar}},
  \bibinfo {author} {\bibfnamefont {M.~E.}\ \bibnamefont {Cates}}, \bibinfo
  {author} {\bibfnamefont {Y.}~\bibnamefont {Kafri}}, \ and\ \bibinfo {author}
  {\bibfnamefont {J.}~\bibnamefont {Tailleur}},\ }\href@noop {} {\bibfield
  {journal} {\bibinfo  {journal} {Physical Review E}\ }\textbf {\bibinfo
  {volume} {97}},\ \bibinfo {pages} {020602(R)} (\bibinfo {year}
  {2016})}\BibitemShut {NoStop}%
\bibitem [{\citenamefont {Turci}\ and\ \citenamefont
  {Wilding}(2021)}]{Turci2021}%
  \BibitemOpen
  \bibfield  {author} {\bibinfo {author} {\bibfnamefont {F.}~\bibnamefont
  {Turci}}\ and\ \bibinfo {author} {\bibfnamefont {N.~B.}\ \bibnamefont
  {Wilding}},\ }\href@noop {} {\bibfield  {journal} {\bibinfo  {journal}
  {Physical Review Letters}\ }\textbf {\bibinfo {volume} {126}},\ \bibinfo
  {pages} {038002} (\bibinfo {year} {2021})}\BibitemShut {NoStop}%
\bibitem [{\citenamefont {Reichhardt}\ and\ \citenamefont {{Olson
  Reichhardt}}(2014)}]{Reichhardt2014}%
  \BibitemOpen
  \bibfield  {author} {\bibinfo {author} {\bibfnamefont {C.}~\bibnamefont
  {Reichhardt}}\ and\ \bibinfo {author} {\bibfnamefont {C.~J.}\ \bibnamefont
  {{Olson Reichhardt}}},\ }\href@noop {} {\bibfield  {journal} {\bibinfo
  {journal} {Soft Matter}\ }\textbf {\bibinfo {volume} {10}},\ \bibinfo {pages}
  {7502} (\bibinfo {year} {2014})}\BibitemShut {NoStop}%
\bibitem [{\citenamefont {Solon}\ \emph {et~al.}(2015)\citenamefont {Solon},
  \citenamefont {Stenhammar}, \citenamefont {Wittkowski}, \citenamefont
  {Kardar}, \citenamefont {Kafri}, \citenamefont {Cates},\ and\ \citenamefont
  {Tailleur}}]{Solon2015d}%
  \BibitemOpen
  \bibfield  {author} {\bibinfo {author} {\bibfnamefont {A.~P.}\ \bibnamefont
  {Solon}}, \bibinfo {author} {\bibfnamefont {J.}~\bibnamefont {Stenhammar}},
  \bibinfo {author} {\bibfnamefont {R.}~\bibnamefont {Wittkowski}}, \bibinfo
  {author} {\bibfnamefont {M.}~\bibnamefont {Kardar}}, \bibinfo {author}
  {\bibfnamefont {Y.}~\bibnamefont {Kafri}}, \bibinfo {author} {\bibfnamefont
  {M.~E.}\ \bibnamefont {Cates}}, \ and\ \bibinfo {author} {\bibfnamefont
  {J.}~\bibnamefont {Tailleur}},\ }\href@noop {} {\bibfield  {journal}
  {\bibinfo  {journal} {Physical Review Letters}\ }\textbf {\bibinfo {volume}
  {114}},\ \bibinfo {pages} {198301} (\bibinfo {year} {2015})}\BibitemShut
  {NoStop}%
\bibitem [{\citenamefont {Liu}\ and\ \citenamefont {Nagel}(1998)}]{Liu1998}%
  \BibitemOpen
  \bibfield  {author} {\bibinfo {author} {\bibfnamefont {A.~J.}\ \bibnamefont
  {Liu}}\ and\ \bibinfo {author} {\bibfnamefont {S.~R.}\ \bibnamefont
  {Nagel}},\ }\href@noop {} {\bibfield  {journal} {\bibinfo  {journal}
  {Nature}\ }\textbf {\bibinfo {volume} {81}},\ \bibinfo {pages} {226}
  (\bibinfo {year} {1998})}\BibitemShut {NoStop}%
\bibitem [{\citenamefont {Henkes}\ \emph {et~al.}(2011)\citenamefont {Henkes},
  \citenamefont {Fily},\ and\ \citenamefont {Marchetti}}]{Henkes2011}%
  \BibitemOpen
  \bibfield  {author} {\bibinfo {author} {\bibfnamefont {S.}~\bibnamefont
  {Henkes}}, \bibinfo {author} {\bibfnamefont {Y.}~\bibnamefont {Fily}}, \ and\
  \bibinfo {author} {\bibfnamefont {M.~C.}\ \bibnamefont {Marchetti}},\
  }\href@noop {} {\bibfield  {journal} {\bibinfo  {journal} {Physical Review E
  - Statistical, Nonlinear, and Soft Matter Physics}\ }\textbf {\bibinfo
  {volume} {84}},\ \bibinfo {pages} {040301(R)} (\bibinfo {year}
  {2011})}\BibitemShut {NoStop}%
\bibitem [{\citenamefont {Digregorio}\ \emph {et~al.}(2018)\citenamefont
  {Digregorio}, \citenamefont {Levis}, \citenamefont {Suma}, \citenamefont
  {Cugliandolo}, \citenamefont {Gonnella},\ and\ \citenamefont
  {Pagonabarraga}}]{Digregorio2018}%
  \BibitemOpen
  \bibfield  {author} {\bibinfo {author} {\bibfnamefont {P.}~\bibnamefont
  {Digregorio}}, \bibinfo {author} {\bibfnamefont {D.}~\bibnamefont {Levis}},
  \bibinfo {author} {\bibfnamefont {A.}~\bibnamefont {Suma}}, \bibinfo {author}
  {\bibfnamefont {L.~F.}\ \bibnamefont {Cugliandolo}}, \bibinfo {author}
  {\bibfnamefont {G.}~\bibnamefont {Gonnella}}, \ and\ \bibinfo {author}
  {\bibfnamefont {I.}~\bibnamefont {Pagonabarraga}},\ }\href@noop {} {\bibfield
   {journal} {\bibinfo  {journal} {Physical Review Letters}\ }\textbf {\bibinfo
  {volume} {121}},\ \bibinfo {pages} {98003} (\bibinfo {year}
  {2018})}\BibitemShut {NoStop}%
\bibitem [{\citenamefont {Merrigan}\ \emph {et~al.}(2020)\citenamefont
  {Merrigan}, \citenamefont {Ramola}, \citenamefont {Chatterjee}, \citenamefont
  {Segall}, \citenamefont {Shokef},\ and\ \citenamefont
  {Chakraborty}}]{Merrigan2019}%
  \BibitemOpen
  \bibfield  {author} {\bibinfo {author} {\bibfnamefont {C.}~\bibnamefont
  {Merrigan}}, \bibinfo {author} {\bibfnamefont {K.}~\bibnamefont {Ramola}},
  \bibinfo {author} {\bibfnamefont {R.}~\bibnamefont {Chatterjee}}, \bibinfo
  {author} {\bibfnamefont {N.}~\bibnamefont {Segall}}, \bibinfo {author}
  {\bibfnamefont {Y.}~\bibnamefont {Shokef}}, \ and\ \bibinfo {author}
  {\bibfnamefont {B.}~\bibnamefont {Chakraborty}},\ }\href@noop {} {\bibfield
  {journal} {\bibinfo  {journal} {Physical Review Research}\ }\textbf {\bibinfo
  {volume} {2}},\ \bibinfo {pages} {013260} (\bibinfo {year}
  {2020})}\BibitemShut {NoStop}%
\bibitem [{\citenamefont {Mandal}\ \emph {et~al.}(2020)\citenamefont {Mandal},
  \citenamefont {Bhuyan}, \citenamefont {Chaudhuri}, \citenamefont {Dasgupta},\
  and\ \citenamefont {Rao}}]{Mandal2020}%
  \BibitemOpen
  \bibfield  {author} {\bibinfo {author} {\bibfnamefont {R.}~\bibnamefont
  {Mandal}}, \bibinfo {author} {\bibfnamefont {P.~J.}\ \bibnamefont {Bhuyan}},
  \bibinfo {author} {\bibfnamefont {P.}~\bibnamefont {Chaudhuri}}, \bibinfo
  {author} {\bibfnamefont {C.}~\bibnamefont {Dasgupta}}, \ and\ \bibinfo
  {author} {\bibfnamefont {M.}~\bibnamefont {Rao}},\ }\href@noop {} {\bibfield
  {journal} {\bibinfo  {journal} {Nature Communications}\ }\textbf {\bibinfo
  {volume} {11}},\ \bibinfo {pages} {2581} (\bibinfo {year}
  {2020})}\BibitemShut {NoStop}%
\bibitem [{\citenamefont {Zhang}\ \emph {et~al.}(2021)\citenamefont {Zhang},
  \citenamefont {Alert}, \citenamefont {Yan}, \citenamefont {Wingreen},\ and\
  \citenamefont {Granick}}]{Zhang2021}%
  \BibitemOpen
  \bibfield  {author} {\bibinfo {author} {\bibfnamefont {J.}~\bibnamefont
  {Zhang}}, \bibinfo {author} {\bibfnamefont {R.}~\bibnamefont {Alert}},
  \bibinfo {author} {\bibfnamefont {J.}~\bibnamefont {Yan}}, \bibinfo {author}
  {\bibfnamefont {N.~S.}\ \bibnamefont {Wingreen}}, \ and\ \bibinfo {author}
  {\bibfnamefont {S.}~\bibnamefont {Granick}},\ }\href@noop {} {\bibfield
  {journal} {\bibinfo  {journal} {Nature Physics}\ ,\ \bibinfo {pages}
  {s41567}} (\bibinfo {year} {2021})}\BibitemShut {NoStop}%
\bibitem [{\citenamefont {Gershenson}\ and\ \citenamefont
  {Helbing}(2011)}]{Gershenson2011}%
  \BibitemOpen
  \bibfield  {author} {\bibinfo {author} {\bibfnamefont {C.}~\bibnamefont
  {Gershenson}}\ and\ \bibinfo {author} {\bibfnamefont {D.}~\bibnamefont
  {Helbing}},\ }\href@noop {} {\bibfield  {journal} {\bibinfo  {journal}
  {Complexity}\ }\textbf {\bibinfo {volume} {16}},\ \bibinfo {pages} {10}
  (\bibinfo {year} {2011})}\BibitemShut {NoStop}%
\bibitem [{\citenamefont {Chepizhko}\ \emph {et~al.}(2013)\citenamefont
  {Chepizhko}, \citenamefont {Altmann},\ and\ \citenamefont
  {Peruani}}]{Chepizhko2013}%
  \BibitemOpen
  \bibfield  {author} {\bibinfo {author} {\bibfnamefont {O.}~\bibnamefont
  {Chepizhko}}, \bibinfo {author} {\bibfnamefont {E.~G.}\ \bibnamefont
  {Altmann}}, \ and\ \bibinfo {author} {\bibfnamefont {F.}~\bibnamefont
  {Peruani}},\ }\href@noop {} {\bibfield  {journal} {\bibinfo  {journal}
  {Physical Review Letters}\ }\textbf {\bibinfo {volume} {110}},\ \bibinfo
  {pages} {238101} (\bibinfo {year} {2013})}\BibitemShut {NoStop}%
\bibitem [{\citenamefont {Azimzade}\ and\ \citenamefont
  {Mashaghi}(2017)}]{Azimzade2017}%
  \BibitemOpen
  \bibfield  {author} {\bibinfo {author} {\bibfnamefont {Y.}~\bibnamefont
  {Azimzade}}\ and\ \bibinfo {author} {\bibfnamefont {A.}~\bibnamefont
  {Mashaghi}},\ }\href@noop {} {\bibfield  {journal} {\bibinfo  {journal}
  {Physical Review E}\ }\textbf {\bibinfo {volume} {96}},\ \bibinfo {pages}
  {062415} (\bibinfo {year} {2017})}\BibitemShut {NoStop}%
\bibitem [{\citenamefont {Volpe}\ and\ \citenamefont
  {Volpe}(2017)}]{Volpe2017}%
  \BibitemOpen
  \bibfield  {author} {\bibinfo {author} {\bibfnamefont {G.}~\bibnamefont
  {Volpe}}\ and\ \bibinfo {author} {\bibfnamefont {G.}~\bibnamefont {Volpe}},\
  }\href@noop {} {\bibfield  {journal} {\bibinfo  {journal} {Proceedings of the
  National Academy of Sciences of the United States of America}\ }\textbf
  {\bibinfo {volume} {114}},\ \bibinfo {pages} {11350} (\bibinfo {year}
  {2017})}\BibitemShut {NoStop}%
\bibitem [{\citenamefont {Bertrand}\ \emph {et~al.}(2018)\citenamefont
  {Bertrand}, \citenamefont {Zhao}, \citenamefont {B{\'{e}}nichou},
  \citenamefont {Tailleur},\ and\ \citenamefont {Voituriez}}]{Bertrand2018}%
  \BibitemOpen
  \bibfield  {author} {\bibinfo {author} {\bibfnamefont {T.}~\bibnamefont
  {Bertrand}}, \bibinfo {author} {\bibfnamefont {Y.}~\bibnamefont {Zhao}},
  \bibinfo {author} {\bibfnamefont {O.}~\bibnamefont {B{\'{e}}nichou}},
  \bibinfo {author} {\bibfnamefont {J.}~\bibnamefont {Tailleur}}, \ and\
  \bibinfo {author} {\bibfnamefont {R.}~\bibnamefont {Voituriez}},\ }\href@noop
  {} {\bibfield  {journal} {\bibinfo  {journal} {Physical Review Letters}\
  }\textbf {\bibinfo {volume} {120}},\ \bibinfo {pages} {198103} (\bibinfo
  {year} {2018})}\BibitemShut {NoStop}%
\bibitem [{\citenamefont {Raible}\ and\ \citenamefont
  {Engel}(2004)}]{Raible2004}%
  \BibitemOpen
  \bibfield  {author} {\bibinfo {author} {\bibfnamefont {M.}~\bibnamefont
  {Raible}}\ and\ \bibinfo {author} {\bibfnamefont {A.}~\bibnamefont {Engel}},\
  }\href {\doibase 10.1002/aoc.757} {\bibfield  {journal} {\bibinfo  {journal}
  {Applied Organometallic Chemistry}\ }\textbf {\bibinfo {volume} {18}},\
  \bibinfo {pages} {536} (\bibinfo {year} {2004})}\BibitemShut {NoStop}%
\bibitem [{\citenamefont {Winkler}\ \emph {et~al.}(2015)\citenamefont
  {Winkler}, \citenamefont {Wysocki},\ and\ \citenamefont
  {Gompper}}]{Winkler2015}%
  \BibitemOpen
  \bibfield  {author} {\bibinfo {author} {\bibfnamefont {R.~G.}\ \bibnamefont
  {Winkler}}, \bibinfo {author} {\bibfnamefont {A.}~\bibnamefont {Wysocki}}, \
  and\ \bibinfo {author} {\bibfnamefont {G.}~\bibnamefont {Gompper}},\
  }\href@noop {} {\bibfield  {journal} {\bibinfo  {journal} {Soft Matter}\
  }\textbf {\bibinfo {volume} {11}},\ \bibinfo {pages} {6680} (\bibinfo {year}
  {2015})}\BibitemShut {NoStop}%
\bibitem [{\citenamefont {Redner}\ \emph {et~al.}(2013)\citenamefont {Redner},
  \citenamefont {Hagan},\ and\ \citenamefont {Baskaran}}]{Redner2013}%
  \BibitemOpen
  \bibfield  {author} {\bibinfo {author} {\bibfnamefont {G.~S.}\ \bibnamefont
  {Redner}}, \bibinfo {author} {\bibfnamefont {M.~F.}\ \bibnamefont {Hagan}}, \
  and\ \bibinfo {author} {\bibfnamefont {A.}~\bibnamefont {Baskaran}},\
  }\href@noop {} {\bibfield  {journal} {\bibinfo  {journal} {Physical Review
  Letters}\ }\textbf {\bibinfo {volume} {110}},\ \bibinfo {pages} {055701}
  (\bibinfo {year} {2013})}\BibitemShut {NoStop}%
\bibitem [{\citenamefont {Jaeger}\ and\ \citenamefont
  {Nagel}(1992)}]{Jaeger1992}%
  \BibitemOpen
  \bibfield  {author} {\bibinfo {author} {\bibfnamefont {H.~M.}\ \bibnamefont
  {Jaeger}}\ and\ \bibinfo {author} {\bibfnamefont {S.~R.}\ \bibnamefont
  {Nagel}},\ }\href@noop {} {\bibfield  {journal} {\bibinfo  {journal}
  {Science}\ }\textbf {\bibinfo {volume} {255}},\ \bibinfo {pages} {1523}
  (\bibinfo {year} {1992})}\BibitemShut {NoStop}%
\bibitem [{\citenamefont {Kumar}\ and\ \citenamefont
  {Kumaran}(2005)}]{Kumar2005}%
  \BibitemOpen
  \bibfield  {author} {\bibinfo {author} {\bibfnamefont {V.~S.}\ \bibnamefont
  {Kumar}}\ and\ \bibinfo {author} {\bibfnamefont {V.}~\bibnamefont
  {Kumaran}},\ }\href@noop {} {\bibfield  {journal} {\bibinfo  {journal}
  {Journal of Chemical Physics}\ }\textbf {\bibinfo {volume} {123}},\ \bibinfo
  {pages} {074502} (\bibinfo {year} {2005})}\BibitemShut {NoStop}%
\bibitem [{\citenamefont {Steinhardt}\ \emph {et~al.}(1983)\citenamefont
  {Steinhardt}, \citenamefont {Nelson},\ and\ \citenamefont
  {Ronchetti}}]{Steinhardt1983}%
  \BibitemOpen
  \bibfield  {author} {\bibinfo {author} {\bibfnamefont {P.~J.}\ \bibnamefont
  {Steinhardt}}, \bibinfo {author} {\bibfnamefont {D.~R.}\ \bibnamefont
  {Nelson}}, \ and\ \bibinfo {author} {\bibfnamefont {M.}~\bibnamefont
  {Ronchetti}},\ }\href@noop {} {\bibfield  {journal} {\bibinfo  {journal}
  {Physical Review B}\ }\textbf {\bibinfo {volume} {28}},\ \bibinfo {pages}
  {784} (\bibinfo {year} {1983})}\BibitemShut {NoStop}%
\bibitem [{\citenamefont {Mickel}\ \emph {et~al.}(2013)\citenamefont {Mickel},
  \citenamefont {Kapfer}, \citenamefont {Schr{\"{o}}der-Turk},\ and\
  \citenamefont {Mecke}}]{Mickel2013}%
  \BibitemOpen
  \bibfield  {author} {\bibinfo {author} {\bibfnamefont {W.}~\bibnamefont
  {Mickel}}, \bibinfo {author} {\bibfnamefont {S.~C.}\ \bibnamefont {Kapfer}},
  \bibinfo {author} {\bibfnamefont {G.~E.}\ \bibnamefont
  {Schr{\"{o}}der-Turk}}, \ and\ \bibinfo {author} {\bibfnamefont
  {K.}~\bibnamefont {Mecke}},\ }\href@noop {} {\bibfield  {journal} {\bibinfo
  {journal} {Journal of Chemical Physics}\ }\textbf {\bibinfo {volume} {138}},\
  \bibinfo {pages} {044501} (\bibinfo {year} {2013})}\BibitemShut {NoStop}%
\bibitem [{\citenamefont {{Van Damme}}\ \emph {et~al.}(2020)\citenamefont {{Van
  Damme}}, \citenamefont {Coli}, \citenamefont {{Van Roij}},\ and\
  \citenamefont {Dijkstra}}]{VanDamme2020}%
  \BibitemOpen
  \bibfield  {author} {\bibinfo {author} {\bibfnamefont {R.}~\bibnamefont {{Van
  Damme}}}, \bibinfo {author} {\bibfnamefont {G.~M.}\ \bibnamefont {Coli}},
  \bibinfo {author} {\bibfnamefont {R.}~\bibnamefont {{Van Roij}}}, \ and\
  \bibinfo {author} {\bibfnamefont {M.}~\bibnamefont {Dijkstra}},\ }\href@noop
  {} {\bibfield  {journal} {\bibinfo  {journal} {ACS Nano}\ }\textbf {\bibinfo
  {volume} {14}},\ \bibinfo {pages} {15144} (\bibinfo {year}
  {2020})}\BibitemShut {NoStop}%
\bibitem [{\citenamefont {Majumdar}\ and\ \citenamefont
  {Orland}(2015)}]{Majumdar2015}%
  \BibitemOpen
  \bibfield  {author} {\bibinfo {author} {\bibfnamefont {S.~N.}\ \bibnamefont
  {Majumdar}}\ and\ \bibinfo {author} {\bibfnamefont {H.}~\bibnamefont
  {Orland}},\ }\href@noop {} {\bibfield  {journal} {\bibinfo  {journal}
  {Journal of Statistical Mechanics: Theory and Experiment}\ }\textbf {\bibinfo
  {volume} {2015}},\ \bibinfo {pages} {P06039} (\bibinfo {year}
  {2015})}\BibitemShut {NoStop}%
\bibitem [{\citenamefont {Szavits-Nossan}\ and\ \citenamefont
  {Evans}(2015)}]{Szavits-Nossan2015}%
  \BibitemOpen
  \bibfield  {author} {\bibinfo {author} {\bibfnamefont {J.}~\bibnamefont
  {Szavits-Nossan}}\ and\ \bibinfo {author} {\bibfnamefont {M.~R.}\
  \bibnamefont {Evans}},\ }\href@noop {} {\bibfield  {journal} {\bibinfo
  {journal} {Journal of Statistical Mechanics: Theory and Experiment}\ }\textbf
  {\bibinfo {volume} {2015}},\ \bibinfo {pages} {P12008} (\bibinfo {year}
  {2015})}\BibitemShut {NoStop}%
\bibitem [{\citenamefont {Redner}(2001)}]{Redner2001}%
  \BibitemOpen
  \bibfield  {author} {\bibinfo {author} {\bibfnamefont {S.}~\bibnamefont
  {Redner}},\ }\href@noop {} {\emph {\bibinfo {title} {{A Guide to
  First-Passage Processes}}}}\ (\bibinfo {year} {2001})\BibitemShut {NoStop}%
\bibitem [{\citenamefont {B{\'{e}}nichou}\ \emph {et~al.}(2010)\citenamefont
  {B{\'{e}}nichou}, \citenamefont {Chevalier}, \citenamefont {Klafter},
  \citenamefont {Meyer},\ and\ \citenamefont {Voituriez}}]{Benichou2010}%
  \BibitemOpen
  \bibfield  {author} {\bibinfo {author} {\bibfnamefont {O.}~\bibnamefont
  {B{\'{e}}nichou}}, \bibinfo {author} {\bibfnamefont {C.}~\bibnamefont
  {Chevalier}}, \bibinfo {author} {\bibfnamefont {J.}~\bibnamefont {Klafter}},
  \bibinfo {author} {\bibfnamefont {B.}~\bibnamefont {Meyer}}, \ and\ \bibinfo
  {author} {\bibfnamefont {R.}~\bibnamefont {Voituriez}},\ }\href@noop {}
  {\bibfield  {journal} {\bibinfo  {journal} {Nature Chemistry}\ }\textbf
  {\bibinfo {volume} {2}},\ \bibinfo {pages} {472} (\bibinfo {year}
  {2010})}\BibitemShut {NoStop}%
\bibitem [{\citenamefont {Grebenkov}(2015)}]{Grebenkov2015}%
  \BibitemOpen
  \bibfield  {author} {\bibinfo {author} {\bibfnamefont {D.~S.}\ \bibnamefont
  {Grebenkov}},\ }\href@noop {} {\bibfield  {journal} {\bibinfo  {journal}
  {Journal of Physics A: Mathematical and Theoretical}\ }\textbf {\bibinfo
  {volume} {48}} (\bibinfo {year} {2015})}\BibitemShut {NoStop}%
\bibitem [{\citenamefont {Agranov}\ and\ \citenamefont
  {Meerson}(2017)}]{Agranov2017}%
  \BibitemOpen
  \bibfield  {author} {\bibinfo {author} {\bibfnamefont {T.}~\bibnamefont
  {Agranov}}\ and\ \bibinfo {author} {\bibfnamefont {B.}~\bibnamefont
  {Meerson}},\ }\href@noop {} {\bibfield  {journal} {\bibinfo  {journal}
  {Physical Review E}\ }\textbf {\bibinfo {volume} {95}},\ \bibinfo {pages}
  {062124} (\bibinfo {year} {2017})}\BibitemShut {NoStop}%
\bibitem [{\citenamefont {Agranov}\ \emph {et~al.}(2019)\citenamefont
  {Agranov}, \citenamefont {Krapivsky},\ and\ \citenamefont
  {Meerson}}]{Agranov2019}%
  \BibitemOpen
  \bibfield  {author} {\bibinfo {author} {\bibfnamefont {T.}~\bibnamefont
  {Agranov}}, \bibinfo {author} {\bibfnamefont {P.~L.}\ \bibnamefont
  {Krapivsky}}, \ and\ \bibinfo {author} {\bibfnamefont {B.}~\bibnamefont
  {Meerson}},\ }\href@noop {} {\bibfield  {journal} {\bibinfo  {journal}
  {Physical Review E}\ }\textbf {\bibinfo {volume} {99}},\ \bibinfo {pages}
  {52102} (\bibinfo {year} {2019})}\BibitemShut {NoStop}%
\bibitem [{\citenamefont {Mej{\'{i}}a-monasterio}\ \emph
  {et~al.}(2020)\citenamefont {Mej{\'{i}}a-monasterio}, \citenamefont
  {Nechaev}, \citenamefont {Oshanin},\ and\ \citenamefont
  {Vasilyev}}]{MejiaMonasterio2020}%
  \BibitemOpen
  \bibfield  {author} {\bibinfo {author} {\bibfnamefont {C.}~\bibnamefont
  {Mej{\'{i}}a-monasterio}}, \bibinfo {author} {\bibfnamefont {S.}~\bibnamefont
  {Nechaev}}, \bibinfo {author} {\bibfnamefont {G.}~\bibnamefont {Oshanin}}, \
  and\ \bibinfo {author} {\bibfnamefont {O.}~\bibnamefont {Vasilyev}},\
  }\href@noop {} {\bibfield  {journal} {\bibinfo  {journal} {New Journal of
  Physics}\ }\textbf {\bibinfo {volume} {22}},\ \bibinfo {pages} {033024}
  (\bibinfo {year} {2020})}\BibitemShut {NoStop}%
\bibitem [{\citenamefont {Basu}\ \emph {et~al.}(2018)\citenamefont {Basu},
  \citenamefont {Majumdar}, \citenamefont {Rosso},\ and\ \citenamefont
  {Schehr}}]{Basu2018}%
  \BibitemOpen
  \bibfield  {author} {\bibinfo {author} {\bibfnamefont {U.}~\bibnamefont
  {Basu}}, \bibinfo {author} {\bibfnamefont {S.~N.}\ \bibnamefont {Majumdar}},
  \bibinfo {author} {\bibfnamefont {A.}~\bibnamefont {Rosso}}, \ and\ \bibinfo
  {author} {\bibfnamefont {G.}~\bibnamefont {Schehr}},\ }\href@noop {}
  {\bibfield  {journal} {\bibinfo  {journal} {Physical Review E}\ }\textbf
  {\bibinfo {volume} {98}},\ \bibinfo {pages} {062121} (\bibinfo {year}
  {2018})}\BibitemShut {NoStop}%
\end{thebibliography}%

\end{document}